\documentclass[aps,onecolumn,preprint,superscriptaddress,nofootinbib,floats]{revtex4}

\usepackage{amsmath,amssymb,color,mathrsfs, graphicx,verbatim,epsfig, bbm, wasysym, axodraw}
\usepackage[hyperfootnotes=false]{hyperref}
\usepackage{slashed}
\usepackage{caption}

\allowdisplaybreaks

\makeatletter
\def\l@subsection#1#2{}
\def\l@subsubsection#1#2{}
\makeatother

\def\lsim{\mathrel{\rlap{\lower4pt\hbox{\hskip1pt$\sim$}}
    \raise1pt\hbox{$<$}}}
\def\gsim{\mathrel{\rlap{\lower4pt\hbox{\hskip1pt$\sim$}}
    \raise1pt\hbox{$>$}}}

\newcommand{\be}{\begin{eqnarray}}
\newcommand{\ee}{\end{eqnarray}}

\def\addresses#1#2{\hbox to \hsize{\@tablebox{#1}\hfil\@tablebox{#2}}}
\def\@tablebox#1{\vtop{\hsize=5in \begin{flushleft} #1 \end{flushleft}}}

\def\beq{\begin{equation}}
\def\eeq{\end{equation}}
\def\bit{\begin{itemize}}
\def\eit{\end{itemize}}
\def\beqa{\begin{eqnarray}}
\def\eeqa{\end{eqnarray}}

\def\mph{m_\phi^2}
\def\mpb{m_{\bar{\phi}}^2}
\def\mcb{m_{\bar{\chi}}^2}

\def\mc{m_\chi^2}
\def\bp{b_{\phi}}
\def\bc{b_{\chi}}

\def\bphi{\bar{\phi}}
\def\bchi{\bar{\chi}}
\def\bxi{\bar{\xi}}
\def\lx{\lambda_\xi}

\def\msc{m_{\rm sc}} 
\def\lf{\frac{1}{16 \pi^2}}
\def\phit{\tilde{\phi}}
\def\Phit{\tilde{\Phi}}
\def\qt{\tilde{q}}
\def\ut{\tilde{u}}
\def\dt{\tilde{d}}
\def\lamu{\lambda_U}
\def\lamub{\bar{\lambda}_U}
\def\lamd{\lambda_D}
\def\lamdb{\bar{\lambda}_D}

\newcommand{\mo}{\mathcal{O}}

\newcommand\blfootnote[1]{%
  \begingroup
  \renewcommand\thefootnote{}\footnote{#1}%
  \addtocounter{footnote}{-1}%
  \endgroup
}

\begin{document}

\baselineskip 0.6cm

\begin{titlepage}

\thispagestyle{empty}

\begin{flushright}
CERN-PH-TH/2014-047
\end{flushright}

\begin{center}

\vskip 2cm

{\Large \bf  Split SUSY Radiates Flavor }

\vskip 1.0cm
{\large  Matthew Baumgart$^{1}$, Daniel Stolarski$^{2,3,4}$, and Thomas Zorawski$^{2}$}
\vskip 0.4cm
{\it $^1$ Department of Physics, Carnegie Mellon University, Pittsburgh, PA 15213} \\
{\it $^2$ Department of Physics and Astronomy, Johns Hopkins University, Baltimore, MD 21218} \\
{\it $^3$ Maryland Center for Fundamental Physics, University of Maryland, \\ College Park, MD 20742} \\
{\it $^4$Theory Division, Physics Department, CERN, CH-1211 Geneva 23, Switzerland}\\
\vskip 1.2cm

\end{center}

\noindent Radiative flavor models where the hierarchies of Standard Model (SM) fermion masses and mixings are explained via loop corrections are elegant ways to solve the SM flavor puzzle. Here we build such a model in the context of Mini-Split Supersymmetry (SUSY) where both flavor and SUSY breaking occur at a scale of 1000 TeV. This model is consistent with the observed Higgs mass, unification, and WIMP dark matter.  The high scale allows large flavor mixing among the sfermions, which provides part of the mechanism for radiative flavor generation. In the deep UV, all flavors are treated democratically, but at the SUSY breaking scale, the 3$^{\rm rd}$, 2$^{\rm nd}$, and 1$^{\rm st}$ generation Yukawa couplings are generated at tree level, one loop, and two loops, respectively. Save for one, all the dimensionless parameters in the theory are $\mathcal{O}(1)$, with the exception being a modest and technically natural tuning that explains both the smallness of the bottom Yukawa coupling and the largeness of the Cabibbo angle. 

\blfootnote{baumgart@cmu.edu \\ daniel.stolarski@cern.ch \\ tz137@pha.jhu.edu}

\end{titlepage}

\tableofcontents

\setcounter{page}{1}

%%%%%%%%%%%%%%%%%%%%%%
\section{Introduction}
\label{sec:intro}
%%%%%%%%%%%%%%%%%%%%%%

\begin{figure}[ht]
\centering
\includegraphics[width=0.5\textwidth]{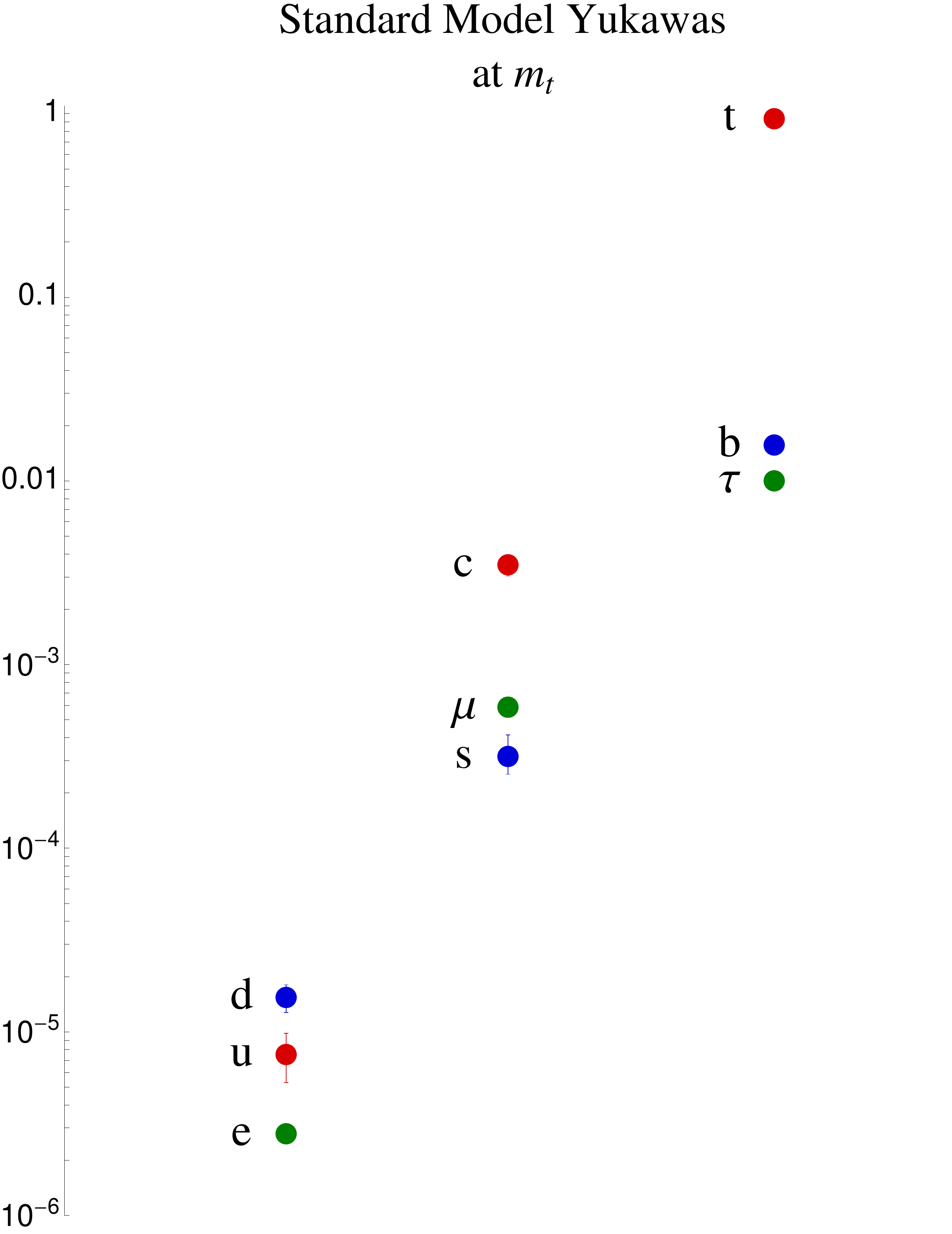}
\captionsetup{justification=raggedright,
singlelinecheck=false}
\caption{We take the running-mass values at the top pole-mass reported in~\cite{Xing:2011aa} and divide by $v$ = 174.1 GeV, as used in \cite{Giudice:2011cg}.}
\label{fig:sm}
\end{figure}

Unlike the gauge sector, the Standard Model (SM) flavor sector has a complicated menagerie of dimensionless parameters whose values differ by orders of magnitude. Furthermore, the patterns of masses and mixings of the SM fermions do not appear random, even on a logarithmic scale; there is a hint of structure that emerges upon close inspection ({\it cf.~}Fig.~\ref{fig:sm}). For example, the masses of the 3$^{\rm rd}$ generation fermions are all much larger than the masses of the 2$^{\rm nd}$ generation fields with the same quantum numbers, which in turn are all much heavier than the corresponding 1$^{\rm st}$ generation fermions. The Standard Model offers no explanation for any of this structure, with the Yukawa couplings simply given as dimensionless inputs.	

One possible explanation for the flavor structure stems from the following observation about, for example, the up, charm, and top quarks
\be
\frac{m_c}{m_t} \simeq \frac{m_u}{m_c} \simeq\mathcal{O}(1) \times \frac{1}{16\pi^2} \; .
\ee
This leads to the idea of radiative flavor breaking~\cite{Ibanez:1981nw,Balakrishna:1987qd,Balakrishna:1988ks,Balakrishna:1988xg,Balakrishna:1988bn,Babu:1989fg,Nilles:1989pd,He:1989er,Rattazzi:1990wu,Babu:1990fr,Appelquist:2006ag,Barr:2007ma,Dobrescu:2008sz,Hashimoto:2009xi,Ibarra:2014fla}, where only the 3$^{\rm rd}$ generation Yukawa couplings are generated at tree level, while the 2$^{\rm nd}$ generation Yukawas are generated as one-loop effects, and the 1$^{\rm st}$ generation is a two-loop effect. This is an old idea that has its origins in trying to explain the electron mass as a loop effect of the muon mass~\cite{Georgi:1972mc,Weinberg:1972ws,Georgi:1972hy,Barr:1976bk}. This framework not only explains the cascading down of the masses in different generations, but it can also easily be embedded into a UV theory where all SM fields are treated democratically so that different symmetry charges need not be given to the different fields. 

Supersymmetric  theories of radiative flavor generation~\cite{Ibanez:1982xg,Hall:1985dx,Banks:1987iu,Kagan:1989fp,Babu:1989tv,Dobrescu:1995gz,ArkaniHamed:1995fq,ArkaniHamed:1996zw,Liu:2005rs,Graham:2009gr,Crivellin:2011sj,Conlon:2011ac} can incorporate many of the usual advantages to supersymmetry (SUSY), including a natural dark matter candidate and improved gauge coupling unification. In the context of radiative flavor generation, SUSY has additional advantages. The non-renormalization theorems for the superpotential~\cite{Salam:1974jj,Grisaru:1979wc} mean that radiative corrections cannot generate new operators such as 1$^{\rm st}$ and 2$^{\rm nd}$ generation Yukawa couplings. This forces flavor and SUSY breaking to be tied together, likely giving a common scale to both phenomena.  In addition, SUSY requires the theory to include an additional set of particles which transform under flavor, the sfermions. While non-supersymmetric theories of radiative flavor generation require introducing a host of new fields, SUSY models are potentially more economical because the sfermions can contribute to generating the flavor hierarchy. 

In order to use the sfermions to generate flavor, there must be large flavor breaking in the sfermion sector. Unfortunately, if sfermions are at the weak scale, low energy flavor tests require them to be nearly flavor diagonal~\cite{Martin:1997ns}, a difficulty encountered by many of the early attempts to build such a model~\cite{Ibanez:1982xg,Hall:1985dx,ArkaniHamed:1996zw}. Because the Yukawa couplings are dimensionless parameters, they are quite insensitive to the scale at which they are generated. On the other hand, the flavor observables that constrain the flavor breaking in the sfermion sector correspond to higher dimension operators, so they decouple quickly with heavier sfermion masses. Therefore, spectra where the sfermions are much above the weak scale such as Split~\cite{Wells:2004di,ArkaniHamed:2004fb,Giudice:2004tc} and Supersplit~\cite{Fox:2005yp,Hall:2009nd,Giudice:2011cg} Supersymmetry can be used for radiative flavor generation with sfermions potentially as  heavy as the GUT or Planck scale~\cite{Graham:2009gr}. 

\begin{figure}
\centering
\includegraphics[width=0.5\textwidth]{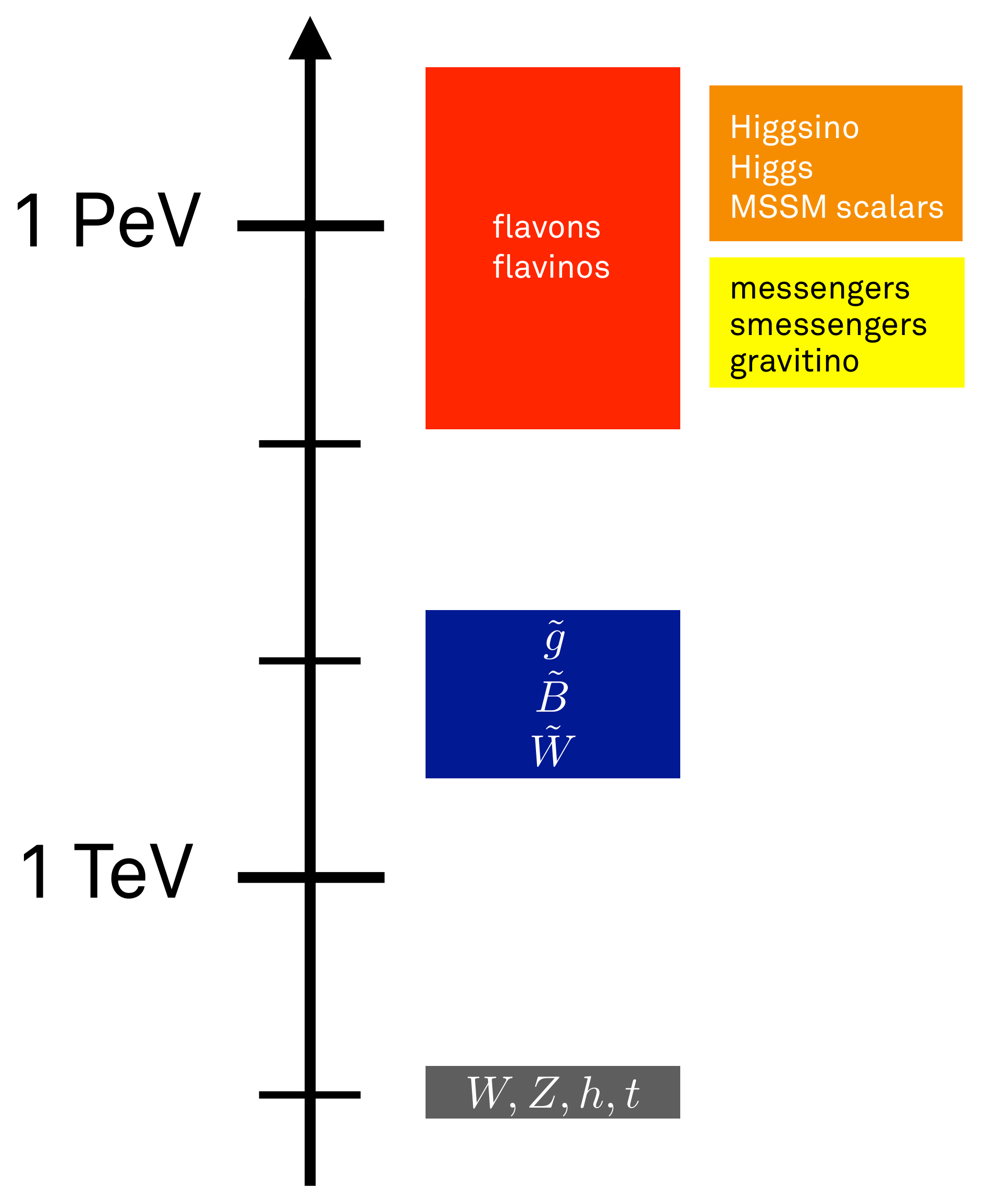}
\captionsetup{justification=raggedright,
singlelinecheck=false}
\caption{The spectrum of the model presented on a log scale. The heaviest known SM particles are at the bottom around 100 GeV. The gauginos are at the 10 TeV scale with the gluino typically heaviest and the Wino typically lightest and closer to 3 TeV. The rest of the spectrum is roughly at the PeV (= 1000 TeV) scale, but they are typically spread out over a couple of decades in mass. As discussed in Sec.~\ref{sec:details}, the messengers mix with the squarks and sleptons.}
\label{fig:broadspectrum}
\end{figure}

Motivated by the lack of evidence for SUSY at the LHC and the discovery of a Higgs with mass near 125 GeV~\cite{Aad:2012tfa,Chatrchyan:2012ufa}, theories with sfermions much above the weak scale have received renewed interest~\cite{Hall:2011jd,Kane:2011kj,Ibe:2011aa,Ibe:2012hu,Bhattacherjee:2012ed,Arvanitaki:2012ps,Hall:2012zp,ArkaniHamed:2012gw}. For reasons that will be reviewed in Sec.~\ref{sec:split}, the data points to a sfermion mass scale of $m_{\rm sc} \sim 1000$ TeV, and we refer to this framework as Mini-Split SUSY~\cite{Arvanitaki:2012ps}. In this work, we build a model where the SM fermion masses are generated radiatively in a Mini-Split setup. The spectrum is outlined in Fig.~\ref{fig:broadspectrum}: the MSSM scalars as well as all the additional ingredients needed for the model are at the scale $m_{\rm sc}$, while gauginos are significantly lighter, around 10 TeV. 

\begin{figure}
\centering
\includegraphics[width=0.5\textwidth]{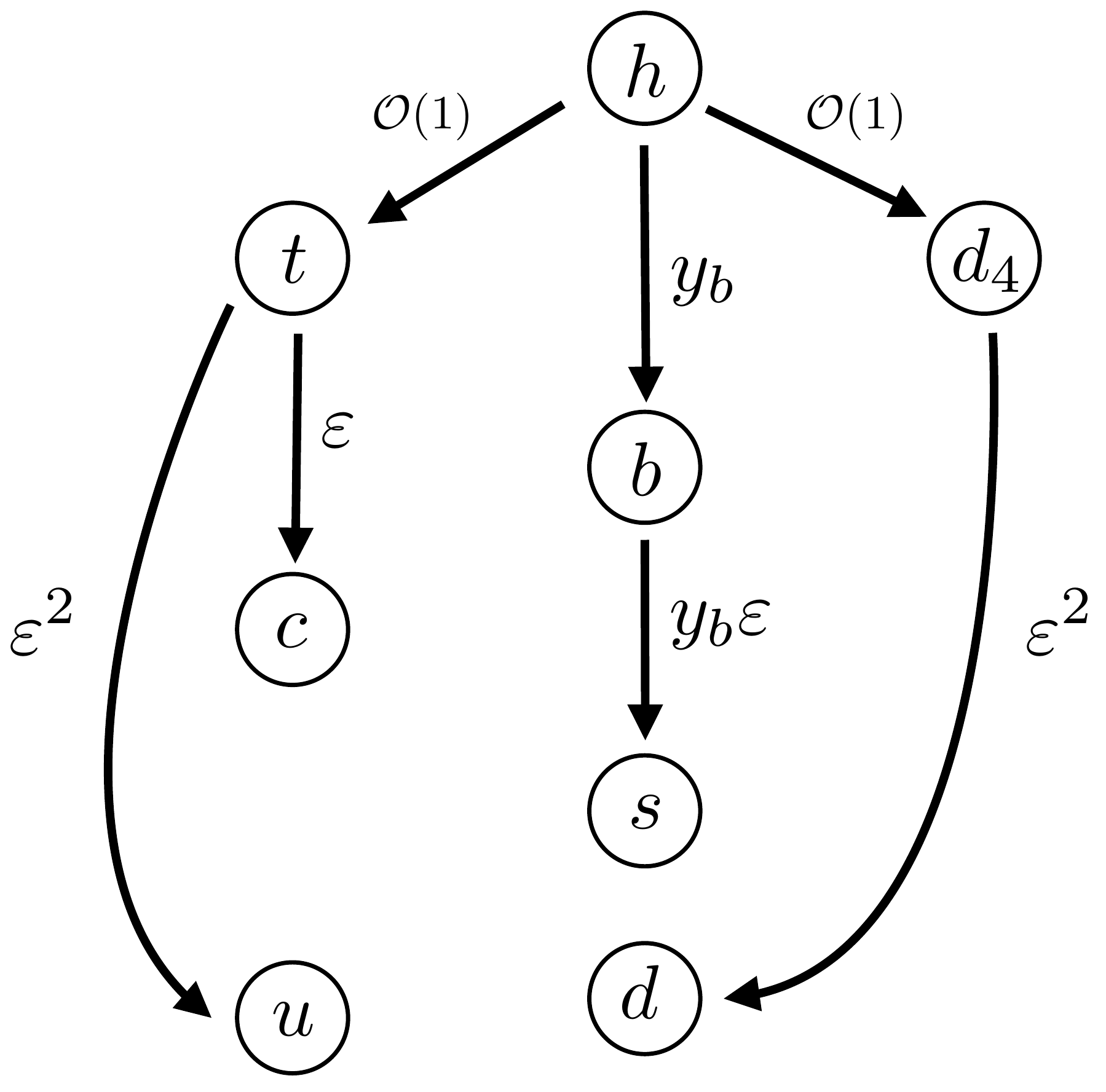}
\captionsetup{justification=raggedright,
singlelinecheck=false}
\caption{A schematic representation of the model given in this work. The top and $d_4$ fields have $\mathcal{O}(1)$ couplings to the Higgs, while the coupling of the $b$ is somewhat smaller. The 2$^{\rm nd}$ generation gets one-loop couplings from the 3$^{\rm rd}$ generation with $\varepsilon$ being a loop factor. The top and $d_4$ seed Yukawa couplings for the up and down which are parametrically two-loop size.  }
\label{fig:schematic}
\end{figure}

The flavor model has a $U(1)_F$ symmetry in the UV which forbids the Yukawa couplings.  However, unlike previous models, all of the SM matter multiplets are neutral under this symmetry, with only the Higgs fields being charged. Therefore, the UV theory treats all the SM fields democratically, and no special charges are needed for the different generations. SUSY breaking occurs at the scale $m_{\rm sc}$ and seeds spontaneous $U(1)_F$ breaking. This allows tree level Yukawa couplings to be generated for the 3$^{\rm rd}$ generation fermions. The relative smallness of the bottom and tau Yukawa couplings to the top Yukawa comes from a modest and technically natural tuning, but this is the only hierarchy not automatically explained by this model. Radiative corrections from the $U(1)_F$ breaking sector generate one-loop Yukawa couplings for the 2$^{\rm nd}$ generation. Finally, the 1$^{\rm st}$ generation Yukawas are generated by two-loop diagrams of sfermions which have large flavor breaking in their SUSY-breaking masses. A schematic representation of the fermion mass hierarchies is given in Fig.~\ref{fig:schematic}. The CKM matrix also has the right structure, with the small parameter required for a small bottom Yukawa being the reason that the Cabibbo angle is larger than a loop factor. Finally, this model preserves the predictions of gauge coupling unification and dark matter of Mini-Split SUSY. 

The organization of this paper is as follows. In Sec.~\ref{sec:split}, we review the motivation and spectrum of the Mini-Split SUSY framework. In Sec.~\ref{sec:model} we describe our model and give the parametric sizes of elements of the Yukawa matrices and translate those into the SM fermion masses and mixing angles. In Sec.~\ref{sec:details}, we calculate the predictions of the model in detail including the spectrum of fields at $m_{\rm sc}$ as well as the SM fermion masses and mixings. We also present a benchmark point in parameter space which faithfully reproduces SM flavor observables ({\it cf.}~Figs.~\ref{fig:smVsUs} and \ref{fig:ckm}). In Sec.~\ref{sec:constraints} we describe the constraints on the model and potential future phenomenology, and we conclude in Sec.~\ref{sec:conc}.

%%%%%%%%%%%%%%%%%%%%%%
\section{Review of Mini-Split SUSY}
\label{sec:split}
%%%%%%%%%%%%%%%%%%%%%%

The discovery of a Higgs-like state with a mass near 125~GeV~\cite{Aad:2012tfa,Chatrchyan:2012ufa} and the absence of any direct evidence for superpartners at the LHC has led to a reconsideration of the paradigm of weak-scale supersymmetry~\cite{Hall:2011jd,Kane:2011kj,Ibe:2011aa,Ibe:2012hu,Bhattacherjee:2012ed,Arvanitaki:2012ps,Hall:2012zp,ArkaniHamed:2012gw}. Raising the scalar masses far above the weak scale introduces significant tuning in the Higgs mass relative to weak-scale SUSY. In exchange for the loss of naturalness, we get a much simpler explanation of various phenomena, including SUSY breaking and its communication to the MSSM fields. In this section we will briefly review the basic setup envisioned in, {\it e.g.},~\cite{ArkaniHamed:2012gw}, and describe why this leads to more straightforward models.

We consider a SUSY breaking scale parametrized by a gravitino mass $m_{3/2}$. If the field that breaks SUSY is denoted as $X$, then the K\"{a}hler potential contains Planck-suppressed higher dimensional operators of the form 
\begin{equation}
\int d^4\theta \frac{1}{M_{\rm Pl}^2}  X^\dagger X \Phi^\dagger \Phi,
\label{eq:soft-scal}
\end{equation} 
where $\Phi$ is a MSSM matter superfield.  These yield scalar masses which are parametrically 
\begin{equation}
m_{\rm sc} \sim m_{3/2}.
\end{equation} 
In generic models of SUSY breaking, $X$ is not a total singlet and carries either gauge or global charge. Therefore, the gaugino mass operator $X W_\alpha W^\alpha/M_{\rm pl}$ is forbidden and the leading contribution to gaugino masses comes from anomaly mediation~\cite{Giudice:1998xp,Randall:1998uk}. This gives gaugino masses that are parametrically 
\be
m_{1/2} \sim \frac{g^2}{16\pi^2} m_{3/2},
\label{eq:gaugino}
\ee
where $g$ is the relevant gauge coupling. Similar arguments show that $\mathbf{a}$-terms are only generated at loop level and are thus insignificant for computation of the spectrum. 

For the Higgs sector, we can write down operators of the form 
\be
\int d^4\theta \left(1+\frac{X^\dagger X}{M_{\rm Pl}^2}\right) (H_u H_d + \rm{h.c.}) \,.
\label{eq:mu-bmu}
\ee
 This generates not only the SUSY breaking $B_\mu$-term but also the supersymmetric $\mu$-term through the Giudice-Masiero mechanism~\cite{Giudice:1988yz}. They are of order 
\be
\mu^2 \sim B_\mu \sim m_{3/2}^2 .
\label{eq:muB}
\ee
This differs from the original Split SUSY construction~\cite{Wells:2004di,ArkaniHamed:2004fb,Giudice:2004tc} where the $\mu$-term and thus the Higgsinos had mass of order the gaugino masses. In this model, all the parameters in the Higgs potential are $\mathcal{O}(m_{3/2})$, and among them one tuning is required to get the Higgs vacuum expectation value (vev) and the mass of the lightest physical scalar to be of order the weak scale. The remainder of the Higgs states and the Higgsinos all have mass $\mathcal{O}(m_{3/2})$.

The spectrum described above is shown in Fig.~\ref{fig:broadspectrum}, but we have chosen the scale $m_{3/2} \sim 1000$ TeV.  Having scalars at this scale gives a number of interesting results. First, radiative corrections from the heavy stops raise the Higgs mass above the tree level bound of $m_Z$. The Higgs mass is logarithmically sensitive to the scalar masses, so there is a wide range of stop mass which can give the observed Higgs mass, but for $\tan\beta$ being $\mathcal{O}($a few), the stop loop can raise the Higgs mass to its experimentally measured value, and a detailed computation is shown in Fig.~3 of~\cite{ArkaniHamed:2012gw}. This size for $\tan\beta$ follows naturally from having all soft scalar masses come from the same source, as in Eq.~\eqref{eq:soft-scal}.  We thus expect $m_{H_u}^2 \sim m_{H_d}^2$, which predicts the moderate $\tan\beta$ needed for the measured Higgs mass to be consistent with our chosen mass scale for the stops. 

A second feature of scalars around 1000 TeV is that gaugino masses are of order 10 TeV, with their spectrum well predicted by anomaly mediation plus the threshold corrections that arise from the messenger and Higgsino sectors. This scale is mostly unexplored by current collider searches, but is within reach of the next generation of energy frontier experiments. The LSP in this framework is the lightest gaugino, and, because $R$-parity is conserved, it is stable. Pure anomaly mediation predicts a wino LSP, which is a viable candidate for WIMP dark matter with its relic abundance matching the observed value for dark matter if its mass is around 3 TeV~\cite{Hisano:2006nn}. This dark matter candidate is becoming constrained by indirect detection experiments~\cite{Cohen:2013ama,Fan:2013faa,Hryczuk:2014hpa}, but it is still a possibility. In Sec.~\ref{sec:gaugino}, we will discuss the full gaugino mass spectrum in the presence of threshold corrections, and we will explore the possibilities for dark matter in more detail.

A third advantage of this scale is that gauge coupling unification works as well as in the MSSM. As the scalars come in complete SU(5) multiplets, $m_{\rm sc}$ has little effect on unification. On the other hand, the gauginos and Higgsinos are not in complete SU(5) representations, so their masses can have a strong effect. Split SUSY with a light $\mu$ was previously shown to unify well~\cite{ArkaniHamed:2004fb}, and raising $\mu$ in accordance with Eq.~\eqref{eq:muB} is still consistent with unification~\cite{ArkaniHamed:2012gw}.

The fourth feature of scalars at the  PeV scale (PeV = $10^{15}$ eV) is that the SUSY flavor problem is nearly nonexistent. The soft masses for $Q, U, D, L, E$ in the MSSM are $3\times 3$ matrices in flavor space, but if those matrices have generic weak scale entries then the model will be badly ruled out by low energy flavor constraints. A similar statement can be made for $\mathbf{a}$-terms. In order for weak-scale SUSY to be viable, the soft mass matrices must either be nearly proportional to the unit matrix~\cite{Martin:1997ns}, or approximately aligned with the SM Yukawa matrices~\cite{Nir:1993mx}. Models such as gauge mediation~\cite{Dine:1981gu,AlvarezGaume:1981wy,Dimopoulos:1982gm,Dine:1994vc,Dine:1995ag} solve this SUSY flavor problem for weak-scale sfermions, but when the mass of the scalars is raised, it is ameliorated as well. This is because the effects of squarks and sleptons decouple from low energy flavor experiments like $1/m_{\rm sc}^{n}$, where $n$ is a positive integer that depends on the process. Detailed studies~\cite{McKeen:2013dma,Moroi:2013sfa,Eliaz:2013aaa,Altmannshofer:2013lfa} have recently confirmed that PeV scale SUSY is safe from nearly all low energy processes, with Kaon mixing and proton decay being notable exceptions whose treatment requires more attention. We will discuss the bounds in detail in Sec.~\ref{sec:constraints}, but we will take that the soft masses for the matter partners as anarchic in flavor space. 

Because of these advantages, we find that Mini-Split SUSY is an interesting laboratory. In particular, the allowed large flavor mixing in the scalar sector provides a mechanism to build a model which explains the SM flavor structure through radiative corrections. While the anarchic flavor structure of the scalars can generate the 1$^{\rm st}$ generation masses at loop level from those of the 3$^{\rm rd}$ generation \cite{ArkaniHamed:2012gw}, more structure is necessary to generate the full SM spectrum. In the following sections, we will present such a model.

%%%%%%%%%%%%%%%%%%%%%%
\section{A Model of Flavor}
\label{sec:model}
%%%%%%%%%%%%%%%%%%%%%%

In this section we give a schematic description of the model and describe the parametric sizes of the SM flavor parameters. We show the full spectrum in Fig.~\ref{fig:broadspectrum}, present a benchmark in Sec.~\ref{sec:details}, and the details of the calculations in the Appendices. We begin by describing the dynamics needed for the up sector, and we will cover the rest of the SM fermions in subsequent sections.

\subsection{Up Sector}

Our model is an extension of the MSSM with the spectrum broadly described in Sec.~\ref{sec:split}. The basic premise is that the hierarchy of masses between generations is a hierarchy in the number of loops.  Crucial to the setup is a means to forbid tree-level Higgs Yukawas to all but the 3$^{\rm rd}$ generation.  Satisfying this criteria, we must then ensure the remaining chiral symmetries are broken in stages to parametrically separate the first two generations.  Furthermore, the different couplings of the generations occur solely as a consequence of linear algebra.  We make no {\it ad hoc} or symmetry-based distinctions between them.  To prevent Yukawa couplings at tree level, we add a new $U(1)_F$ gauge group under which the Higgs superfields are charged, but all other MSSM fields are neutral. We discuss complications associated with a new gauge group such as anomalies in App.~\ref{app:field}. For the up sector, we also introduce one additional generation of vectorlike messenger quarks, $Q$, $\bar Q$ and $U$, $\bar U$ which have $U(1)_F$ charges such that a primordial Yukawa coupling $\lambda_U Q U H_u$ can be written down.  The set of fields needed for the up sector as well as their charges are given in Tab.~\ref{tab:ufields}.

\begin{table}[tp]
\captionsetup{justification=raggedright,
singlelinecheck=false}
\caption{Charge assignments of the Higgs and up-sector messenger fields. Here $R_p$ denotes the usual $R$-parity.  Note that the MSSM fields $q$ and $u$ fields are neutral under $U(1)_F$.}
\begin{center}
\begin{tabular}{|c|c|c|c|} \hline
Field & U(1)$_F$ & $SU(3)\times SU(2)\times U(1)$ & $R_p$\\ \hline
$H_u,\, H_d$ & $\mp2$ & $ (\mathbf 1, \mathbf 2)_{1/2}+ (\mathbf 1, \mathbf 2)_{-1/2} $ & $+$ \\ \hline
$Q$, $\bar Q$ & $\pm 1$ & $ (\mathbf 3, \mathbf 2)_{1/6}+ (\bar{\mathbf 3}, \mathbf 2)_{-1/6}$ & $-$ \\ 
$U$, $\bar U$ & $\pm 1$ & $ (\bar{\mathbf 3}, \mathbf 1)_{-2/3}+ (\mathbf 3, \mathbf 1)_{2/3}$  & $-$ \\ \hline
\end{tabular}
\end{center}
\label{tab:ufields}
\end{table}

In order to generate any Yukawa couplings, we need to spontaneously break $U(1)_F$. This requires the introduction of ``flavon'' fields shown in Tab.~\ref{tab:flavons}. As described in App.~\ref{app:flavon}, we need each of these fields in order to get a potential that generates the flavon vevs required for SM Yukawas. The flavons get soft masses from the same mechanisms as the MSSM matter. Taking $\phi$ as an example, the soft terms are given by
\be
V_{\phi}^{\rm soft} = \frac{1}{2} (m_\phi^2)_{ij}\, \phi_i^\dagger \phi_j +  \frac{1}{2} (m_{\bar \phi}^2)_{ij}\, {\bar \phi}_i^\dagger {\bar \phi}_j 
                              - (b^\phi_{ij}\, \phi_i \bar \phi_j \, +\, {\rm h.c}) \, .
\label{eq:Vsoft}
\ee
Once we include the $D$-terms arising from $U(1)_F$, the flavon scalar potential is analogous to the Higgs potential in the MSSM, so there is a large region of parameter space that can be chosen such that all the $\phi$ fields acquire vevs. Since all the dimensional parameters in the potential are of the same order, we naturally get $\langle \phi_i \rangle \sim m_{\rm sc}$. 
The potential minimization is described in more detail in App.~\ref{app:flavon}. 

\begin{table}[tp]
\captionsetup{justification=raggedright,
singlelinecheck=false}
\caption{The set of flavon fields needed to break $U(1)_F$, along with their charge assignments. }
\begin{center}
\begin{tabular}{|c|c|c|c|} \hline
Field & U(1)$_F$ & $SU(3)\times SU(2)\times U(1)$ & $R_p$\\ \hline
$\phi_{1,2}$, $\bar \phi_{1,2}$ & $\pm1$ & $ (\mathbf 1, \mathbf 1)_{0}$ & $+$ \\ 
$\chi_{1,2}$, $\bar \chi_{1,2}$ & $\mp3$ & $ (\mathbf 1, \mathbf 1)_{0}$  & $+$ \\ 
$\xi$, $\bar \xi$ & $\mp2$ & $ (\mathbf 1, \mathbf 1)_{0}$  & $+$ \\ \hline
\end{tabular}
\end{center}
\label{tab:flavons}
\end{table}

From the field content of Tabs.~\ref{tab:ufields} and~\ref{tab:flavons}, we can write down a general superpotential
\begin{eqnarray}
W_{\rm up} &=& \lambda_U Q\, U\, H_u + \bar{\lambda}_U \bar{Q}\, \bar{U}\, H_d +  f^q_{ij} \,q_i \,\bar Q \,\phi_j +  f^u_{ij} \,u_i \,\bar U \,\phi_j \nonumber\\
&& + \, \mu_Q\, Q \bar Q + \mu_U\, U \bar U + \mu \,H_u H_d + \mu^\phi_{ij} \phi_i \bar \phi_j \, ,
\label{eq:Wup}
\end{eqnarray}
where we have ignored the interactions of the $\chi$ and $\xi$ flavons for now. The $f$ couplings have flavor indices, but the $\lambda$ couplings to the Higgs are just numbers. The $\mu$-terms are all of order $m_{\rm sc}$ and are generated via the dynamics of Eq.~\eqref{eq:mu-bmu}, so all the states described in Tabs.~\ref{tab:ufields} and~\ref{tab:flavons} will have mass $\mathcal{O}(m_{\rm sc})$.

\subsubsection{Top Yukawa}
\label{subsubsec:top}

\begin{figure}
\centering
\includegraphics[width=0.5\textwidth]{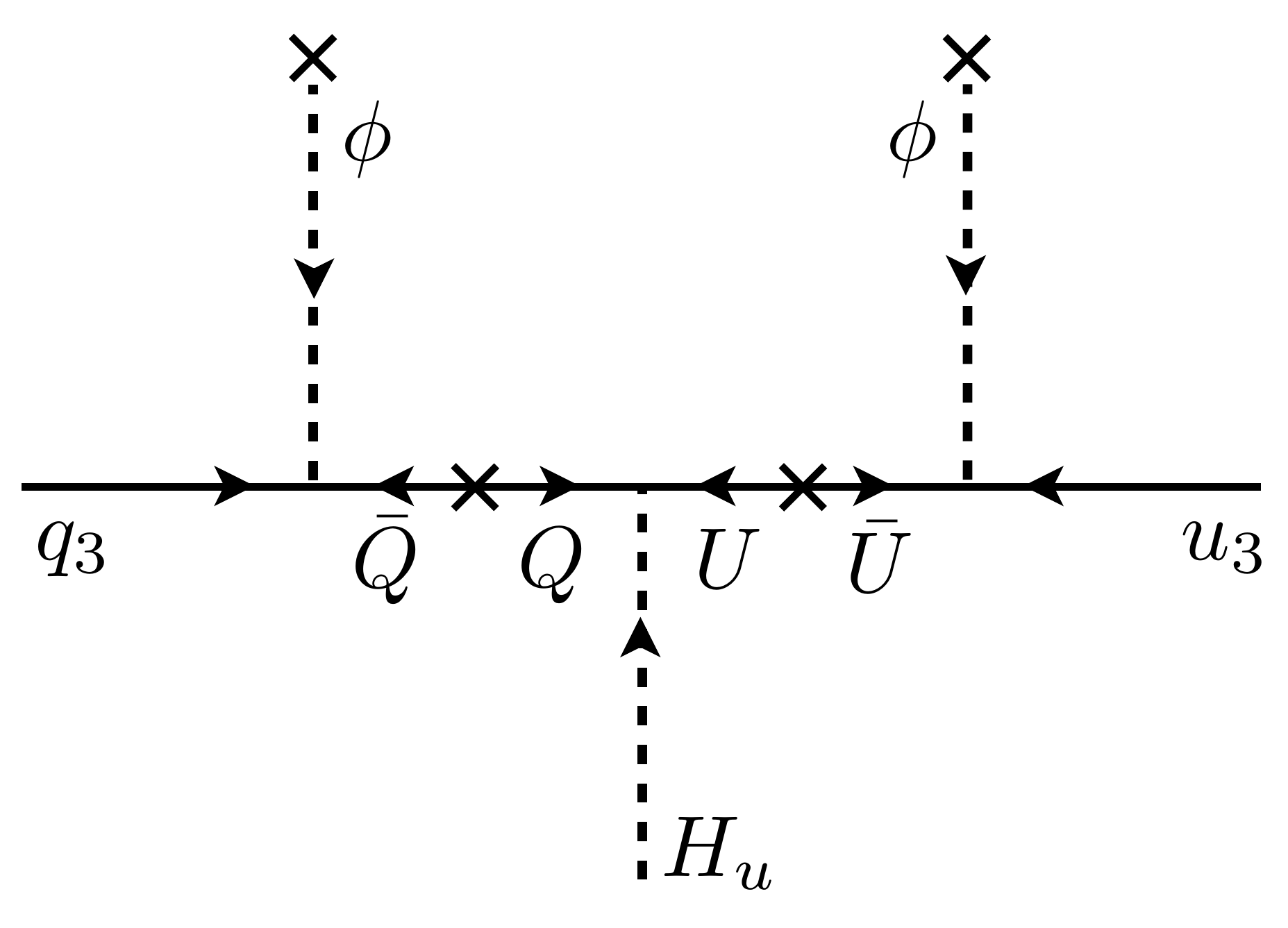}
\caption{Feynman diagram for generating the top Yukawa coupling.}
\label{fig:top}
\end{figure}

With these ingredients, we can generate a top Yukawa coupling at tree level, with all other Yukawas still zero. This arises from the messenger exchange diagram in Fig.~\ref{fig:top}. When $U(1)_F$ is broken by the vev of $\phi$, the $f$ couplings in Eq.~\eqref{eq:Wup} generate a mixing between the MSSM-like fields and the heavy vectorlike fields. The $f$ couplings have an index in $\phi$ doublet space as well as an index in flavor space. We can choose our $\phi$ basis such that only $\phi_1$ gets a vev and $\langle \phi_2 \rangle=0$. If we set $\phi$ to its vev and ignore interactions of the propagating $\phi$ for now, we see that $f^q$ and $f^u$ are just column vectors, so they are both rank 1. We can thus choose bases for $q_i$ and $u_i$ such $f^q_{ij}$ and $f^u_{ij}$ are only non-zero in the ``3'' direction in flavor space. This basis now defines the top quark. It is the only up-type quark to mix with the vectorlike quarks.  Thus, Fig.~\ref{fig:top} only generates a top Yukawa coupling. This mechanism, which is similar to that of previous works such as~\cite{Balakrishna:1988ks,Balakrishna:1988xg,Balakrishna:1988bn,Kagan:1989fp,Dobrescu:2008sz}, allows a UV theory where all the SM quarks are treated democratically to generate only a top Yukawa coupling at tree level. 

To calculate the top Yukawa from the interactions given in Eq.~\eqref{eq:Wup}, we need to rotate the fields as described above.  We can make the schematic argument of the previous paragraph more rigorous as follows: without loss of generality, we can use the $U(3)_q,\, U(3)_u$ symmetries that exist in the limit of zero $f^q,\, f^u$ couplings to remove any interaction between the 1$^{\rm st}$ generation quarks and the flavons. In a generic basis,  both $\phi_1$ and $\phi_2$ get vevs and we use the residual $U(2)$ symmetries to decouple the 2$^{\rm nd}$ generation $q$ and $u$ fields from them.  For example, from the original $q_2$ and $q_3$, we get
\be
q'_3 = \frac{f^q_{22}\, \phi_2 \,q_2 + (f^q_{31}\, \phi_1 + f^q_{32} \,\phi_2) q_3}{\sqrt{(f^q_{22}\, \phi_2)^2 + (f^q_{31} \,\phi_1 + f^q_{32}\, \phi_2)^2}},
\ee
and $q'_2$ is the orthogonal linear combination. Here and throughout we will use the name of a field to represent a vev when the context is clear.   We can now define the following two matrices $R$ and $F$
\be
q_i = R_{ij} q'_j \qquad F^q_{ij} = R^\dag_{ik} f^q_{kj},
\label{eq:Fdefine}
\ee
where $R$ is the matrix that rotates between the interaction and mass eigenbases for the quarks, and $F$ is the rotation of the $f$ couplings into the mass basis.  We make an analogous rotation for the $u$ fields and their couplings. By construction, only $q'_3$ couples to the flavon vevs.  Since this defines the 3$^{\rm rd}$ generation, we drop the $'$ notation for this post-rotation state hereafter.

Having performed the appropriate rotations on the quarks and $f$ couplings, we are at last in position to calculate the contribution to the top Yukawa from Fig.~\ref{fig:top}, getting
\beq
y_t = \frac{\lambda_U \, F^q_{3i} F^u_{3j} \, \phi_i \phi_j}{\mu_Q \mu_U}.
\label{eq:topyukpre}
\eeq
We see that for dimensionless factors of $\mo(1)$ and all dimensionful factors of the same order, $\sim \msc$, we get an $\mo(1)$ top Yukawa.  Of course, if there are no hierarchies in the parameters in Eq.~\eqref{eq:topyukpre}, then calculating $y_t$ requires us to go beyond the double-vev insertion approximation of Fig.~\ref{fig:top}.  Rather, after $U(1)_F$ symmetry breaking, we need to fully diagonalize the $q_3-Q$ and $u_3-U$ mass eigenstates.  We save the details of this discussion for Sec.~\ref{subsec:wfrn}, but we stress that a full treatment of the top Yukawa maintains its $\mo(1)$ parametric size.

\subsubsection{Charm Yukawa}
\label{subsubsec:charm}

\begin{figure}
\centering
\includegraphics[width=0.9\textwidth]{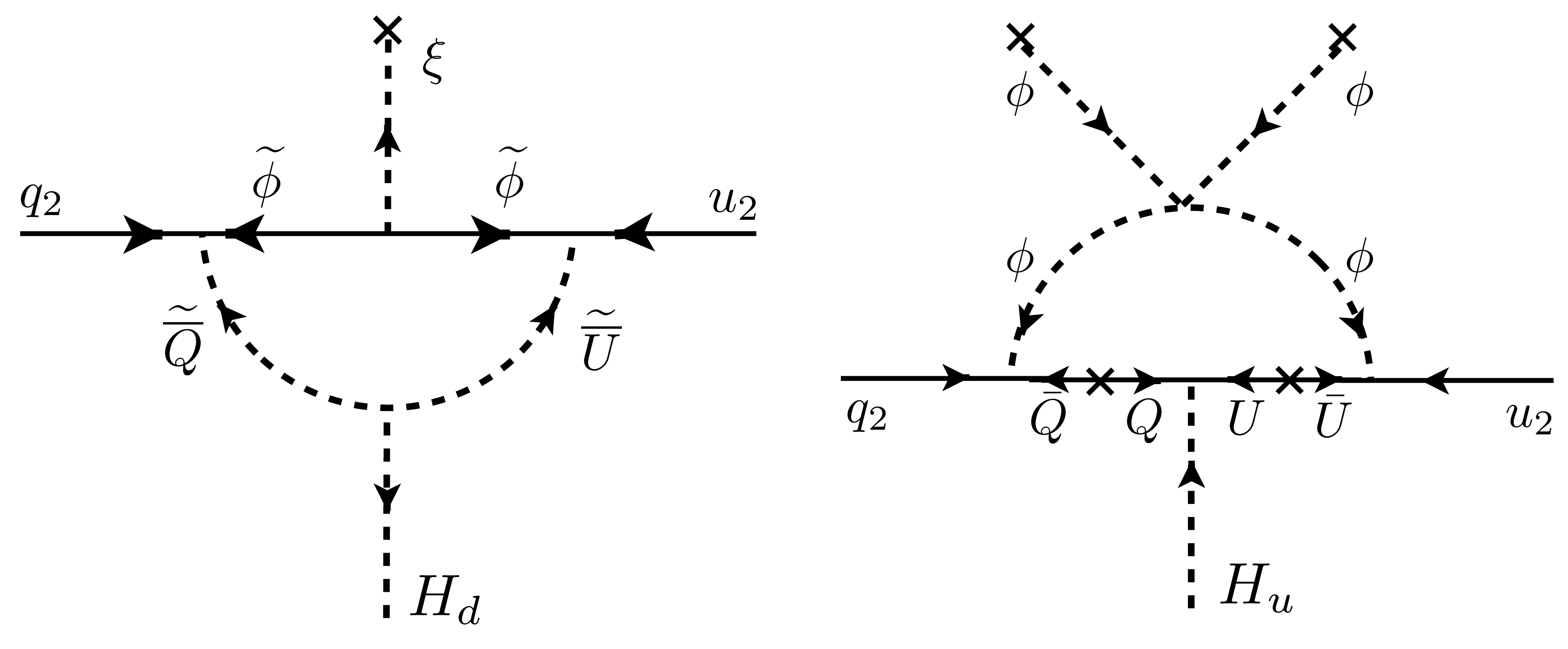}
\captionsetup{justification=raggedright,singlelinecheck=false}
\caption{Feynman diagrams for generating the charm Yukawa coupling. We use the convention that fields which get vevs such as $\phi$ and $H_u$ have tildes over their fermions, while fields which do not get vevs such as $Q$ and $u$ have tildes over their scalar components. }
\label{fig:charm}
\end{figure}

The $U(1)_F$-breaking dynamics which generate a tree-level top Yukawa coupling also generate a charm Yukawa at one loop. This occurs through the two processes shown in Fig.~\ref{fig:charm}.  These two diagrams contain the same superpotential $f^q$ and $f^u$ couplings from Eq.~\eqref{eq:Wup}, but we must perform a SUSY rotation to get from the flavon-messenger diagram to the flavino-smessenger diagram.  While the $\langle \phi \rangle$ can be rotated so it only points in one direction, there are still two propagating fields, so we can define the 2$^{\rm nd}$ generation of quarks as the linear combination that does not couple to $\langle \phi \rangle$ but does couple to the propagating $\phi$. This defines the 1$^{\rm st}$ generation as the quark which does not couple to $\phi$ at all. 

Because of the $U(1)_F$ gauge symmetry, there is a $D$-term of the form 
\be
V^D = \frac{g_F^2}{2} \left( \phi_i^\dagger \phi_i - \bar \phi_i^\dagger \bar\phi_i + ... \right)^2 \, .
\label{eq:phiD}
\ee
This generates a flavon four-point coupling, allowing us to draw the diagram on the right side of Fig.~\ref{fig:charm}. This diagram must connect a $\phi$ that does not get a vev to one that does in order to generate a charm mass. This can only happen if there is misalignment between the basis where the vev points in a single direction and the basis where the mass matrix is diagonal. While this generically occurs for our flavon potential, the size of the flavon-messenger diagram is suppressed by this misalignment. 

Thus, we need to construct the supersymmetrized version of this diagram, a flavino-smessenger diagram, to get a charm Yukawa of the right size. Clearly this is only possible if the flavino $\tilde{\phi}$ has a Majorana mass. As we can see from Tab.~\ref{tab:flavons}, the following superpotential operators are allowed
\be
W = \lambda_{ij} \phi_i\phi_j \xi + \bar{\lambda}_{ij} {\bar \phi}_i {\bar \phi}_j \bar{ \xi} + 
\lambda'_{ij} \phi_i \chi_j \bar{\xi} + \bar{\lambda}'_{ij} \bar{\phi}_i \bar{\chi}_j  \xi\; .
\label{eq:pureflavonsup}
\ee
These generate the desired flavino mass if $\xi$ gets a vev. This is the mechanism shown on the left side of Fig.~\ref{fig:charm}, which turns out to be the dominant contribution to the charm mass and justifies the inclusion of the $\xi$ flavon in the theory. As we will show in App.~\ref{app:flavon}, the $\xi$ and $\bar{\xi}$ flavons serve several other important functions, which explains the flavon content of Tab.~\ref{tab:flavons}. 

The diagrams in Fig.~\ref{fig:charm} also generate Yukawa couplings of the form $q_3 u_2$ and $q_2 u_3$ which are parametrically one loop. They also give small corrections to the top Yukawa coupling.  We will give a detailed description of the computation of the one-loop Yukawa couplings in App.~\ref{app:ylist}, with the dominant contribution to charm given in Eq.~\eqref{eq:finolpup}.

\subsubsection{Up Yukawa}
\label{subsubsec:charm}

\begin{figure}
\centering
\includegraphics[width=0.5\textwidth]{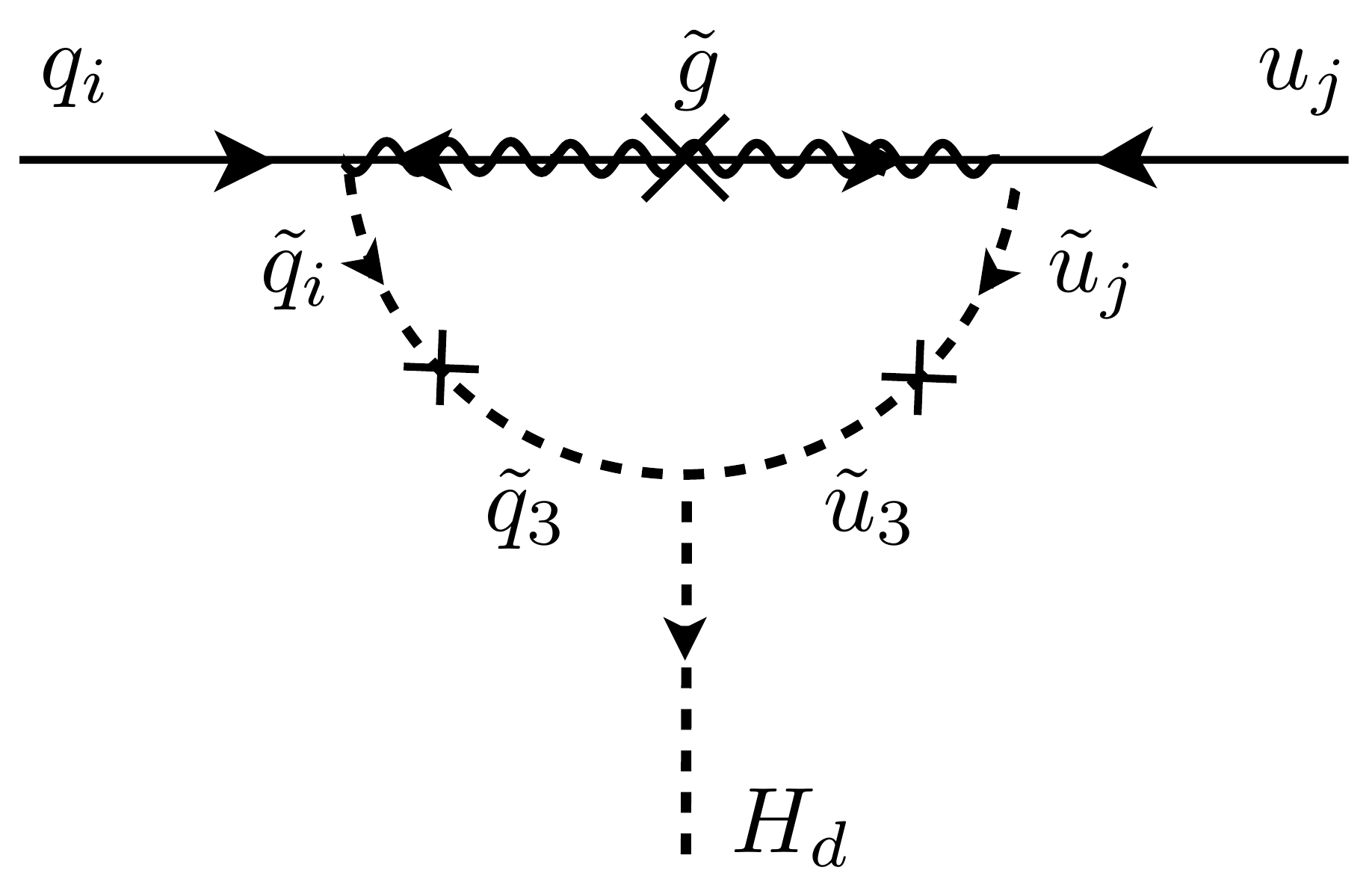}
\caption{Feynman diagram for generating the up Yukawa coupling. }
\label{fig:up}
\end{figure}

Finally, we can generate an up quark Yukawa coupling and fill out the rest of the Yukawa matrix through the diagram in Fig.~\ref{fig:up}. It was pointed out in~\cite{ArkaniHamed:2012gw} that these diagrams have the correct parametric size to generate the up quark mass, and we utilize this here. This diagram is one loop, but it has a chirality flip coming from the gluino mass rather than a primordial Yukawa coupling used in the processes of Figs.~\ref{fig:top} and~\ref{fig:charm}. Therefore, this diagram will be suppressed by $m_{\tilde{g}}/m_{\rm sc}$, which from Eq.~\eqref{eq:gaugino} is a loop factor. Therefore, the up Yukawa coupling generated by the diagram in Fig.~\ref{fig:up} is parametrically of two-loop size. 

The coupling to the Higgs still comes from the top Yukawa coupling, but here we use the fact that the squark soft masses are anarchic in flavor space as the source of flavor breaking. In the mass insertion approximation~\cite{Hall:1985dx}, one can imagine $\tilde{q}_3$ and $\tilde{u}_3$ coupling to the Higgs, and then each being converted to a different flavor by an $\mathcal{O}(1)$ mass insertion. Fig.~\ref{fig:up} is drawn in this way, but for truly anarchic mixing, a better picture is that the squarks that couple  to the Higgs have couplings between the gluino and all three flavors of quarks. Here we see that it is crucial that the mass of the squarks be much above the weak scale, because if not, the mass insertion would be constrained to be small, and the loop diagram in Fig.~\ref{fig:up} would be too small to generate the up mass.  The expression for the up Yukawa and related mixing diagrams is given in Eq.~\eqref{eq:ginolpup}, where flavon vev insertions are summed to all orders to diagonalize the squark-smessenger masses.  

Thus we see that with the fields introduced in Tabs.~\ref{tab:ufields} and~\ref{tab:flavons}, we can get an up-type Yukawa matrix which is parametrically of the form
\be  
\mathbf{y^u} \sim \left( \begin{array}{ccc}
\varepsilon^2 & \varepsilon^2 & \varepsilon^2 \\
\varepsilon^2 & \varepsilon     & \varepsilon \\
\varepsilon^2 & \varepsilon     & 1 \end{array} \right),
\label{eq:yup}
\ee 
where $\varepsilon$ is a loop factor. This matrix gives quark masses $(m_t, m_c, m_u) \sim v (1,\epsilon,\epsilon^2)$, which is the right power counting to match the measured quark masses. The structure of the model is shown diagrammatically in Fig.~\ref{fig:schematic}. In Secs.~\ref{subsec:wfrn}, ~\ref{subsec:benchmark}, and App.~\ref{app:loop} we will give more explicit computations of the quark Yukawa couplings and show how the SM can be numerically reproduced.

\subsection{Down and Lepton Sectors}
\label{sec:down-lep}

Because unification is a feature of SUSY even in the split regime, we build a model that is manifestly consistent with SU(5) unification.\footnote{We do not attempt to solve the doublet-triplet splitting problem for the Higgs that is ubiquitous in all SUSY GUT constructions.} Therefore, we must add a vectorlike $E$ messenger field which has the same SM quantum numbers as the MSSM right handed electron, and the same $U(1)_F$ charge as $Q$ and $U$ to complete the $\mathbf{10}$ representation. In order to generate down and lepton type Yukawa couplings, we must also add a vectorlike $\mathbf{\bar 5}$ representation. Thus we have a full vectorlike generation of messengers charged under $U(1)_F$. The additional particle content needed to generate the down and lepton Yukawa couplings is given in Tab.~\ref{tab:dfields}, while the full particle content of our model is given in Tab.~\ref{tab:fields} in the Appendix. 

\begin{table}[tp]
\captionsetup{justification=raggedright,
singlelinecheck=false}
\caption{Fields needed to generate the down and lepton Yukawa couplings in addition to those in Tabs.~\ref{tab:ufields} and~\ref{tab:flavons}, as well as their charges.}
\begin{center}
\begin{tabular}{|c|c|c|c|} \hline
Field & U(1)$_F$ & $SU(3)\times SU(2)\times U(1)$ & $R_p$\\ \hline
$E$, $\bar E$ & $\pm 1$ & $ (\mathbf 1, \mathbf 1)_{1}+ (\mathbf 1, \mathbf 1)_{-1}$ & $-$ \\ 
$D$, $\bar D$ & $\mp 3$ & $ (\bar{\mathbf 3}, \mathbf 1)_{1/3}+ (\mathbf 3, \mathbf 1)_{-1/3}$  & $-$ \\ 
$L$, $\bar L$ & $\mp 3$ & $ (\mathbf 1, \mathbf 2)_{-1/2}+ (\mathbf 1, \mathbf 2)_{1/2}$ & $-$ \\ \hline
$\ell_4$, $\bar\ell$ & 0 & $ (\mathbf 1, \mathbf 2)_{-1/2}+ (\mathbf 1, \mathbf 2)_{1/2}$ & $-$ \\ 
$d_4$, $\bar d$ & 0 & $ (\bar{\mathbf 3}, \mathbf 1)_{1/3}+ (\mathbf 3, \mathbf 1)_{-1/3}$  & $-$ \\ \hline
\end{tabular}
\end{center}
\label{tab:dfields}
\end{table}

The up-type field content can be described in SU(5) language as $\mathbf{10}_i \mathbf{10}_j \mathbf{ 5}_H$ where $i$ and $j$ are SM flavor indices. Similarly, both the down and lepton type Yukawas can be described as $\mathbf{10}_i \mathbf{\bar 5}_j \mathbf{\bar 5}_H$. Therefore, in the rest of this section we describe the generation of down-type Yukawa couplings, but the leptons can be derived by trivial replacements within $SU(5)$ representations. 

As described in Sec.~\ref{sec:split}, the Mini-Split SUSY scenario works for $\tan\beta$ of moderate size, so the bottom and $\tau$ Yukawa couplings are parametrically smaller than that of the top quark. Therefore, if we were to use the same dynamics as we used for the up-type quarks, we would expect, for example, $m_d/m_u \,\sim\, m_b/m_t$. Because it is of critical importance that the down quark be comparable in mass or heavier than the up quark, we enhance the structure of the model to fix this relation. We add an additional vectorlike down-type quark pair: $d_4$ and its conjugate partner $\bar d$, which are neutral under $U(1)_F$. Unlike the $D$, $d_4$ can mix with the SM $d_i$ because they have the same (trivial) $U(1)_F$ charge, and we have an additional ``barred'' version of the flavon coupling, as $\bar{d}$ couples to $\bar{\chi}$.

With this field content, we can write the following superpotential
\begin{eqnarray}
W_{\rm down} &=& \lambda_D \, Q\, D\, H_d +\bar{\lambda}_D\, \bar{Q}\, \bar{D}\, H_u  +  f^d_{ij} \,d_i \,\bar D \,\chi_j + \bar{f}_i \,\bar d \,D \,\bar{\chi}_i  \nonumber\\
&& + \, \mu_D\, D \bar D + \mu^d_{i} \,d_i \bar d + \mu^\chi_{\,ij} \,\chi_i \bar \chi_j \, ,
\label{eq:Wdown}
\end{eqnarray}
where again all the dimensionless couplings are $\mathcal{O}(1)$ and all the dimensionful terms are $\mathcal{O}(m_{\rm sc})$. We can now choose a basis in flavor space such that $\mu^d$ only points in one direction, and this direction picks out the fourth generation of $d$. This shows that the fourth generation $d_4$ and $\bar d$ will be heavy while the remaining three generations will be massless before electroweak symmetry breaking. 

After choosing $\mu^d$ to point only in the `4' direction, there is still a residual $U(3)_d$ flavor symmetry in the absence of the $f^d$ coupling. This symmetry exists even if $f^d$ has an $\mathcal{O}(1)$ entry in the `4' direction in $d$ flavor space. Thus, it is technically natural for all the $f^d$ couplings to the SM-like $d$ triplet to be small. This is the scenario we take in this model, namely
\be  
f^d \sim \left( \begin{array}{cc}
y_b & y_b \\
y_b & y_b \\
y_b & y_b \\
1 & 1 \end{array} \right),
\label{eq:fd}
\ee
in a generic basis where only $\mu^d$ has been rotated into the fourth component.  We have dropped $\mathcal{O}(1)$ coefficients in each entry. Here $y_b$ is parametrically the size of the bottom (or $\tau$) Yukawa coupling, and the choice of the coupling of the form of $f_d$ represents a technically natural tuning of order 10\%. This is the only parametric hierarchy in the flavor sector not explained by our model. 

\begin{figure}
\centering
\includegraphics[width=0.5\textwidth]{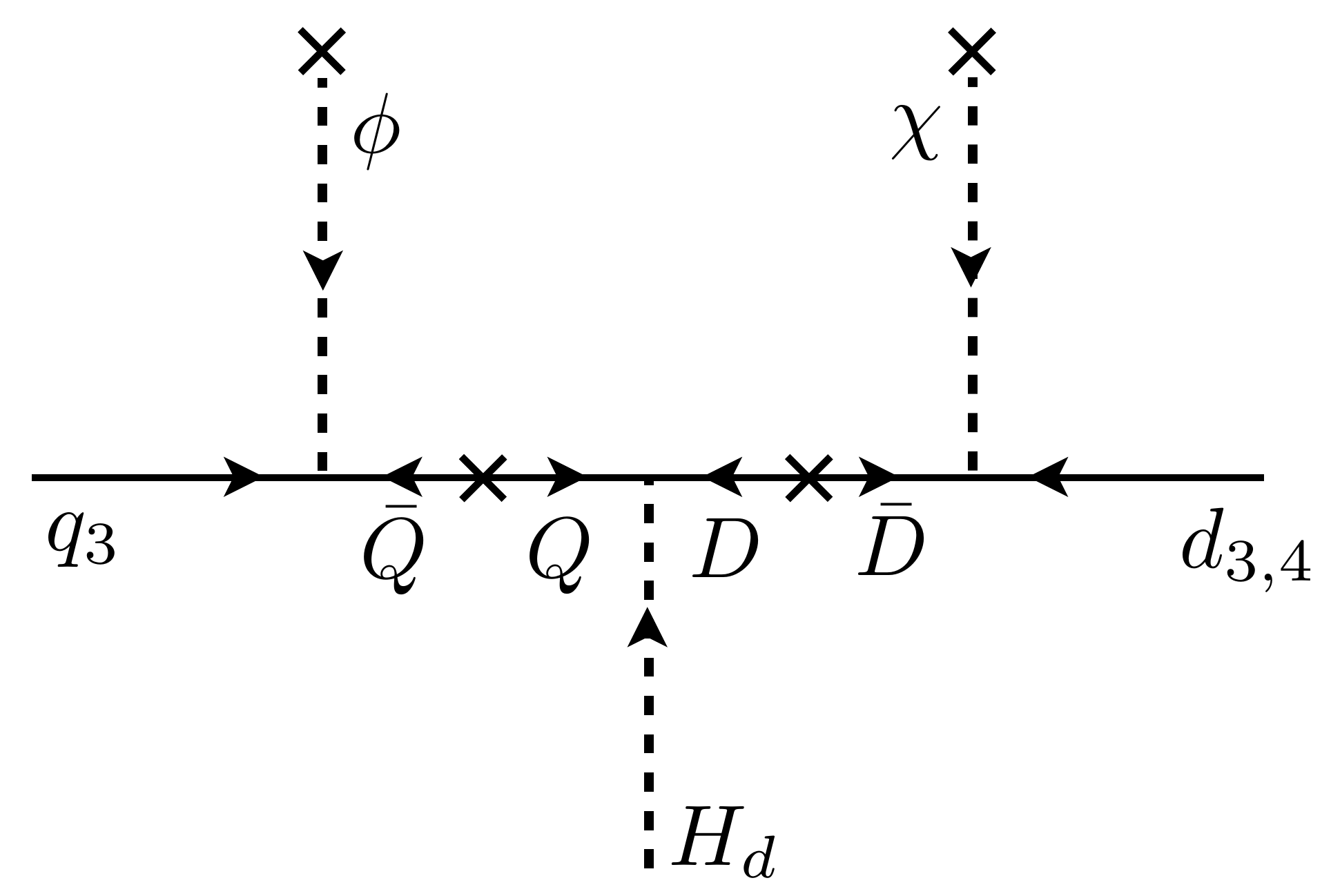}
\caption{Feynman diagram for generating the bottom and $d_4$ Yukawa couplings.}
\label{fig:bottom}
\end{figure}

We now see that there is a process analogous to that of Fig.~\ref{fig:top} for the bottom and $d_4$ quarks shown in Fig.~\ref{fig:bottom}. As in the top case, we can pick a basis where the $\chi$ vev is only in one direction, and then we can use the $U(3)_d$ symmetry to make the $f^d$ coupling to the $\chi$ vev parametrically $f^d_{\langle\chi\rangle} \sim (0,0,y_b,1)$. Since we have a fourth generation, the down Yukawa matrix at the scale of $U(1)_F$ breaking is now $3\times 4$ and it is given by the outer product of $f^d_{\langle\chi\rangle}$ with the corresponding coupling from the $q$ doublet $f^q_{\langle\phi\rangle} \sim (0,0,1)$ computed in the previous section. 

One-loop 2$^{\rm nd}$ generation masses proceed in nearly the same fashion as in the up sector through the diagrams in Fig.~\ref{fig:charm} with up-type quarks replaced by down-type, and $\chi$ replacing $\phi$ where necessary. The flavino-smessenger diagram requires the use of the $\chi\phi\bar{\xi}$ coupling given in Eq.~\eqref{eq:pureflavonsup}. This shows that the $U(1)_F$ charge assignments given in Tab.~\ref{tab:flavons} are optimal for this model because they allow the generation of both up- and down-type flavino diagrams.  The one-loop strange mass diagrams require $d_2$ to couple to $\chi$, so they are parametrically of size $y_b \, \varepsilon$, where $\varepsilon$ is again the loop factor. This is because the parametrization of Eq.~\eqref{eq:fd}  is natural only if all SM-like couplings to $\chi$ are $\mathcal{O}(y_b)$, thus the one-loop diagram has a small coupling. The parametrics of this model then predict that $m_s/m_b \sim m_c/m_t$, a relation that is good to within a factor of a few in nature. 

Finally, we can fill out the rest of the Yukawa matrix with the process analogous to that shown in Fig.~\ref{fig:up}. Besides the obvious substitution of $u$ with $d$, the main difference is that the coupling to the Higgs now comes from the fourth generation down squark instead of the sbottom. That Yukawa coupling is $\mathcal{O}(1)$ instead of $\mathcal{O}(y_b)$. Because  $d_4$ has the same quantum numbers as the SM $d_i$, we expect that SUSY breaking soft terms mix $\tilde{d}_4$ strongly with all the SM down-type squarks. Therefore the fourth generation Yukawa coupling can be transmitted to all the other down-type squarks parametrically at two-loop order. Thus, we see that adding this fourth generation changes the incorrect relation of $m_d/m_b \sim m_u/m_t$, to the much more accurate one $m_d/m_t \sim m_u/m_t$ because the fourth generation Yukawa and that of the top Yukawa are the same parametric size.

Putting all the results together, the Yukawa matrix in the down sector is parametrically of the form 
\be  
\mathbf{y^d} \sim \left( \begin{array}{cccc}
\varepsilon^2 & \varepsilon^2 & \varepsilon^2 & \varepsilon^2\\
\varepsilon^2 & y_b\, \varepsilon     & y_b\, \varepsilon & \varepsilon \\
\varepsilon^2 &  y_b\, \varepsilon     & y_b & 1 \end{array} \right) \, .
\label{eq:ydown}
\ee 
Here, $y_b$ is the approximate bottom Yukawa coupling, which is somewhat larger than $\varepsilon \simeq g^2/16\pi^2$, the loop factor. While they are not so different in size, we keep track of the parametrics separately so the different physical mechanisms can be more easily understood. This Yukawa matrix is $3\times 4$ because it describes the coupling of 3 $q$'s to 4 $d$'s. This matrix holds at the scale $m_{\rm sc}$ where $U(1)_F$ is broken. At lower scales, the $d_4$ can be integrated out because it has a large supersymmetric mass, and the fourth column of the matrix can simply be truncated at leading order.  

After this truncation, we have a $3\times3$ matrix which gives the quark masses as $(m_b, m_s, m_d) \sim v (y_b, y_b\,\varepsilon,\varepsilon^2)$, and we have $\varepsilon < y_b < 1$. This shows that the down sector has a parametrically different hierarchy than the up sector. Instead of equal steps going down in generation, this model explains why the ratio of the strange to bottom mass is smaller than down to strange. The full cascading structure of the quark masses in this model is shown in Fig.~\ref{fig:schematic}. 

As explained above, the structure of the leptons is nearly identical with $q$ replaced by $e$ and $d$ replaced by $\ell$. The most important change is that the diagram analogous to Fig.~\ref{fig:up} for the leptons has a bino exchange instead of a gluino. Thus we get that $m_e/m_d \sim g_1^4/g_3^4\simeq 0.03$, where two of the factors of the gauge coupling come from the coupling to the gaugino, and two more come from the gaugino mass in Eq.~\eqref{eq:gaugino}.  Here we have run the gauge couplings up to $m_{\rm sc} \simeq 1000$ TeV where this diagram is generated. The parametric estimate for the relative size of the electron and down is somewhat small, but it is not too far off. We now see that our model successfully predicts the masses of the SM fermions at the parametric level, and all that remains is the mixing angles between the quarks.

\subsection{CKM Matrix}

At scales well below $m_{\rm sc}$, we have the following parametric Yukawa matrices taken from Eqs.~\eqref{eq:yup} and~\eqref{eq:ydown}:
\be  
\mathbf{y^u} \sim \left( \begin{array}{ccc}
\varepsilon^2 & \varepsilon^2 & \varepsilon^2 \\
\varepsilon^2 & \varepsilon     & \varepsilon \\
\varepsilon^2 & \varepsilon     & 1 \end{array} \right), \;\;\;\;\;\;\;\;\;\;\;\;\;\;\;\;\;\;
\mathbf{y^d} \sim \left( \begin{array}{ccc}
\varepsilon^2 & \varepsilon^2 & \varepsilon^2 \\
\varepsilon^2 & y_b\, \varepsilon     & y_b\, \varepsilon  \\
\varepsilon^2 &  y_b\, \varepsilon     & y_b  \end{array} \right) \, .
\label{eq:yukawas}
\ee 
In order to compute the CKM matrix, we use the standard procedure of finding the matrices which diagonalize $\mathbf{y^u}$ and $\mathbf{y^d}$. In particular, we have 
\be
V_u^\dagger \,\mathbf{y^{u\dagger} y^u}\, V_u=1/v^2 \,{\rm diag}(m_u^2,m_c^2,m_t^2)\, ,
\ee
where $V_u$ acts on the ``$q$'' indices of $\mathbf{y^u}$. There is an analogous formula for $\mathbf{y^d}$. From Eq.~\eqref{eq:yukawas} we can compute
\be  
V_u \sim \left( \begin{array}{ccc}
1 & \varepsilon & \varepsilon^2 \\
\varepsilon & 1     & \varepsilon \\
\varepsilon^2 & \varepsilon     & 1 \end{array} \right), \;\;\;\;\;\;\;\;\;\;\;\;\;\;\;\;\;\;
V_d \sim \left( \begin{array}{ccc}
1 & \varepsilon/y_b & \varepsilon^2/y_b \\
\varepsilon/y_b & 1     &  \varepsilon  \\
\varepsilon^2/y_b &   \varepsilon     & 1  \end{array} \right) \, ,
\label{eq:diag-mat}
\ee 
where we have taken $\varepsilon \ll y_b \ll 1$. In reality, we will soon see that $\varepsilon/ y_b \simeq \sin\theta_c \simeq 0.2$ where $\theta_c$ is the Cabibbo angle and not that much smaller than 1. 

To compute the CKM matrix, we simply take
\be
V_{\rm CKM} = V_u^\dagger V_d \simeq V_d,
\ee
where the second relation comes from the fact that $V_u$ more closely approximates the unit matrix than does $V_d$. This parametric relation, predicts, for example, $|V_{us}||V_{cb}| \simeq |V_{td}|$, which holds very well in nature. 

The above discussion is only applicable to the absolute value of the elements of the rotation matrices, but in general, we expect every element of $\mathbf{y^u}$ and $\mathbf{y^d}$ to have independent phases. Taking $(\mathbf{y^u})_{33}$ from Eq.~\eqref{eq:topyukpre} as an example, we see that all the $\lambda$, $F$ and $\mu$ couplings will have phases, so the total coupling will also have a phase. Similar arguments can be made about the other elements of the Yukawa matrices, with different couplings entering the computations so they will have independent phases. Therefore, in the absence of cancellation, the physical phase of the CKM matrix will also be $\mathcal{O}(1)$. In Sec.~\ref{subsec:benchmark} we will describe a point in the parameter space of this model which reproduces the Standard Model more accurately, but, just from the parametric estimates of this section, we see that we have succeeded in explaining nearly all the hierarchies of the SM flavor sector.

%%%%%%%%%%%%%%%%%%%%%%
\section{Computing the Spectrum}
\label{sec:details}
%%%%%%%%%%%%%%%%%%%%%%

In this section we give the details of the computation of the masses of the various states in the theory, including the gauginos, the light Higgs, and of course the SM fermions. We also describe a benchmark point in parameter space so that we can give definite numbers for every effect for at least one point in parameter space. The details of the benchmark including the reproduction of the SM flavor parameters is described Sec.~\ref{subsec:benchmark}.

\subsection{Gaugino Spectrum, Unification, and Dark Matter}
\label{sec:gaugino}

In our framework, the gauginos are the only states that are relatively light and could be probed in the near future, so it is important to have a precise understanding of the mass hierarchy for phenomenological reasons. As stated in Sec.~\ref{sec:split}, the gaugino masses are on the anomaly-mediated trajectory above the messenger scale $\mu_M \sim \msc$. Because SUSY is broken at the messenger scale, integrating out the messengers will induce threshold corrections that will deflect them from their anomaly-mediated values. The Higgs states will also shift the gaugino masses, but they must be treated with care because one of the states remains light. 

Our flavor model requires one set of $\mathbf{10} + \mathbf{\overline{10}}$, containing $Q$ and $U$ and their conjugates, and two sets of $\mathbf{5} + \mathbf{\bar{5}}$, one containing $D$ and the other $d_4$.  The soft masses and $b$-terms for the messengers are generated by the Giudice-Masiero (GM) mechanism~\cite{Giudice:1988yz}, as in Eq.~\eqref{eq:mu-bmu}. The $b$-term generated by the GM operator is opposite in sign to that obtained from a superpotential mass term which explains why the messengers do not decouple. As described in~\cite{Gupta:2012gu}, the threshold correction due to each messenger pair depends on the supersymmetric messenger mass $\mu_M$, the holomorphic SUSY breaking mass, $b_M$, and the soft mass $m_M^2$. We define the following dimensionless ratios for a given messenger pair $M$:
\be
r_M = |b_M|/|\mu_M|^2 \qquad c_M^2 = m_M^2/|\mu_M|^2 \, .
\ee
We can compute the threshold correction for a given messenger pair with Dynkin index $C_M$ defined as $1/2$ for a fundamental of $SU(N)$ and $Y^2$ for hypercharge. The threshold correction is then given by
\beq
\Delta m_{\tilde{i}} = -2e^{i \theta_M} C_M \frac{\alpha_i}{2 \pi} f_t(y_1,y_2)  \frac{|b_M|}{|\mu_M|}, \qquad f_t(y_1,y_2) = \frac{y_1 \log y_1 - y_2 \log y_2 - y_1 y_2  \log(y_1/y_2)}{(y_1 - 1) (y_2 - 1) (y_2 - y_1)},
\eeq
\noindent with $y_i = M_i^2/|\mu_M|^2$, where $M_{1,2}^2$ are the eigenvalues of the scalar messenger mass-squared matrix with $M_1 > M_2$ and are given by
\beq
y_1= 1+ c_M^2 + r_M, \qquad y_2= 1+ c_M^2 - r_M.
\eeq
The phase is defined as $\theta_M = \arg(b_M/\mu_M)$ and vanishes if the contact terms in Eq.~\eqref{eq:mu-bmu} are absent, which is the pure GM limit, since in that case both $b_M$ and $\mu_M$ arise from the same operator. In general, the phase will be non-zero, and we work in a convention where $\mu_M$ is real. 

For the Higgs doublets, we are taking\footnote{Note that our notation differs here slightly from Sec.~\ref{sec:split} to make it clear that $\mu_H$ and $b_H$ are the parameters in the Higgs potential, but, in this context, the Higgs multiplet is another messenger.} $\mu_H \sim \msc$, so they act as an additional messenger pair that contributes its own threshold correction. Because one linear combination of the doublets is tuned to be light, the form of the threshold correction is different:
\beq
\Delta m_{\tilde{i}} =- \frac{\alpha_i}{4 \pi} |\mu_H | e^{i \theta_H} \sin 2 \beta \frac{m_A^2}{|\mu_H|^2-m_A^2} \log \frac{|\mu_H|^2}{m_A^2}, \qquad m_A^2 = \frac{2 r_H }{\sin 2 \beta} |\mu_H|^2,
\eeq
where $r_H = b_H / |\mu_H|^2$ and $m_A$ is the physical pseudoscalar mass which is approximately degenerate with the rest of the heavy Higgses. Note that with our conventions, there is an overall sign here relative to expressions found elsewhere in the literature~\cite{Harigaya:2013asa}. Here we work in the convention where $b_H$ is real, so $\theta_H =\arg(\mu_H)$. Since $\tan \beta = \mathcal{O}(1)$ and $\mu_H \sim m_A = \mathcal{O}(m_{sc})$, the Higgsino threshold corrections are comparable in size to those of the messengers. Furthermore, as emphasized in~\cite{Harigaya:2013asa}, the phase freedom allows for a rich spectrum of gaugino masses, since interference between the various contributions can lead to wino, bino, or gluino LSP.  

We now describe the parameters of the gaugino sector for our benchmark point. The spectrum contains a 3.0 TeV wino LSP for suitable dark matter phenomenology, which we will discuss below.  For consistency with the SM flavor analysis, we integrate out all heavy states at 1000 TeV. The threshold corrections can then be calculated as described above, using the Dynkin indices in App.~\ref{app:mess}. We then run down all the masses to the TeV scale and include any appropriate pole mass corrections. Since we have not considered the lepton sector in any detail, we simply assume the parameters are the same as those for the quarks in the same GUT multiplet. For simplicity, we have taken all of the phases in the messenger sector to be $\pi$ (except for $d_4$, $l_4$), and take $\theta_H$ = 0, which means the Higgsino threshold correction is opposite in sign to the contribution from AMSB. Generalizing to $\mathcal{O}(1)$ phases does not change the picture significantly. To obtain our benchmark spectrum with a wino LSP and a decently-sized gluino mass (needed for 1$^{\rm st}$ generation Yukawas), we take $m_{3/2} = 1100$ TeV. Tab.~\ref{tab:gaugein} contains all the messenger and Higgs sector input parameters relevant for calculating the gaugino spectrum in the way described above. 

The discussion above was predicated on the assumption of no mixing between quark and messenger fields. However, as described in Sec.~\ref{subsubsec:diag}, once the flavons get vevs there is mixing between squarks and smessengers as well as mixing between 3$^{\rm rd}$ generation quarks and messengers. In computing the gaugino spectrum, we take this mixing into account, and the detailed formula is given in App.~\ref{app:mess}. In fact, proper accounting of mixing decreases the messenger threshold corrections by a factor of a few, since these are dominated by $Q$ and we have large $q_3$-$Q$ mixing. We find that the messenger corrections are about an order of magnitude smaller than the AMSB soft masses. The gaugino pole masses are $m_{\tilde{W}} = 3.0$ TeV, $m_{\tilde{B}} = 13.3$ TeV, and $m_{\tilde{g}} = 20.9$ TeV.

\begin{table}[tp]
\captionsetup{justification=raggedright,
singlelinecheck=false}
\caption{Benchmark parameters for the messenger and Higgs sectors. $c_H$ is fixed by the requirement of a light Higgs state. The $c$ column for $D$ contains two values because here we take different soft masses for $D$ and $\bar{D}$; similarly for $d_4$.}

\begin{center}
\begin{tabular}{|c|c|c|c|c|} \hline
Messenger & $\mu_M$ & $c_M$ & $r_M$ & $\theta_M$ \\ \hline 
$Q$ & 1000 & 1.17 & 1.1 & $\pi$ \\ 
$U$, $E$ & 1000 & 1.58 & 1.15 & $\pi$ \\ 
$D$, $L$ & 750 & 3.0, 3.46 & 2.0 & $\pi$ \\ 
$d_4$, $l_4$ & 728 & 3.36, 3.81 & 0.5 & 0 \\ 
$H$ & 2400 & fixed & 7.8 & 0 \\ \hline
\end{tabular}
\end{center}
\label{tab:gaugein}
\end{table}

Taking these gaugino masses, we can examine gauge coupling unification. The Mini-Split framework differs from regular Split SUSY only in that $\mu_H$ is large, i.e. at the same scale as the sfermions. In an analysis~\cite{ArkaniHamed:2012gw} carried out with a similar gaugino spectrum and $\mu_H = m_{sc} = 1000$ TeV, it was shown that raising $\mu_H$ results in good unification with no messengers, with a predicted $\alpha_3(m_Z) = 0.111$ as compared to the experimental value of  $\alpha_3(m_Z) = 0.118\pm 0.003$~\cite{Beringer:1900zz}. This is consistent with unification because there are in general unknown threshold corrections at the GUT scale of $\mathcal{O}(1/4\pi)$.\footnote{In~\cite{Wang:2013rba} it was argued that this spectrum is inconsistent with unification, but that work requires that the gauge couplings unite much more precisely than the parametric size of the threshold corrections.} Therefore, we see that with our field content, the model is consistent with unification. The cases of $N = 1$ and $N = 4$ messengers were also studied, with sfermions, messengers, Higgsinos and heavy Higgses all introduced into the two-loop running at a common scale of $m_{sc}$. Unification still works well and occurs at a slightly larger scale, with a larger coupling at unification and a slightly smaller predicted $\alpha_3(m_Z)$, as $N$ is increased. Our extra matter charged under the SM, i.e.~the messenger sector and fourth down-type generation, corresponds to $N = 5$. The gauge couplings do not blow up because the messengers are heavy. In fact, because of the high messenger scale, perturbative control is retained even for $N = 6$. For our benchmark with $N=5$, the unification scale is $1.1 \times 10^{16}$ GeV, $\alpha^{-1} = 9.3$ at unification, and $g_3 - g_2 = 0.05$ at the GUT scale corresponding to a predicted $\alpha_3(m_Z) = 0.109$.  

Finally, we can summarize the dark matter scenario, which is qualitatively very similar to that described in~\cite{ArkaniHamed:2012gw}.  Because the $\mu$-term for the Higgs is so much larger than the gaugino masses, the wino and bino do not mix with one another or with the Higgsino and are very nearly pure states.  If the wino is the LSP, then it has a weak scale annihilation cross section and will behave as a usual WIMP. It will have the right relic abundance if it has a mass around 3 TeV~\cite{Hisano:2006nn}. In this case, there would be WIMP annihilations in regions of high dark matter density such as the galactic center, and these could be looked for as indirect dark matter detection signals. Results from various telescopes~\cite{Cohen:2013ama,Fan:2013faa,Hryczuk:2014hpa} have placed stringent constraints on thermal wino dark matter which are in tension with this scenario for the standard dark matter halo profiles. On the other hand, for profiles that are less steep or cored near the galactic center, this scenario is still viable. 

One could imagine many other dark matter scenarios consistent with the Mini-Split framework and the model presented here. For example, if the wino is lighter than 3 TeV, then it will only make up some of the dark matter, but the rest could be made up of another particle such as an axion. Alternatively, if the LSP is produced non-thermally~\cite{Moroi:1999zb,Gherghetta:1999sw,Grajek:2008jb}, then its mass could be heavier than 3 TeV. Our model could also produce a bino LSP with different choices of the parameters in Tab.~\ref{tab:gaugein}. While a thermal bino would overclose the universe, it could be nonthermal, or its abundance could be reduced by the co-annihilation mechanism~\cite{Ibe:2013pua}. From this analysis we see that while the flavor dynamics do not directly affect the dark matter story, the two sectors fit well together in the framework of Mini-Split Supersymmetry.

\subsection{Higgs Mass and Quartic}
\label{subsec:cons}
Since SUSY is broken well above the scale of electroweak symmetry breaking,  $SU(2)$ is preserved to a very good approximation, and thus, as in previous Split SUSY models, the tuning in the Higgs sector produces one light doublet, which includes the SM-like Higgs, and one heavy doublet, with degenerate scalars of mass $m_A$. The leading contributions to the mass of the light Higgs are the usual ones in Split SUSY models.   At tree level, there is a contribution from the $D$-term of the SM $SU(2)\times U(1)$ gauge group, and there are loop contributions arising from the large splitting between the top and stop quarks. These are analyzed in detail for a 125 GeV Higgs in~\cite{Giudice:2011cg, ArkaniHamed:2012gw}.\footnote{We thank Gian Giudice for helpful discussion on the results of~\cite{Giudice:2011cg} concerning these effects.} If the scalars all have a common mass $m_{sc}$, then the Higgs mass essentially depends only on $m_{sc}$ and $\tan\beta$ (aside from a very slight dependence on the wino and gluino masses). A 125.7 GeV Higgs mass~\cite{CMS-PAS-HIG-13-005,ATLAS-CONF-2013-014} implies $\lambda = 0.26$ at the weak scale. Running this up to a scale of  $m_{sc}=1000$ TeV and taking only the gauginos to be below $m_{sc}$ gives $\lambda=0.058$. Both the tree level and one loop contributions to the quartic depend on $\tan\beta$, and a quartic of the right size can be obtained in the MSSM with $m_{sc}=1000$ TeV if $\tan\beta=2.2$.

\begin{figure}
\centering
\includegraphics[width=0.6\textwidth]{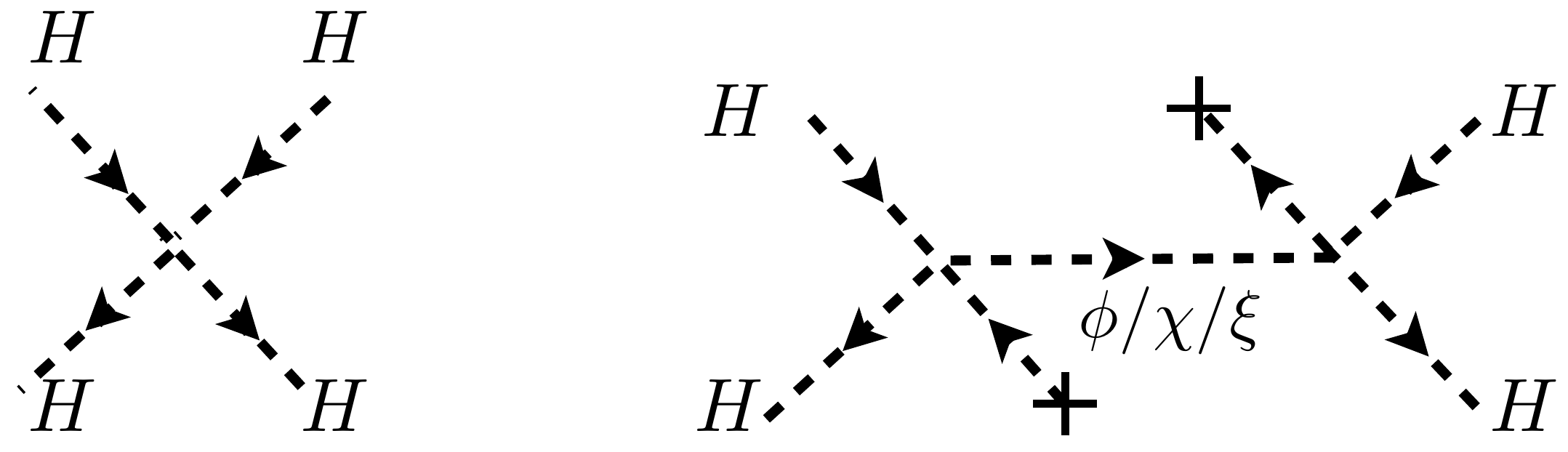}
\caption{Feynman diagrams the Higgs quartic generated by the $U(1)_F$ $D$-term.}
\label{fig:dterm}
\end{figure}

In this model, there are additional subdominant contributions to the Higgs quartic, so the relationship between the Higgs mass and $\tan\beta$ will be modified. The first of these arises from ``non-decoupling $D$-terms''~\cite{Batra:2003nj,Maloney:2004rc} from the new $U(1)_F$. The $D$-term is of the form
\be
V^D = \frac{g_F^2}{2} \left( \phi_i^\dagger \phi_i - \bar \phi_i^\dagger \bar\phi_i -3 \chi_i^\dagger \chi_i +3 \bar \chi_i^\dagger \bar\chi_i 
                    - 4 \xi^\dagger \xi + 4 \bar{\xi}^\dagger \bar{\xi} - 2 H_u^\dagger H_u + 2 H_d^\dagger H_d + ... \right)^2 \, ,
\label{eq:Vd}
\ee
where the ellipses include terms with the messenger fields which do not get vevs.  Expanding this out generates a Higgs quartic. We can also integrate out the flavons and use the fact that they get vevs to generate additional Higgs quartics. These contributions are shown in Fig.~\ref{fig:dterm}. Thus, we generate the coupling
\be
\frac{\lambda_F'}{2} \left(H_u^\dagger H_u - H_d^\dagger H_d \right)^2 \rightarrow \frac{\lambda_F }{2} \, (H^\dagger H)^2
\ee
\be
\lambda_F=4 g_F^2 \cos^2 2\beta\left(1 - \frac{1}{2} g_F^2 \sum\limits_{\phi,\chi,\xi} (q v)_i (m^2)^{-1}_{ij} (q v)_j\right),
\label{eq:quartic}
\ee
where the sum is over real and imaginary components of all flavon species.  Here, $(q v)$ is a vector of the flavon vevs multiplied by their $U(1)_F$ charges, where the charge is the same for both real and imaginary components. The matrix $(m^2)^{-1}$ is the inverse of the flavon mass squared matrix in the vacuum, and we take all the flavon vevs to be well above the electroweak scale.  
In the limit where $U(1)_F$ is Higgsed supersymmetricly, $\lambda_F$ must go to zero, which will occur as a perfect cancellation between the two terms in Eq.~\eqref{eq:quartic}. 
In a general region of parameter space where the soft masses and the supersymmetric mass are comparable, there is still a partial cancellation between the two terms in Eq.~\eqref{eq:quartic}, with $\lambda_F$ about an order of magnitude smaller than $4 g_F^2$. For the benchmark described in this section, $\lambda_F$, which comes from the $U(1)_F$ $D$-term, is 0.013, compared to the tree-level MSSM value of 0.037. 

The new vectorlike states in this model have large couplings to the Higgs, so they will contribute to the Higgs quartic through loops. As these loops must vanish in the supersymmetric limit, they are sensitive to the splitting between scalar and fermion masses. Therefore, the effects of the stops are parametrically larger than those of the new vectorlike states. We here compute the full one-loop contribution to the Higgs quartic in the unbroken electroweak theory. Because the (s)tops mix with messengers, it is difficult to disentangle the different effects, and we compute all the one-loop threshold corrections in the mass basis.  The diagrams are scalar bubble, triangle, and box diagrams, as well as fermion box diagrams and external line corrections from Higgs wavefunction renormalization. Since our Higgsino mass is $\mo(\msc)$, we also consider the MSSM contributions from mixed gaugino-Higgsino boxes and contributions to the Higgs field-strength renormalization.  In the benchmark, the up-type new generation contributes 0.014 while the new down-type fields contribute 9$\times 10^{-4}$ to the Higgs quartic. In a realistic model, there would also be contributions from the lepton sector, which we estimate to be 1/$N_c$ of the down contribution.  Once we sum up all the tree level and one-loop contributions, we obtain the right Higgs quartic and mass with $\tan\beta=1.8$. Therefore, we see that while the effects from the model are indeed subdominant, they need to be taken into account to properly compute the spectrum.

\subsection{Mass Eigenstates and Wavefunction Renormalization}
\label{subsec:wfrn}
Before computing the SM flavor parameters in detail, it is necessary to address effects that can induce $\mo(1)$ changes to the basic arguments of Sec.~\ref{sec:model}.  They are the full diagonalization of the (s)quark-(s)messenger fields after $U(1)_F$ breaking and the one-loop wavefunction renormalization.  We stress that the parametric hierarchies given by loop counting are left intact by these considerations, but they can have important numerical effects.  We consider them in turn.

\subsubsection{Diagonalization}
\label{subsubsec:diag}

Once the flavons get vevs, the UV distinction between quark and messenger superfields breaks down. In the fermion sector, only the 3$^{\rm rd}$ generation mixes at tree level.  For $q_3$ and $u_3$, we need only consider the 2$\times$2 mixing with the $Q$, $U$ messengers. We denote the mass eigenstates as $q_3'$ and $Q'$ with the lower case $q'$ representing SM states, while the capital $Q'$ is a state with mass $\sqrt{\mu_Q^2 + |F^q_{3i}\, \phi_i|^2}$; $\mu_Q$ is the supersymmetric mass for the messengers defined in Eq.~\eqref{eq:Wup}, and $F$ is a rotation of the superpotential coupling defined in Eq.~\eqref{eq:Fdefine}. We here take the convention where $\mu_Q$ is real. The mixing is then parametrized as 
\beq
\label{qmatrix}
\left( \begin{array}{c} q_3     \\
           Q \end{array} \right)~ = \left( \begin{array}{cc} c_q &  s_q^*      \\
          -s_q & c_q \end{array} \right)~ \left( \begin{array}{c} q_3'     \\
           Q'
 \end{array} \right)~,
 \eeq
with
\be
c_q = \frac{\mu_Q}{\sqrt{\mu_Q^2 + |F^q_{3i}\, \phi_i|^2}}, \qquad s_q = \frac{F^q_{3i}\, \phi_i}{\sqrt{\mu_Q^2 + |F^q_{3i}\, \phi_i|^2}} ,
\label{eq:qrot}
\ee
and analogous expressions for $u$ and $d$. For notational compactness, in this section we often use the same notation for both the scalar field and its vev.  In the case that $F^q_{3i}\, \phi_i$ is real, $c_q, s_q$ just become cosine and sine of a rotation angle. After rotating to mass eigenstate basis, Eq.~\eqref{eq:topyukpre} for the top Yukawa is modified to
\be
y_t = \lambda_U s_q s_u,
\label{eq:yt}
\ee
where we recover our earlier formula in the limit $\mu \gg F\, \phi$. 

Since $d_3$, $d_4$, and $\bar d$ couple to flavon vevs, the diagonalization in the down sector is more complicated. The fermion mass matrix takes the form 
\beq
 \label{dmatrix}
 (\bar d ~ \bar D)~ \left( \begin{array}{ccc} 0 & \mu_d & \bar{f}_i \bar{\chi}_i        \\
          F^d_{3i} \chi_i  & F^d_{4i} \chi_i & \mu_D \end{array} \right)~ \left( \begin{array}{c} d_3     \\
           d_4 \\ D
 \end{array} \right)~.
 \eeq
To find the SM down quark eigenstate $d_3'$, we solve for the null space of the matrix above, yielding
\be
d_3' = \frac{\left(\mu_d \mu_D - F^d_{4i} \chi_i \bar{f}_j \bar{\chi}_j \right)^* d_3 + \left(F^d_{3i} \chi_i \bar{f}_j \bar{\chi}_j \right)^* d_4  -  \left(\mu_d F^d_{3i} \chi_i  \right)^* D}{\sqrt{|\mu_d \mu_D - F^d_{4i} \chi_i \bar{f}_j \bar{\chi}_j |^2 + |F^d_{3i} \chi_i \bar{f}_j \bar{\chi}_j |^2 +  |\mu_d F^d_{3i} \chi_i|^2}},
\label{eq:d3p}
\ee
where $\bar{f}$ is defined in Eq.~\eqref{eq:Wdown}, and $F^d$ is analogous to $F^q$, derived from $f^d$ in Eq.~\eqref{eq:Wdown}. Our expression for the bottom Yukawa is thus replaced by
\be
y_b = \lambda_D \, s_q  \frac{ \mu_d F^d_{3i} \chi_i }
                                                               {\sqrt{|\mu_d \mu_D - F^d_{4i} \chi_i \bar{f}_j \bar{\chi}_j |^2 + |F^d_{3i} \chi_i \bar{f}_j \bar{\chi}_j |^2 +  |\mu_d F^d_{3i} \chi_i|^2}}
                                                               .
\label{eq:yb}
\ee
Once we perform this rotation, $d_3'$ decouples, and we are left with a 2 $\times$ 2 Dirac mass matrix that we diagonalize in the usual way. We summarize the product of rotations as
\begin{eqnarray}
d^i_{\rm mass} &=& \Gamma^{\dag ij}_d \, d^j_{\rm gauge}, \nonumber \\
\bar{d}^i_{\rm mass} &=& \Gamma^{\dag ij}_{\bar{d}} \, \bar{d}^j_{\rm gauge}, 
\label{eq:rotconvd}
\end{eqnarray}
where $d \equiv( d^3,\,d^4,\, D )$ contains both quark and messenger fields and $\bar{d} \equiv( \bar{d},\,\bar{D})$.

For scalars, due to the anarchic mixing among the squarks from their soft masses, diagonalization is more involved, resulting in 5$\times$5 matrices for $\tilde{q},\, \tilde{u}$ and 7$\times$7 for $\tilde{d}$.  Just as with the fermions, the 3$^{\rm rd}$ generation mixes directly with the messengers via the $\phi$ or $\chi$ vevs.  Additionally, the 2$^{\rm nd}$ generation also has a tree-level coupling to the messengers through the $\bar \phi$ or $\bar \chi$ vevs.\footnote{This provides an interesting example of the importance of supersymmetry to our model.  Without the holomorphicity and non-renormalization properties of supersymmetric theories, we would expect to generate tree-level Yukawas for the 2$^{\rm nd}$ generation from the vev, $\langle \bar \phi^* \rangle$, and we would need special potentials in the flavon sector that only broke $U(1)_F$ symmetry with unbarred fields.  Supersymmetry allows us to take more generic flavon potentials, while forbidding the barred-flavon vev Yukawa coupling to SM fermions.}  For example, in the $q$ sector, we have
\beq
\mathcal{L} \supset \tilde{\bar Q} \, \tilde{q}_i \, F^q_{ij} \, \mu^{{\phi}^*}_{jk} \, \langle \bar \phi_k^* \rangle \,+\, {\rm h.c.} \, .
\eeq
We do not attempt an analytic diagonalization of the scalar sector, but we perform rotations numerically for the analysis of our benchmark that recovers the Standard Model.  For future reference, our convention for rotation matrices is the following ({\it e.g.} for the $q$ sector):
\beq
\tilde{q}^i_{\rm mass} \,=\, \Gamma^{\dag ij}_{\tilde{q}} \, \tilde{q}^j_{\rm gauge}, 
\label{eq:rotconvq}
\eeq
where $\tilde{q} \equiv ( \tilde{q}^1,\,\tilde{q}^2,\,\tilde{q}^3,\,\tilde{Q},\, - \epsilon \tilde{\bar Q }^*)$ contains both squark and smessenger fields. There is an analogous expression for the down sector that involves $\tilde{d} \equiv ( \tilde{d}^1,\,\tilde{d}^2,\,\tilde{d}^3,\,\tilde{d}^4, \,\tilde{D}, \, \tilde{\bar d }^*,\,\tilde{\bar D }^*)$.

The flavon sector contains its own nontrivial rotations after $U(1)_F$ breaking.  The flavino matrix is 11$\times$11 and the flavon matrix 20$\times$20, since CP is generically broken and the real and imaginary scalar components mix.  For the former, the $U(1)_F$ gaugino, $\tilde{Z}'$, mixes strongly with the flavinos, and thus after symmetry breaking we simply count it among their number.  Let $\tilde{\Phi}_{\rm gauge} \equiv ( \phi,\,\chi,\,\xi,\,\bar \phi,\, \bar \chi,\, \bar \xi,\, \tilde{Z}')$ be the fermion components of the superfields appearing in Eqs.~\eqref{eq:Wup}, \eqref{eq:pureflavonsup}, and~\eqref{eq:Wdown} plus the gaugino, with $\Phi_{\rm gauge}$ the corresponding scalars, arranged with the ten real-component fields followed by the ten imaginary ones.  Then, in analogy with Eq.~\eqref{eq:rotconvq}, we write
\be
\Phi^i_{\rm mass} &=& \Gamma^{\dag ij}_\Phi \, \Phi^j_{\rm gauge}, \nonumber \\
\tilde{\Phi}^i_{\rm mass} &=& \Gamma^{\dag ij}_{\tilde\Phi} \, \tilde{\Phi}^j_{\rm gauge}.
\label{eq:rotconvf}
\ee
The one additional subtlety in the flavon sector is that one must identify the zero-mass eigenstate that corresponds to the longitudinal mode of the heavy $U(1)_F$ boson.  We work in a unitary gauge where this state never appears in calculations with flavons in mass eigenstate basis.

\begin{figure}
\centering
\includegraphics[width=0.9\textwidth]{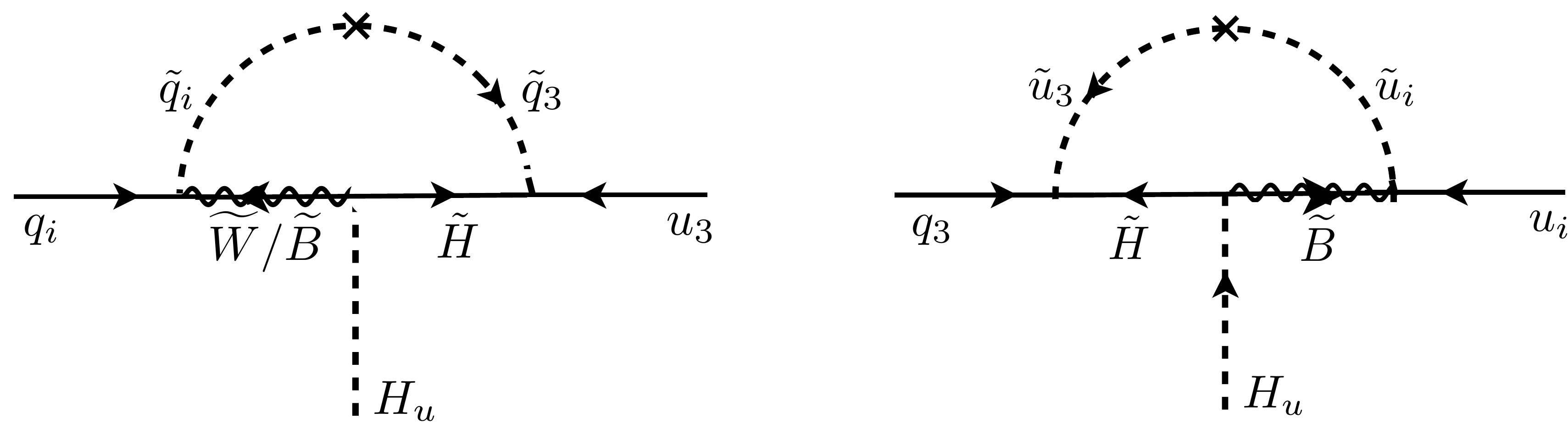}
\captionsetup{justification=raggedright,singlelinecheck=false}
\caption{One-loop electroweak contribution to off-diagonal Yukawa couplings. These are potentially important for `13' and `31' entries of the Yukawa matrix. }
\label{fig:GauginoHiggsino}
\end{figure}

In addition to the one-loop contributions to the Yukawa couplings discussed in Sec.~\ref{sec:model}, there are additional contributions from loops of Higgsinos and electroweak gauginos shown in Fig.~\ref{fig:GauginoHiggsino}. Unlike the gaugino contribution to the 1$^{\rm st}$ generation mass from Fig.~\ref{fig:up}, there is no gaugino mass insertion in this diagram and thus no parametric suppression. Therefore, one would expect that these diagrams are important, but they turn out to be small. One needs the full treatment of rotation matrices above to understand why they are suppressed.  Taking the full fermion and scalar rotations, we get the following contribution to the up-type Yukawa matrix:
\beq
y^u_{3i} \, = \, \lambda_U \, s_u \, r^q_i \, \Gamma^{{ij}^*}_{\tilde{q}} \, \Gamma^{4j}_{\tilde{q}} \, C_{Fk} \frac{\alpha_k}{\pi} \, \mathcal{G}(\mu_H,\, m_{{\rm{ino}_k}},\, m_{\tilde{q}_j}),
\label{eq:ghup}
\eeq
where $\mathcal{G}$ is a dimensionless loop function given in Eq.~\eqref{eq:fonfct}, the $j$ index sums over $q$-type scalar mass eigenstates, and $k$ sums over the gauginos that couple to $q$ and the Higgs, $SU(2)_L$ and $U(1)_Y$.  The factor $r^q_i \equiv (1,1, c_q )$ accounts for rotations in the fermion sector.  A more explicit expression for $y^u_{3i}$ is given in Eq.~\eqref{ghlpd3i}. To get the analogous $y^u_{i3}$ contribution, we would replace $q \leftrightarrow u$, and only the bino would contribute.  

The additional suppression for these terms comes from the initial product of rotation matrices.  By the convention set below Eq.~\eqref{eq:rotconvq}, the ``4$^{\rm th}$'' gauge index corresponds to the $\tilde{Q}$ smessenger, while the $i$ index goes from 1-3 over the MSSM fields.  Thus, in the limit that the $\tilde{q}$ scalars are all mass degenerate, Eq.~\eqref{eq:ghup} vanishes exactly.  That is not the generic situation, but since $\mathcal{G}$ has only logarithmic dependence on mass, there is still a large residual cancellation.\footnote{Interestingly, the diagram formally diverges and requires regularization.  In dimensional regularization, the 1/$\epsilon$ pole replaces the finite loop function $\mathcal{G}$ in the divergent term.  However, this removes any dependence on the mass eigenstates, and the rotation factor multiplies to zero.  Thus, the contribution is actually finite and has no dependence on the renormalization scale.  There is, however, a renormalization scale dependent contribution to $y^u_{33}$ coming from the $q_3'$ portion of $Q$.  In this case, the rotation matrix prefactor does not cancel upon summing over mass eigenstates.  However, for our numerical analysis, we do not include one-loop corrections to $y_t$, and therefore drop this contribution as well as the finite one to $y^u_{33}$ from Eq.~\eqref{eq:ghup}.}  In practice, these ``gaugino-Higgsino'' loops are suppressed compared to any other one-loop contribution, and are even typically smaller than the parametrically two-loop contributions that generate 1$^{\rm st}$ generation masses.  In the `13' and `31' entries though, they can have important subleading effects, and so we include them in our computations.

\subsubsection{Wavefunction renormalization}
\label{subsubsec:wfrn}

\begin{figure}
\centering
\includegraphics[width=\textwidth]{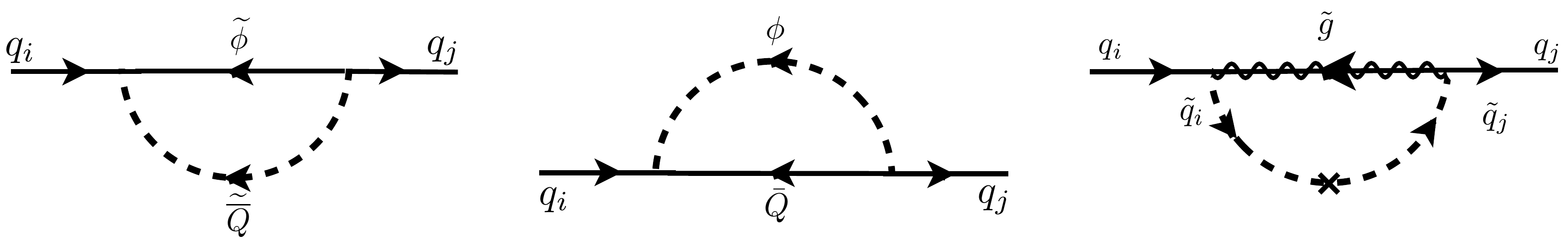}
\captionsetup{justification=raggedright,singlelinecheck=false}
\caption{Diagrams that induce flavor-violating wavefunction renormalization for the fermions. The left two diagrams only contribute to the 2$^{\rm nd}$ and 3$^{\rm rd}$ generations, while the one on the right is present for all fermions.}
\label{fig:wfcnRen}
\end{figure}

It is well-known that in radiative flavor models we can get wavefunction renormalization at one loop from the same dynamics that generates masses.  In our case, the SM quark superfields are renormalized by flavon-messenger, flavino-smessenger, and gaugino contributions as shown in Fig.~\ref{fig:wfcnRen}.\footnote{We also include the renormalization of $u_3', q_3'$ due to Higgs superfields.}  The Higgs also receives wavefunction renormalization from the fields which it has large couplings to, the messenger and 3$^{\rm rd}$ generation quark superfields. In our computations here we will neglect flavon and flavino loops in the down sector because they are suppressed by $y_b^2$ in the 2$^{\rm nd}$ and 3$^{\rm rd}$ generations.  Effects involving $d_4$ and $D$ can be larger, but since $m_b \ll \msc$, to a good approximation we neglect kinetic mixing between the vector-like and the SM $d$ quarks.  We find that with the benchmark parameters presented in Sec.~\ref{subsec:benchmark}, including the heavy fermions in wavefunction renormalization only changed our SM quark predictions at the 1\% level, and is thus below our working precision.  Furthermore, calculating the one-loop shift in the mass of the $d_4$ and $D$-like quarks themselves is beyond our scope.  

The one-loop wavefunction renormalization diagrams contribute to the usual $Z_i$ factors for all the SM fields. Taking the up-type as an example, the Yukawa coupling  $ q\,\mathbf{y_u}\, u\, h$, is modified to
\be
\mathbf{y_u} \rightarrow (\boldsymbol{Z_q})^{1/2}\,  \mathbf{y_u} (\boldsymbol{Z_u})^{1/2} \,(Z_h)^{1/2},
\label{eq:wfrn}
\ee
where $Z_i=1-\Sigma^i$ with $\Sigma^i$ being the possibly divergent loop contributions whose one-loop expressions are given in App.~\ref{app:wfrnlist}. For fermions, we will use the conventions and notation of~\cite{Dreiner:2008tw}.  We evaluate the divergent contributions at a common scale $\mu = 1$ PeV because that is where the heavy fields are integrated out. Errors induced from the fact that not all the heavy fields are exactly at 1 PeV are logarithmic in the change in mass and formally of higher loop order. In Eq.~\eqref{eq:wfrn}, we have bolded the terms which are matrices in flavor space. The $\boldsymbol{Z}$ factors for the quarks will in general have off diagonal components, particularly the gluino contribution because of the large squark mixing. Thus we see that wavefunction renormalization is a potentially important effect that not only rescales individual elements of the Yukawa matrices, but also rotates among them. The approximate size of the effects is an increase in the Yukawa couplings of 5-15\%.

\subsection{Standard Model Flavor Parameters}
\label{subsec:benchmark}

As laid out in Sec.~\ref{sec:model}, our model has the right parametric behavior to explain the generational hierarchy of the Standard Model fermion Yukawas and the parameters of the CKM matrix.  Using the equations of Sec.~\ref{subsec:wfrn} and App.~\ref{app:loop}, we find a set of parameters that reproduce the SM quark masses, CKM angles, and phase to within 5\% of their values listed in \cite{Xing:2011aa} for the former and \cite{Antusch:2013jca} for the latter.  
\begin{table}[tp]
\captionsetup{justification=raggedright,singlelinecheck=false}
\caption{Classes of contributions we include for up and down-type Yukawa matrix entries. Complete loop-level formulas are given in App.~\ref{app:loop}, along with those for wavefunction renormalization, which we apply to all entries.  The tree-level expressions are found in Eqs.~\eqref{eq:yt} and \eqref{eq:yb} for $y_t,\, y_b$.  For every entry listed, we include the same class of diagrams for its transpose.}
\begin{center}
\begin{tabular}{|c|c|} \hline
$y_{11},\,y_{12},$ & gluino, \\ 
$y_{13}$ & gaugino-Higgsino \\ \hline 
$y_{22},\,y_{23}$ & flavino, flavon, gluino, \\
 & gaugino-Higgsino \\ \hline 
$y_{33}$ & tree-level \\ \hline
\end{tabular}
\end{center}
\label{tab:conts}
\end{table}
We list the contributions computed in Tab.~\ref{tab:conts}.  
Despite the close agreement we have obtained with the SM in the quark sector, it is important to stress that there are sources of uncertainty in our calculation discussed below at the level of $\mo(15\%)$.  The proximity of our current results to the SM values is meant as a demonstration of the control one has in recovering the SM.  Thus, the inclusion of subleading corrections to the results we have obtained will likely provide no fundamental obstacle to precise recovery.  

\begin{figure}
\centering
\includegraphics[width=0.75\textwidth]{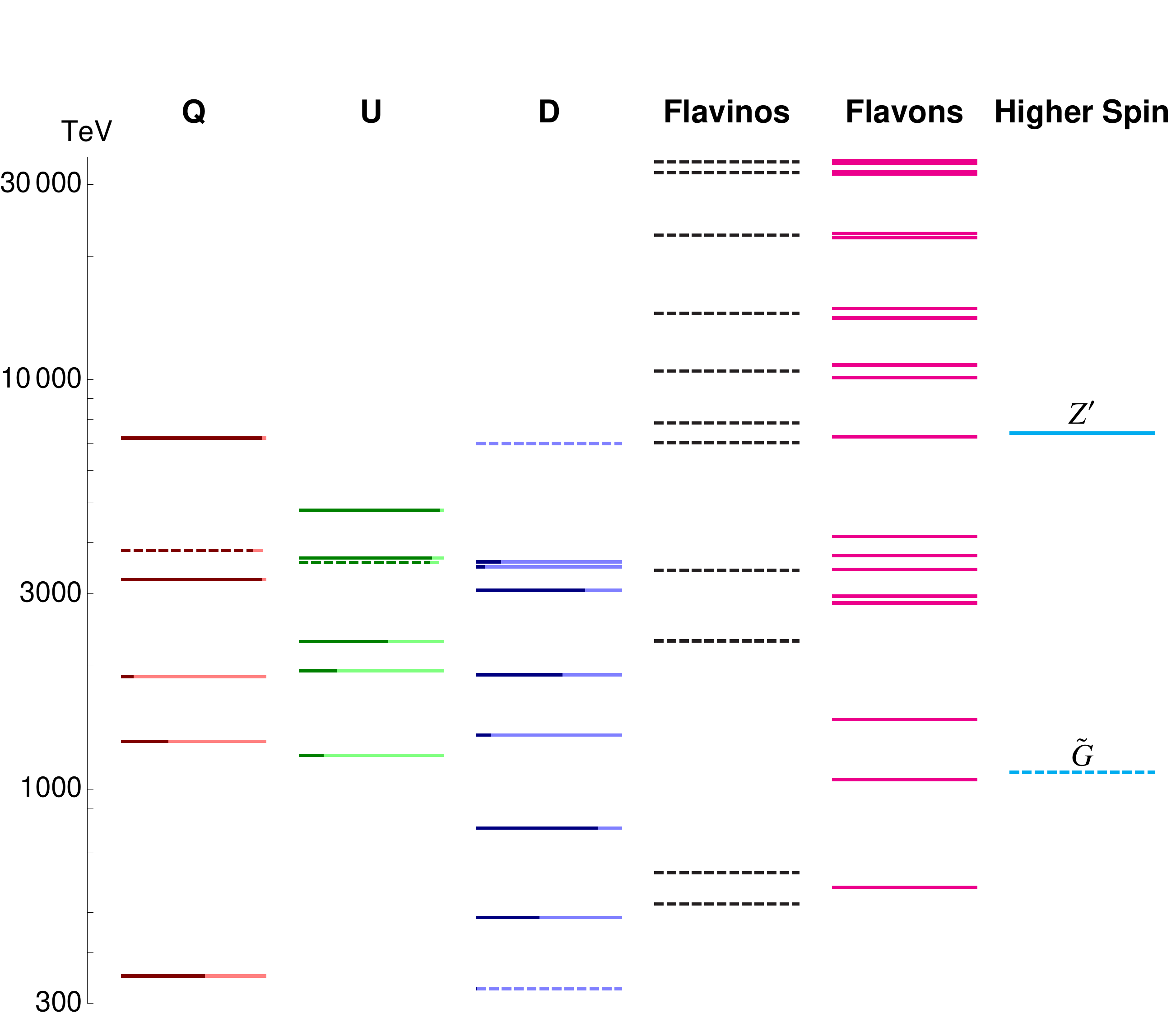}
\captionsetup{justification=raggedright,
singlelinecheck=false}
\caption{Spectrum of non-SM particles for parameters that closely reproduce the SM quark sector.  Solid lines are bosons and dashed lines are fermions.  Shading under $Q,\,U,\,D$ indicates the portion of the mass eigenstate given by MSSM gauge eigenstates (dark) or messenger/$d_4$/$\bar d$ gauge eigenstates (light).  We include only the mass mixing in this quantification.  The flavino states also include the $U(1)_F$ gaugino which strongly mixes with them.  The corresponding $U(1)_F$ gauge boson is shown under ``Higher Spin,'' along with the gravitino. As discussed in the text and shown in Fig.~\ref{fig:broadspectrum}, the gauginos are much lighter than all the fields here.}
\label{fig:detspectrum}
\end{figure}

We now discuss the construction of our benchmark and its properties. The spectrum of new particles for these particular parameters is shown in Fig.~\ref{fig:detspectrum}.  
We generate the parameters of our flavon sector randomly. Scanning over $\mo(1)$ values for dimensionless parameters, $\mo(100-1000)$ TeV values for dimensionful ones, and taking phases in general to be $\mo(1)$, we find a vacuum that is stable and breaks $U(1)_F$ symmetry with vevs that can generate all SM masses. We then use the values of $\lamu,\,\lamub,\,\lamd,\,\lamdb$ from Eqs.~\eqref{eq:Wup} and~\eqref{eq:Wdown} (important for 3$^{\rm rd}$ generation), as well as the $f^{q,u,d}$ couplings (2$^{\rm nd}$ generation) and the squark soft masses (1$^{\rm st}$ generation) plus Higgs and messenger $\mu,\,B_\mu$-terms as handles to recover the SM.  If  $\mu$ is too large, that could potentially lead to deeper vacua that are color breaking~\cite{Frere:1983ag,AlvarezGaume:1983gj,Claudson:1983et}, but we check that this is not a problem for our benchmark.

To have a viable thermal WIMP dark matter particle, we fix the wino mass at 3 TeV and obtain the gaugino spectrum ($m_{\tilde{B}}$ = 13.3 TeV, $m_{\tilde{W}}$ = 3 TeV, $m_{\tilde{g}}$ = 20.9 TeV) as detailed in Sec.~\ref{sec:gaugino}.\footnote{The $U(1)_F$ gaugino mixes strongly with the flavinos and has mass $\mathcal{O}({\rm PeV})$.  We did not compute its full soft mass from anomaly mediation, but the flavino spectrum is highly insensitive to its detailed value if within a few orders of magnitude of the other gauginos.}  The gluino mass offers an additional means to control $m_{u,\,d}$.  In the down sector we subject $f_d^{2,3}$ to the technically natural tuning at $\mo(0.1)$.  Dimensionful values were again $\mo(100-1000)$ TeV.  The only nontrivial constraint comes from kaon physics, further detailed in Sec.~\ref{subsec:mm}, and it favors having $Q$ and $D$ states $\gtrsim$ 1000 TeV.

\begin{figure}[ht]
\centering
\includegraphics[width=0.5\textwidth]{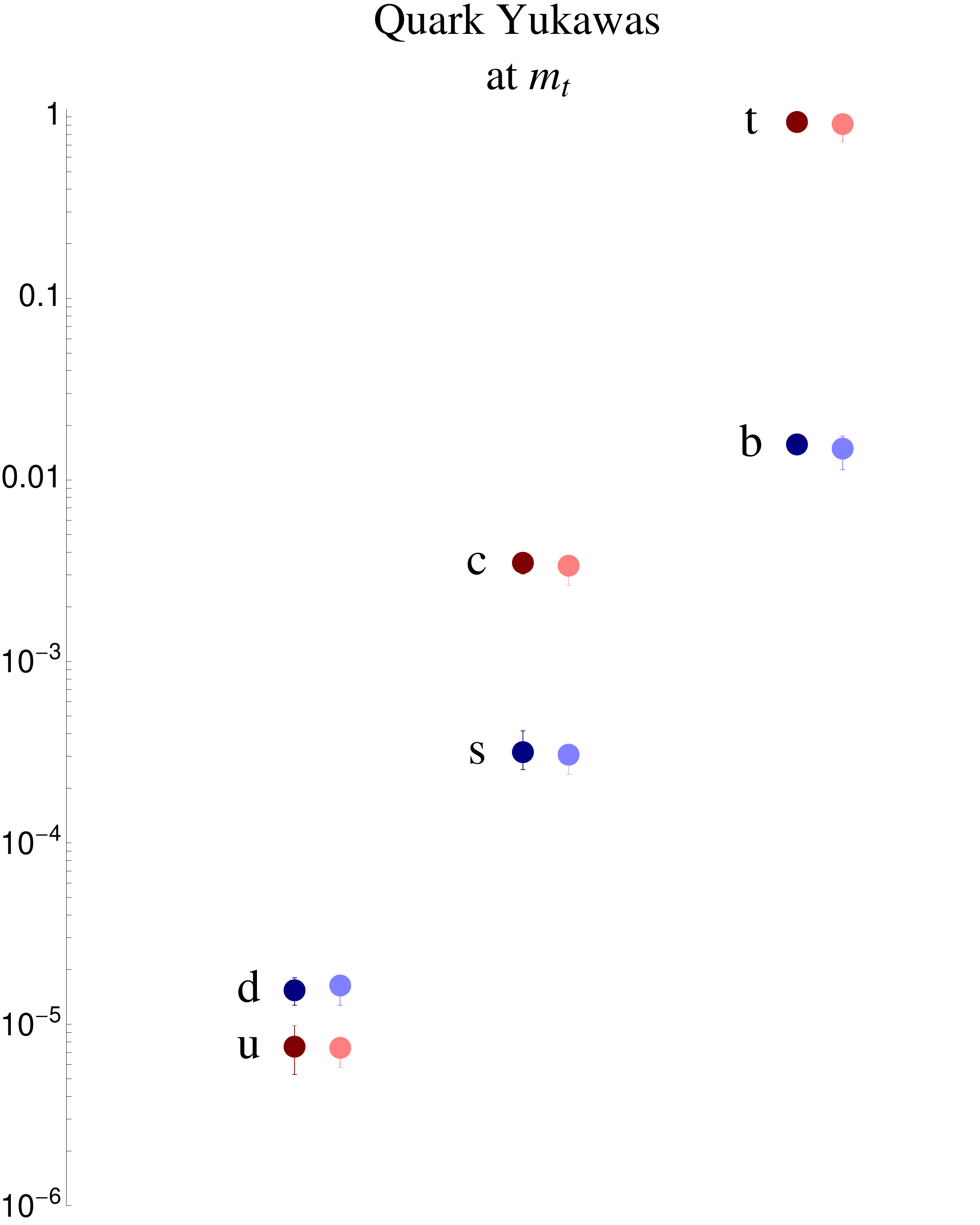}
\captionsetup{justification=raggedright,
singlelinecheck=false}
\caption{SM values from \cite{Xing:2011aa} (dark) and values obtained in our benchmark (light) at the scale $m_t$.  Errors bars for the latter assume uniform shifts in Yukawas by $+15\%, -25\%$ at 1000 TeV, which accounts for a uniform uncertainty of $\pm15\%$ in addition to a 10\% decrease coming from choosing a renormalization scale that is lower than the mass of some of the states (see text).  After applying these uncertainties in the UV, we run the Yukawas to $m_t$.  }
\label{fig:smVsUs}
\end{figure}

\begin{figure}[ht]
\centering
\includegraphics[width=0.6\textwidth]{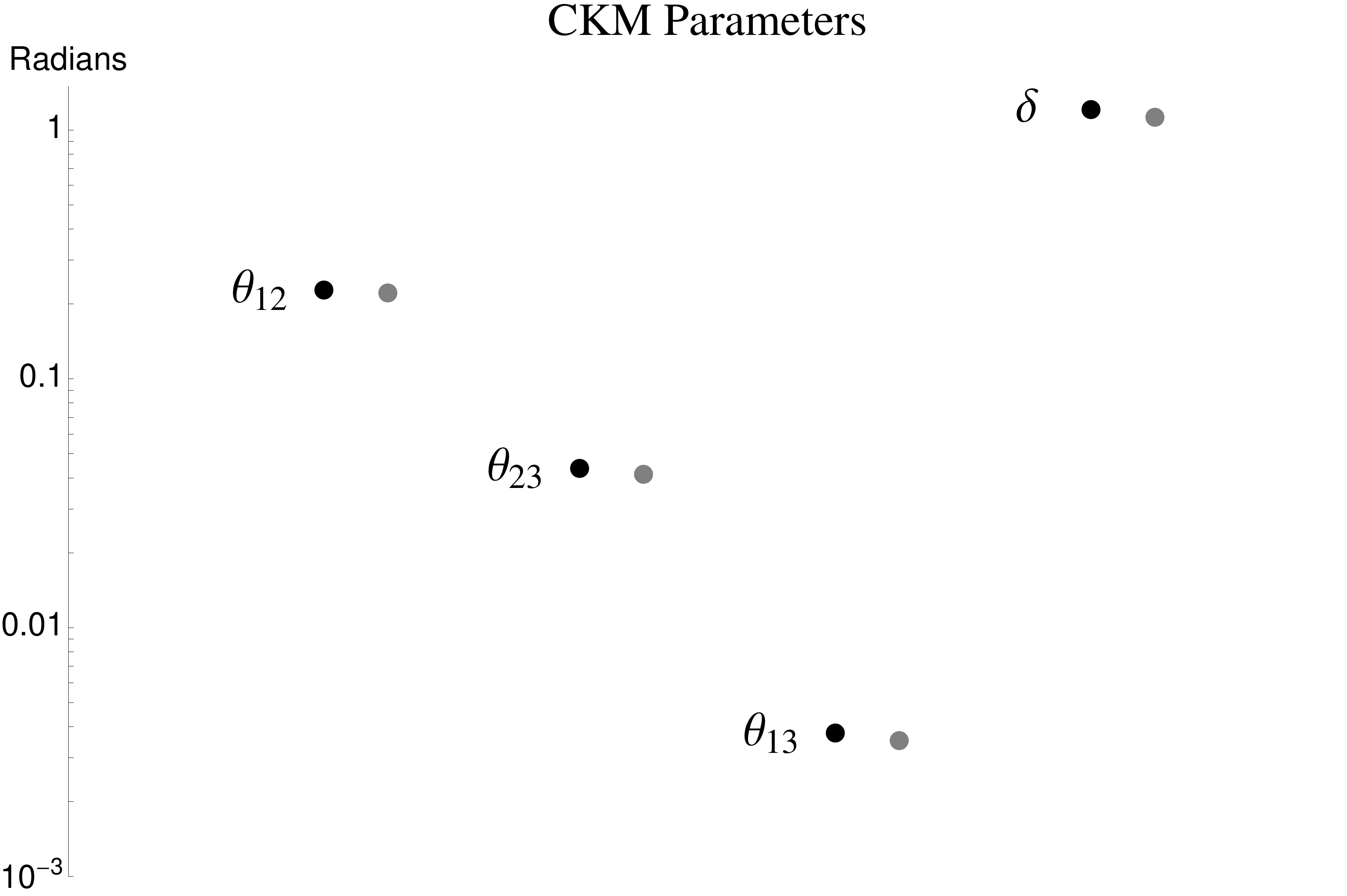}
\captionsetup{justification=raggedright,
singlelinecheck=false}
\caption{SM values from \cite{Antusch:2013jca} (dark) and values obtained in our benchmark (light).  Error bars are smaller than the dot size.  }
\label{fig:ckm}
\end{figure}

For comparison with the SM, we show the values we obtain for our Yukawas at the scale $m_t$ in Fig.~\ref{fig:smVsUs} compared to those depicted earlier for the SM (Fig.~\ref{fig:sm}).  
In Fig.~\ref{fig:ckm} we compare the CKM of our benchmark to that of the SM.\footnote{We have neglected the small running of the CKM parameters, which affects $\theta_{13}$ and $\theta_{23}$ most, at the level of a few percent \cite{Antusch:2013jca}.  We take that reference's SM values at 10 TeV computed in ${\rm \overline{MS}}$ to compare to those in our model evaluated at 1000 TeV.}
For the ten SM quark parameters shown here, the mean discrepancy with the SM is 4\%, though as mentioned above, our results have an uncertainty of $\mo(10-20\%)$. The leading effects that we are currently neglecting include: 1) Some dimensionless couplings are $\approx$ 1.3, leading to $\mo(15\%)$ corrections at next-to-leading order; 2) Including wavefunction renormalization induces scale dependence. We evaluate at a common scale of 1000 TeV before integrating out all non-SM fields besides the gauginos. However, many of our messenger fields are above 1 PeV (with the heaviest at 7.2 PeV), and thus there are $\mo(1)$ logs we are not currently resumming. Changing the renormalization scale from 1000 to 10,000 TeV, decreases our Yukawa values by $\mo(10\%)$;  3) For $y_{33}^{u,d}$, we only include the tree-level values given in Eq.~\eqref{eq:yt} for $y_t$ and Eq.~\eqref{eq:yb} for $y_b$.  The one-loop corrections to these entries could shift them at the level of a few percent; 4) To compute quark masses and the CKM, we just take the 3x3 matrices in the up and down sectors. However, there are additional Yukawas with the messenger fermions, $Q,U,D$ and $d_4$. There will also be kinetic mixing from one-loop wavefunction renormalization. Taking the values for our benchmark in the down sector, where we expect the effects to be strongest due to $d_4$, we found shifts in quark masses at the level of 1-2\% for $y_b$ and $y_d$, with $y_s$ changing negligibly. Thus, we neglect this contribution as well; 5) For our gaugino loops that contribute strongly to 1$^{\rm st}$ generation masses ({\it cf.}~Fig.~\ref{fig:up}), we have treated the threshold correction due to messengers as a mass-insertion, even though these same messengers appear elsewhere dynamically in the loop.  Including the full momentum-dependence of the one-loop correction to the gaugino propagator shifted our masses by $\lesssim 1\%$, which is beyond our precision and we thus ignore this effect. 

Given the agreement in the quark sector, it would be an interesting exercise to reproduce charged leptons as well, something we did not attempt here.  We would wish to maintain consistency with unification, so our $\lambda_{L,E}$ and $f^{l,e}$ couplings would need to be determined for the values we assigned to the quark-sector.  The slepton soft masses and bino mass would offer independent means to control the lepton masses.  In Sec.~\ref{sec:conc}, we sketch a possible model extension that would generate neutrino masses and mixing angles.  Before moving on we note that many of these sources of uncertainty affect the 3$^{\rm rd}$ generation most strongly. Since those Yukawas are dominated by the tree-level contribution, we expect them to be the simplest to adjust once these additional effects are taken into account.

%%%%%%%%%%%%%%%%%%%%%%
\section{Experimental Constraints and Signatures}
\label{sec:constraints}
%%%%%%%%%%%%%%%%%%%%%%

Detailed studies of the low-energy constraints on Mini-Split SUSY have been performed in~\cite{McKeen:2013dma,Moroi:2013sfa,Eliaz:2013aaa,Altmannshofer:2013lfa}. The dominant processes are meson mixing, electric and chromoelectric dipole moments (CEDM), and lepton flavor violation.  In addition to the MSSM fields previously studied, the messengers and flavons contribute to these observables.  The latter leads to large deviations from the predictions of minimal Mini-Split SUSY for processes that only involve the 2$^{\rm nd}$ and 3$^{\rm rd}$ generations of the $\mathbf{10}$ fields, $q$, $u$ and $e$.

The strongest bound comes from $CP$ violation in $K-\overline{K}$ mixing, which requires that the squarks that have large couplings to the gluino and $s$ and $d$ quarks be heavier than a few hundred TeV.  It is the only constraint we needed to compute in detail to test the viability of the benchmark in Sec.~\ref{subsec:benchmark}.  Since it involves the 1$^{\rm st}$ generation, it is essentially a probe of the Mini-Split MSSM, though we account for the presence of messengers.  While the limits from other observables are weaker, we discuss the contributions from flavon dynamics where they contribute strongly and present some detailed formulas in App.~\ref{app:flavobs}.  It will take many generations of future experiments to probe this sector.  However, improved determination on the lattice of kaon parameters could provide evidence for one of the key ingredients of our model, the presence of anarchic squark mixing at several hundred TeV.

\subsection{Meson Mixing}
\label{subsec:mm}

For the case of meson mixing, the MSSM effect is mediated by box diagrams with gluinos and squarks in the loops, but we neglect contributions with gluino mass insertions.  Bounds are independent of the gluino mass as long as it is much lighter than the squarks.  For our benchmark model, we check that the MSSM contribution does not run afoul of kaon constraints.  We use the full mass eigenstate calculation of the squark-gluino box presented in \cite{Altmannshofer:2007cs} to account for the $\mo(1)$ mixing among different squark gauge eigenstates and with the messenger sector.  After matching to the relevant dimension-six operators at 1000 TeV, we run our Wilson coefficients at NLO to 2 GeV using the procedure outlined in \cite{Kersten:2012ed}, from which we also take numerical values for the bag parameters.  For the benchmark in Sec.~\ref{subsec:benchmark}, we get
\be
\epsilon_K^{\rm NP} = 9.4 \times 10^{-5} \, , \nonumber \\
\Delta m^{\rm NP}_{K} = 2 \times 10^{-15} \, {\rm GeV}.
\ee
Our contribution to $\Delta m_{K}$ is safe by three orders of magnitude.  The limit on $\epsilon_K^{\rm NP}$ is 1$\times 10^{-3}$~\cite{Kersten:2012ed,Altmannshofer:2013lfa}, so while our benchmark is safe, there are reasonable regions of parameter space in the model which are excluded by this observable. Thus, an improvement in $\epsilon_K^{\rm SM}$ by an order of magnitude could be the best low energy way of probing the Mini-Split scenario. 

In our model, there is a similar box diagram with flavino and messenger scalars in the loops. The only meson which is precisely measured and does not involve any 1$^{\rm st}$ generation quarks is the $B_s$. Therefore, the operator 
\beq
\mo_1 = (\bar s \gamma_\mu P_L b)( \bar s \gamma^\mu P_L b),
\label{eq:o1}
\eeq
which contributes to $B_s$ mixing will be modified by an $\mathcal{O}(1)$ amount relative to the MSSM, while the operators with other chiralities will be suppressed by powers of $y_b$.   

We calculate the smessenger-flavino as well as the messenger-flavon diagrams that generate $B_s$ mixing.  Obtaining the Wilson coefficient for $\mo_1$, we relate it to quantities in the $B$ meson system following the approach of \cite{Gabbiani:1996hi}, using more recent numerical values from the lattice study in \cite{Carrasco:2013zta}.  The detailed box diagram calculations are given in App.~\ref{app:flavobs}.  We get a contribution to the mass splitting $\Delta M_s = \mo(10^{-20})$ GeV, compared to the SM value, $\Delta M_s = 1.2 \times 10^{-11}$ GeV and a shift in the total CP violating phase of $\mo(10^{-11})$ .  Thus, experimental evidence is beyond the next generation of experiments.

\subsection{(Chromo)Electric Dipole Moments}
\label{subsec:cedm}

\begin{figure}
\centering
\includegraphics[width=0.9\textwidth]{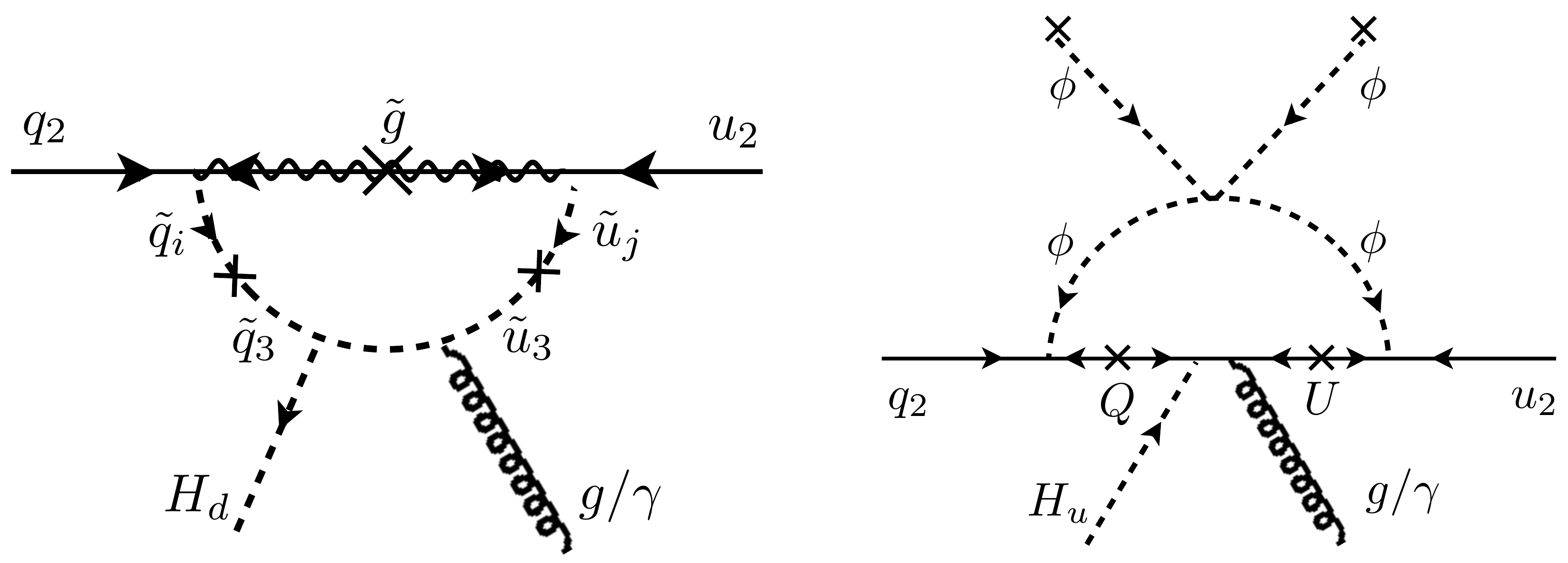}
\captionsetup{justification=raggedright,singlelinecheck=false}
\caption{L: An example diagram of the MSSM gluino contribution to quark (C)EDMs. R: An example of the flavon contribution to quark (C)EDMs.}
\label{fig:edm}
\end{figure}

In the MSSM, (C)EDMs for all up-type quarks come from a one-loop diagram of the type shown on the left side of Fig.~\ref{fig:edm}. This diagram has a gluino mass insertion, so it is proportional to $m_{\tilde{g}}/m_{\tilde{q}} \sim \mo(10^{-2})$. These diagrams are comparable for $u$, $c$, and $t$ if the squarks are anarchic in flavor space, but the strongest experimental bound comes from the up EDM. On the other hand, this model has one-loop diagrams with flavons and messengers going around the loop as shown on the right side of Fig.~\ref{fig:edm}, as well as the supersymmetrized version with flavinos and smessengers.  All the internal fields in this diagram have mass $\mo(m_{\tilde{q}})$, so its effects are enhanced relative to the MSSM. Because the flavons only couple to the 2$^{\rm nd}$ and 3$^{\rm rd}$ generation, these types of diagrams only induce (C)EDMs for top and charm. 

The strongest bounds on these processes come from chromo-EDMs inducing contributions to the neutron EDM. For the top quark, the bound was computed to be $|\tilde{d}_t| \lesssim 1/(100$ TeV)~\cite{Kamenik:2011dk}. This computation uses the fact that there is a separation of scales between the top and $\Lambda_{\rm QCD}$ and runs operators down from the top mass to the QCD scale. We can approximate the bounds on the charm mass by ignoring the running effects besides the scale of $\alpha_s$. Because integrating out a quark generates a finite contribution~\cite{Braaten:1990gq,Chang:1991ry} to the Weinberg operator~\cite{Weinberg:1989dx}, integrating out a lighter quark will lead to a larger contribution to the neutron EDM. Furthermore, the gluon loop that generates this operator is larger because $\alpha_s$ is evaluated at $m_c$ where it is much larger. We approximate $\alpha_s(m_c)\simeq \alpha_s(m_\tau) \simeq 0.35$~\cite{Ackerstaff:1998yj}. Because of these effects, the bounds on $\tilde{d}_c$ are much stronger than on top, and we find $|\tilde{d}_c| \lesssim (6 \times 10^5 \, {\rm TeV})^{-1}$, in rough agreement with the bound of $|\tilde{d}_c| \lesssim (2 \times 10^5 \, {\rm TeV})^{-1}$ from the more detailed study in \cite{Sala:2013osa}. These limits should be taken as accurate to within an order of magnitude because of the uncertainties on the hadronic matrix elements that go into the conversion of the Weinberg operator into the neutron EDM.   

Taking the CEDM of the charm quark as an example, the low energy operator is of the form
\be
-i \frac{\tilde{d}_c}{2} g_s \, \bar{c}  \, \sigma^{\mu\nu} \, \gamma^5 \, t^a  c\, G^a_{\mu\nu}.
\label{eq:edm-op}
\ee
At the scale of electroweak symmetry breaking, this operator matches onto an operator involving the left-handed quark doublet, the right-handed singlet, and the Higgs. This can be seen from the fact that the tensor operator above flips the helicity of the quark, so it must involve a Higgs insertion. In the UV at the scale of SUSY breaking, this operator is generated by diagrams of the type shown in Fig.~\ref{fig:edm}. As discussed above, the diagram on the right is the dominant contribution for the charm and top quarks, and we can estimate its size to be parametrically $\mathcal{O}(v/16\pi^2 m_{\rm mess}^2) \simeq (10^{10}$ TeV$)^{-1}$ for $m_{\rm mess}$ = 3000 TeV.  

We improve on the one-loop estimate by computing 1) the generalization of the right diagram in Fig.~\ref{fig:edm} to flavon and fermion mass-eigenstate basis, 2) an additional flavon-messenger diagram with no mass insertions (besides the SM Higgs vev) proportional to $\bar{\lambda}_{U}$, and 3) the corresponding flavino-smessenger contribution.  We project onto the Dirac structure of a chromoelectric dipole and obtain a numerical value by setting the Higgs to its vev, even though we are formally matching at 1000 TeV.\footnote{Since the calculated values of $\tilde{d}_{c,t}$ are so far below current bounds, the effects of running to the fermion mass scale will not change our conclusions by much beyond an order of magnitude.}  Using the sign and normalization conventions of \cite{Altmannshofer:2013lfa}, we get $|\tilde{d}_t| = (4\times10^{12}$ TeV)$^{-1}$ and $|\tilde{d}_c| =(1.2\times10^{12} \, {\rm TeV})^{-1}$, a bit below our estimate above since couplings and mixing angles are accounted for, and far removed from near future sensitivity.  We give the expressions for the flavon-sector loop contributions in App.~\ref{app:flavobs}. 

Analogues to the operator in Eq.~\eqref{eq:edm-op} for the down and lepton sector will be suppressed by $\mathcal{O}(y_b)\sim \mathcal{O}(y_\tau)$.  This is due to the small coupling of the $\bar{ \mathbf{5}}$ to $\chi$, as explained in Sec.~\ref{sec:down-lep}.  Therefore, the strange and bottom (C)EDMs are enhanced relative to the MSSM diagrams by $y_b m_{\tilde{q}}/m_{\tilde{g}} \sim 10$. For the strange quark, we take the formula from~\cite{Hisano:2012cc} to estimate an experimental bound of $\tilde{d}_s \lesssim (3\times10^{6}$ TeV$)^{-1}$, while the natural size in this model is $ \tilde{d}_s \simeq y_b \, \tilde{d}_c \simeq  (10^{11}$ TeV$)^{-1}$. The CEDM for the $b$-quark can be computed in the same way.  Thus, we see that the model is safe from (C)EDM measurement until several order of magnitude improvement is achieved.  

\subsection{Lepton Flavor Violation}

Lepton flavor violation (LFV) is also a strong constraint on models with anarchic flavor structure, with $\mu \rightarrow e \gamma$ currently imposing the most stringent constraint in the MSSM. The diagrams for LFV have the same structure as those for EDMs shown in Fig.~\ref{fig:edm}, so loops of flavons and messengers are enhanced by $\mathcal{O}(y_\tau m_{\tilde{q}} /m_{\tilde{B}}) \sim 10$ relative to the MSSM diagrams, but only for processes involving only 2$^{\rm nd}$ and 3$^{\rm rd}$ generation leptons. In this case, that means $\tau\rightarrow\mu\gamma$ and other rare $\tau$ decays are enhanced.  In calculating the contribution of our model to charm and top quark (C)EDMs in Sec.~\ref{subsec:cedm}, we also obtained comparable values for the flavor changing dipole operators.  We can use our values in the up quark sector to estimate the contribution to the analogous lepton operator, which is given schematically as 
\beq
\frac{e\, m_\tau}{16\pi^2m_{\rm sc}^2} \, \bar \tau \, \sigma^{\mu\nu} \, \mu\, F_{\mu\nu} \, .
\label{eq:lepedm-op}
\eeq
We expect this to be of similar order as the charm EDM.  Naively, the numerator of the coefficient should be $v$ since the Higgs insertion in Fig.~\ref{fig:edm} is on an internal line which has a large Yukawa coupling. On the other hand, there is a factor of $y_\tau$ coming from the coupling of the left-handed lepton to the flavon, so we can combine that with $v$ to get $m_\tau$. 
Taking into account $\alpha_{\rm EM}$, we estimate BR$(\tau \rightarrow \mu \gamma) \sim  10^{-19}$. The current limit is $\mo(10^{-8})$ with the possibility of a one to two order of magnitude improvement at a future $\tau$ factory.  Thus, this will unfortunately not provide a means to detect the flavor violation in our model in the near future.  The contributions to $\Delta F$ = 1 processes in the quark sector are also significantly below the current experimental limits.

\subsection{Proton Decay}

The problem of proton decay is of a somewhat different nature than the other constraints. None of the terms in the renormalizable Lagrangian induce proton decay, but there are higher dimensional operators allowed by all the symmetries of the theory, such as the dimension five superpotential operator $qqq\ell$, that do. It has long been known that this is a problem in weak-scale SUSY~\cite{Weinberg:1981wj,Sakai:1981pk,Nath:1985ub,Hinchliffe:1992ad}. Raising the scalar masses weakens the bounds, but recent analysis~\cite{McKeen:2013dma,Dine:2013nga,Hisano:2013exa} has shown that this is still a problem, with the cutoff for dimension five operators needing to be higher than the Planck scale to make the proton live long enough. 

One could imagine building a model in the spirit of this one such that $U(1)_F$ forbids the higher dimensional proton decay operators. These kinds of charge assignments tend to be anomalous,\footnote{The $U(1)_F$ could be a spontaneously broken global symmetry with anomalous charges as in~\cite{Dobrescu:2008sz}.  The anomaly will generate a mass for the Goldstone, but additional explicit breaking will likely be needed to raise it higher.  Given the IR issues induced by adding this light state and the need to control corrections to the radiative story from having a merely approximate symmetry, we forego this possibility, though there may be a viable implementation.} 
 so we go in a different direction here by noting that because proton decay is mediated  by higher dimensional operators, it is clearly sensitive to the UV structure of the theory.
Proton decay operators are generically generated by the physics of Grand Unification, but they need not be, as in the case of higher dimensional GUTs~\cite{Hall:2001pg}. As proton decay is a generic problem for all SUSY models and in particular for SUSY GUTs, we simply assume that one of the solutions in the literature, such as~\cite{Hall:2001pg}, is active in the UV but has no impact on scales below the unification scale.

%%%%%%%%%%%%%%%%%%%%%%
\section{Conclusions}
\label{sec:conc}
%%%%%%%%%%%%%%%%%%%%%%

Supersymmetry has been a subject of intense study because of its many interesting theoretical and phenomenological features. As an extension of the Standard Model, it can solve the hierarchy problem, provide a natural WIMP dark matter candidate, and improve gauge coupling unification. The unfortunate lack of evidence for SUSY at the LHC as well as the (fortunate) discovery of a Higgs with mass around 125 GeV has led to a reconsideration of weak-scale SUSY, with Mini-Split SUSY emerging as a framework with many intriguing features. In particular, with scalars around 1000 TeV and gauginos one loop factor lighter, the correct Higgs mass is obtained with dark matter and unification stories being comparably successful. SUSY would then only partially solve the hierarchy problem, leaving us with a meso-tuned picture of the universe.

In this paper, we have explored how an additional feature of Mini-Split SUSY, the automatic solution of the SUSY flavor problem, can be used to address the SM flavor puzzle. In the Standard Model, there is no explanation for the peculiar structure of the masses and mixings of the quarks and leptons. Each generation is substantially lighter than the previous one, and the ratio of 3$^{\rm rd}$ to 2$^{\rm nd}$ generation masses appears remarkably similar to the ratio of 2$^{\rm rd}$ to 1$^{\rm nd}$ generation masses. Thus, one possible explanation of the SM flavor sector is that fermion masses are generated via a hierarchy of loops, with the 3$^{\rm rd}$ generation Yukawa coupling generated at tree level, the 2$^{\rm nd}$ at one loop, and the 1$^{\rm st}$ at two loops: a radiative explanation of flavor. 

In the framework of Mini-Split, the scalars carry flavor quantum numbers and, unlike in weak-scale SUSY, there can be significant mixing between the different flavors of squarks. This mixing can be used in loops to generate the Yukawa couplings. Because new Yukawa couplings cannot be generated by loops in supersymmetric theories, the physics of flavor must be tied to the physics of SUSY breaking. Here we have built a model which radiates flavor around 1000 TeV, the scale which the Higgs mass points to. The full particle content of the model is given in Tab.~\ref{tab:fields}. 

In the UV, this model forbids all Yukawa couplings to the SM matter with a new $U(1)$ symmetry under which the Higgs is charged, but all matter is neutral. SUSY breaking triggers the breaking of $U(1)_F$, and a Yukawa coupling is communicated via a rank 1 messenger sector. This allows only the 3$^{\rm rd}$ generation to get a Yukawa coupling at tree level. The messengers can then generate additional Yukawas at one loop, but because of the size of the messenger sector, these loop contributions only affect the 2$^{\rm nd}$ and 3$^{\rm rd}$ generation. Finally, there is the loop contribution from the sfermions, which is parametrically of two-loop order and involves all generations. This two-loop contribution is only big enough because there is large flavor mixing in the sfermion sector. 

In addition to building a model and giving parametric estimates of the size of all the flavor parameters, we have also computed a detailed spectrum for the quark sector taking into account all leading effects including mixing and wavefucntion renormalization. We have shown that one can get agreement with all the SM flavor parameters to within 5\% at a generic point in parameter space described in Sec.~\ref{sec:details}. We have also computed current constraints and found most of them to be trivially satisfied; however, the constraints from the Kaon system do exclude some of the parameter space. The phenomenology of this model is quite similar to Mini-Split SUSY, but in principle there are deviations in flavor observables involving the 2$^{\rm nd}$ and 3$^{\rm rd}$ generation, such as $B_s$ mixing.

In order to build a complete flavor model, neutrinos must also be included. One can think of neutrino masses as coming from the usual SM dimension five operator. Once the $U(1)_F$ is included, this operator can be generated by either of the following dimension 7 superpotential operators
\be
\frac{1}{M_*^3}  (\ell H_u)(\ell H_u)\bar{\chi}\,\phi \qquad \qquad \frac{1}{M_*^3}  (\ell H_u)(\ell H_u)\xi^2\, ,
\ee
where we have suppressed flavor indices. In this case, the neutrino masses will be given by 
$m_\nu \sim v^2 \langle \bar{\chi} \rangle  \langle \phi \rangle/M_*^3$
for the first operator, and the generalization is clear for the second. Here $v \simeq 174$ GeV is the electroweak scale. In the benchmark given in Sec.~\ref{subsec:benchmark}, the vevs of the flavons are of order $100 - 1000$ TeV, so $M_*$ can be as low as 100 PeV to reproduce the experimentally measured neutrino masses. This scale is somewhat above the scale of the model, but not dramatically. These operators can be UV completed with vectorlike right-handed neutrinos with different $F$ charges, but we leave this analysis including the computation of the neutrino mixing to future work. 

Stepping back, we see that while the lack of evidence for SUSY at the LHC is beginning to close the door on weak-scale SUSY, a window is perhaps opening into the Mini-Split paradigm. Through this window, we have envisioned a solution to the SM flavor puzzle, explaining the many hierarchies we have seen through the physics of radiative corrections. Only the ratio of the bottom to top quark masses is left unexplained, but this ratio is correlated with the size of the Cabibbo angle, giving unexpected agreement in both sectors. All other small numbers in the SM flavor sector are the result of loop corrections and a consequence of linear algebra. The theory does not need to distinguish different generations, yet it generates all the flavor hierarchies we observe in nature.

\section*{Acknowledgments}

We would like to thank Wolfgang Altmannshofer, Nima Arkani-Hamed, Fabrizio Caola, Liang Dai, Jacques Distler, David E. Kaplan, Tracy Slatyer, Andreas Weiler, and Jure Zupan for helpful discussions. We are especially thankful to Gian Giudice for helpful discussion on the results of \cite{Giudice:2011cg}. We are also grateful to Leah Stolarski for helping design the figures. MB was supported by DE-FG-03-91ER40682.  MB and DS are appreciative of the support and hospitality provided by The Galileo Galilei Institute for Theoretical Physics and The Kavli Institute for Theoretical Physics where some of this work was completed.

%%%%%%%%%%%%%%%%%%%%%%
\appendix
\section{Field Content and $U(1)_F$ Gauge Symmetry}
\label{app:field}
%%%%%%%%%%%%%%%%%%%%%%

\begin{table}[tp]
\captionsetup{justification=raggedright,
singlelinecheck=false}
\caption{The full particle content of our model in addition to that of the MSSM. We also give the charges under U(1)$_F$, the SM gauge group, and $R$-parity. Note that the MSSM fields $q,u,d,\ell,e$ are neutral under $U(1)_F$ and negative under $R_p$.}
\begin{center}
\begin{tabular}{|c|c|c|c|} \hline
Field & U(1)$_F$ & $SU(3)\times SU(2)\times U(1)$ & $R_p$\\ \hline
$H_u,\, H_d$ & $\mp2$ & $ (\mathbf 1, \mathbf 2)_{1/2}+ (\mathbf 1, \mathbf 2)_{-1/2} $ & $+$ \\ \hline
$Q$, $\bar Q$ & $\pm 1$ & $ (\mathbf 3, \mathbf 2)_{1/6}+ (\bar{\mathbf 3}, \mathbf 2)_{-1/6}$ & $-$ \\ 
$U$, $\bar U$ & $\pm 1$ & $ (\bar{\mathbf 3}, \mathbf 1)_{-2/3}+ (\mathbf 3, \mathbf 1)_{2/3}$  & $-$ \\ 
$E$, $\bar E$ & $\pm 1$ & $ (\mathbf 1, \mathbf 1)_{1}+ (\mathbf 1, \mathbf 1)_{-1}$ & $-$ \\ \hline
$D$, $\bar D$ & $\mp 3$ & $ (\bar{\mathbf 3}, \mathbf 1)_{1/3}+ (\mathbf 3, \mathbf 1)_{-1/3}$  & $-$ \\ 
$L$, $\bar L$ & $\mp 3$ & $ (\mathbf 1, \mathbf 2)_{-1/2}+ (\mathbf 1, \mathbf 2)_{1/2}$ & $-$ \\ \hline
$\ell_4$, $\bar\ell$ & 0 & $ (\mathbf 1, \mathbf 2)_{-1/2}+ (\mathbf 1, \mathbf 2)_{1/2}$ & $-$ \\ 
$d_4$, $\bar d$ & 0 & $ (\bar{\mathbf 3}, \mathbf 1)_{1/3}+ (\mathbf 3, \mathbf 1)_{-1/3}$  & $-$ \\ \hline
$\phi_{1,2}$, $\bar \phi_{1,2}$ & $\pm1$ & $ (\mathbf 1, \mathbf 1)_{0}$ & $+$ \\ 
$\chi_{1,2}$, $\bar \chi_{1,2}$ & $\mp3$ & $ (\mathbf 1, \mathbf 1)_{0}$  & $+$ \\ 
$\xi$, $\bar \xi$ & $\mp2$ & $ (\mathbf 1, \mathbf 1)_{0}$  & $+$ \\ \hline
\end{tabular}
\end{center}
\label{tab:fields}
\end{table}

In this appendix we review the full field content and address some of the complications associated with introducing a new gauge group. The field content is given in Tabs.~\ref{tab:ufields}, \ref{tab:flavons}, and~\ref{tab:dfields}, and we give the full field content here in Tab.~\ref{tab:fields} for completeness. We begin by noting that all the fields in the theory transforming under $U(1)_F$ are vectorlike, so anomaly cancellation is satisfied trivially. This also allows us to write a supersymmetric mass term for all the fields that are not part of the MSSM. By the logic of Eqs.~\eqref{eq:mu-bmu} and~\eqref{eq:muB}, this mass term is $\mathcal{O}(m_{\rm sc})$, so all the scalars and fermions given in Tab.~\ref{tab:fields} are at the PeV scale. The one exception, of course, is the light Higgs, which is tuned to have a mass around 126 GeV. 

Because the new gauge group is a $U(1)$, a Fayet-Iliopoulos~\cite{Fayet:1974jb} (FI) term is allowed by the gauge symmetry. Fortunately, a high scale FI term is inconsistent with supergravity~\cite{Komargodski:2009pc} and will not be generated. We also assume that any intermediate dynamics between the Planck and PeV scales also does not generate an FI term. Another possibility arising from the abelian nature of the new group is kinetic mixing between hypercharge and $U(1)_F$~\cite{Holdom:1985ag}. If hypercharge is embedded in a GUT, then this operator will be absent at the scale of GUT breaking, but it will be generated by loops of fields charged under both $U(1)$'s such as those in Tab.~\ref{tab:fields}. Because this is a loop effect, we will treat it as a perturbation. 

Once $U(1)_F$ is broken, the gauge fields can be diagonalized by shifting the hypercharge field with component of the $F$ gauge field. This has several effects, but all of them turn out be phenomenologically harmless in the context of Mini-Split SUSY. First, the hypercharged fields acquire some $F$ charge. Because $U(1)_F$ is broken at such a high scale, this has no effect in present experiments. The $D$-term for $U(1)_F$ will also be shifted
\be
D'_F = D_F + \epsilon D_Y,
\ee 
where $\epsilon$ is the coefficient of the kinetic mixing operator. The potential goes as $(D'_F)^2$, which when expanded out contains two different effects. The first is a shift in the coefficient of the $U(1)_Y$ $D$-term by $\mathcal{O}(\epsilon^2)$. The second is an effective FI term for hypercharge coming from the cross term. Both of these modify the scalar potential for the hypercharged scalars, but they have no qualitative effect because all these scalars have large masses from SUSY breaking. Therefore, the effects of kinetic mixing on the $D$-term can be thought of as small corrections to the masses and quartics for these scalars.

%%%%%%%%%%%%%%%%%%%%%%
\section{Flavon Sector Details}
\label{app:flavon}
%%%%%%%%%%%%%%%%%%%%%%

 In this Appendix, we explain the field content and charges of the flavon sector and give a brief description of the potential minimization. In the UV, all SM Yukawa couplings are forbidden by $U(1)_F$, so in order to generate any Yukawas, we need flavons to get vevs and spontaneously break $U(1)_F$. Thus we introduce a set of flavons $\phi$, $\bar{\phi}$ with charges $\pm1$. This determines the charges of $H_u$, $Q$ and $U$. In order to preserve anomaly cancellation and allow a $\mu$-term for the Higgses,  $H_d$ must have opposite $F$ charge to $H_u$. Because this is a supersymmetric theory and Yukawa couplings are superpotential operators, the down-Yukawa coupling must be to $H_d$, so we need a separate set of flavons, $\chi$, $\bar{\chi}$, to generate down-type Yukawa couplings. 

The analysis above shows that for the model to be viable, we need both $\phi$ and $\chi$ to get vevs. Because of the structure of the potential, this turns out to be impossible without introducing additional flavons. Consider the potential for one set of $\phi$, $\bar\phi$, $\chi$, $\bar\chi$, i.e.~ignoring the fact that the flavons are doublets in the full model. The potential is given by 
\be
V &=& \mph |\phi|^2 + \mpb |\bphi|^2 + \mc |\chi|^2 + \mcb |\bchi|^2 - (\bp \phi \bar{\phi} + c. c.) - (\bc \chi \bar{\chi} + c. c.) \nonumber \\
&& +\frac{g_F^2}{2} \left( |\phi|^2- |\bar \phi|^2 -3 |\chi|^2 + 3 |\bar \chi|^2\right)^2 \,.
\ee
The $m_i^2$ are real, and we can do field redefinitions so that the $b$'s are positive and the vevs are real and positive. In the supersymmetric limit, $m^2 = |\mu|^2 > 0$ and $b=0$, so spontaneous $U(1)_F$ breaking is impossible. Once SUSY breaking effects are included, the soft masses can be negative and a $b$-term can be generated, so SUSY breaking can trigger $U(1)_F$ breaking. 

We minimize the potential and get the following conditions:
\begin{eqnarray}
2 \mph \phi - \bp \bphi + 2 \phi D  = 0 \qquad\qquad
2 \mc \chi - \bp \bchi -6 \chi D  = 0 \nonumber\\
2 \mpb \bphi - \bp \phi - 2 \bphi D  = 0 \qquad \qquad
2 \mcb \bchi - \bp \chi +6 \bchi D  = 0,
\end{eqnarray}

\noindent where $D = g_F^2 \left( \phi^2- \bar \phi^2 -3 \chi^2 + 3 \bar \chi^2\right)$ is the $D$-term. Taking linear combinations of the left and right equations such that $D$ cancels out gives us quadratic equations involving only the $\phi$'s or the $\chi$'s:
\be
\bphi^2  - \frac{2}{\bp} \left(\mph +\mpb \right) \phi \bphi + \phi^2 = 0 \qquad \qquad
\bchi^2  - \frac{2}{\bc} \left(\mc +\mcb \right) \chi \bchi + \chi^2 = 0 ,
\ee

\noindent which can be solved for the barred fields in terms of the unbarred ones, $\bphi = r_\phi \phi$, $\bchi = r_\chi \chi$, where $r_\phi$ and $r_\chi$ depend only on the parameters of the potential and not on the fields. Plugging this back into the minimization conditions, we have
\be
(2 \mph - \bp r_\phi + 2 D) \phi = 0& \qquad \qquad
(2 \mc - \bc r_\chi - 6 D) \chi = 0 \nonumber\\
(2 \mpb r_\phi - \bp  - 2 r_\phi D) \phi = 0& \qquad\qquad
(2 \mcb r_\chi - \bc  + 6 r_\chi D) \chi = 0 .
\ee
Since we need nonzero vevs for both the $\phi$'s and $\chi$'s, the expressions in parentheses must all simultaneously be zero. 

We can get a new constraint by taking a linear combination of the first and third equations that eliminates $D$, 
\be
(6 \mph - 3 \bp r_\phi +2 \mc - \bc r_\chi) \phi \chi = 0.
\ee
\noindent Since the expression in parentheses is a function only of parameters and generically does not vanish, we are required to take either $\phi = 0$ or $\chi = 0$. Hence we conclude that with only the $D$-term quartic, either the $\phi$'s or the $\chi$'s will get vevs, but not both. In the case of the full field content where $\phi$ and $\chi$ are doublets, there is less analytic control, but the conclusion still holds and either $\phi$ or $\chi$ will be stable at the origin. 

In order to generate more vevs, we need another potential term that will provide a quartic, so we must introduce another flavon pair, $\xi, \bar \xi$. Choosing the $U(1)_F$ charge of $\xi$ to be $-2$, we can write down the superpotential operators given in Eq.~\eqref{eq:pureflavonsup}, which give the following scalar potential
\begin{eqnarray}
V_F = \lx^2 \left(\phi^2 \bchi^2 + \xi^2 \bchi^2 + \xi^2 \phi^2 + \text{ un-barred} \leftrightarrow \text{ barred} \right) \\
+ 2 \mu \, \lx (\bxi \phi \bchi +\xi \bchi \bphi +\xi \phi \chi + \text{ un-barred} \leftrightarrow \text{ barred} ) , \nonumber
\end{eqnarray}
where for notational simplicity, we have taken a common supersymmetric mass $\mu$ for all the flavons (and continue to assume that all parameters are real). The potential is now quite complicated, but there are generic regions in parameter space where all flat directions are stabilized because of the extra quartic and the origin for all fields is destabilized. Numerical study confirms that generically, if one field gets a vev, then all of them will. 

In the above discussion, we chose the charge of $\xi$ to be $-2$. This turns out to be the unique choice for a viable model. In order to understand this, we examine the most general $U(1)$ that arises from our democratic treatment of SM fields and that is allowed by the flavon couplings required to obtain tree-level Yukawa couplings. We can parametrize the charges of the SM  fields under this U(1) as $X^{SM}_{10} = a$, $X^{SM}_{\bar 5} = b$, and $X_{H_u} = - X_{H_d} = c$. The messenger-Higgs couplings then imply messenger charges of $X^{\rm mess.}_{10} = -c/2$ and $X^{\rm mess.}_{\bar 5} = 3c/2$. These then fix the flavon charges to be $X_{\phi} = -c/2 - a$, and $X_{\chi} = 3 c /2 - b$, so there are three independent $U(1)$ symmetries which allow the Yukawa couplings and mass terms of the theory. Our $U(1)_F$ flavor symmetry corresponds to the case $a = b = 0$. A second independent $U(1)$ can be parametrized by $a = c = 0$, under which $X_{\chi} = -b$ based on the above. Since generically the flavons all get vevs, this global $U(1)$ would be spontaneously broken and would yield a highly problematic massless Goldstone. Rounding out the basis of $U(1)$'s is one under which the flavons are uncharged and is therefore unbroken. Demanding that the flavons are uncharged leads to the conditions $b = -3 a$ and $c = -2a$; this charge assignment is related by a hypercharge rotation to $B-L$, and remains unbroken. 

In order to get a viable spectrum with no Goldstone bosons, we need to explicitly break the second $U(1)$ while leaving $B-L$ and $U(1)_F$ unbroken. By adding an additional vectorlike flavon pair, we gain an additional unconstrained charge. Therefore, in order to break the second $U(1)$, we must assign $U(1)_F$ charges to the new flavon such that two different types of interactions can be written down for the new flavon so that no charge assignment under the second $U(1)$ will be consistent. This uniquely determines the $U(1)_F$ charge to be $\mp2$ because that is the only charge that allows us to write both $\phi\phi\xi$ and $\bar{\phi}\bar{\chi}\xi$. These are the interactions found in Eq.~\eqref{eq:pureflavonsup}, which are needed for the dominant loop contribution to the 2$^{\rm nd}$ generation masses. In particular, the interaction with two $\phi$'s is needed for the charm mass, and the one with $\phi$ and $\chi$ is needed for the strange and muon mass. Therefore, we see that the charge assignment we have chosen for the flavons is not only necessary to get a viable spectrum of flavons, it is also crucial for generating the correct Standard Model Yukawa couplings.

%%%%%%%%%%%%%%%%%%%%%%
\section{Constructing the Yukawa Matrices}
\label{app:loop}
%%%%%%%%%%%%%%%%%%%%%%

\subsection{Radiative Yukawa Generation}
\label{app:ylist}

\noindent In this appendix we give the formulas for the loop effects used to generate the SM Yukawa matrices.  The ``33'' elements are generated at tree-level and are given by $y_t$ in Eq.~\eqref{eq:yt} and $y_b$ in Eq.~\eqref{eq:yb}, and we do not consider loop-level shifts to $y_{33}^{u,d}$ because they are below our numerical precision.    We now proceed to fill out the remainder of the Yukawa matrices with radiative contributions that we list in order of decreasing size. \\
\\
\indent \textbf{a. Flavino/Flavon} \\
\\
The dominant contribution to $y_{ij}$ for $(i,j) = 2,3$  comes from the flavino loop, which effectively sets the size of the 2$^{\rm nd}$ generation masses and is illustrated on the left-hand-side of Fig.~\ref{fig:charm}. In the up sector
\beq
y^u_{ij} = \lf r^u_i r^q_j \, m_{\phit_k} \widetilde{F}^{u}_{ik} \widetilde{F}^{q}_{jk} \Gamma^{{5l}^*}_{\qt} \Gamma^{{5m}^*}_{\ut} H^{QU}_{lm} \mathcal{F}(m_{\phit_k}, \,m_{\qt_l},\, m_{\ut_m}),
\label{eq:finolpup}
\eeq
\beq
\mathcal{F}(m_1, \,m_2,\, m_3) = \frac{m_1^2 \left(m_2^2 \log \frac{m_1^2}{m_2^2} - m_3^2 \log \frac{m_1^2}{m_3^2}\right) + m_2^2 m_3^2 \log \frac{m_2^2}{m_3^2}}{\left(m_1^2-m_2^2\right) \left(m_1^2-m_3^2\right)
\left(m_2^2-m_3^2\right)} .
 \label{eq:finofct}
\eeq
Here $r$ accounts for the fact that the original quark fields appearing at the vertices to which the external lines connect might not be mass eigenstates, for example 
\beq
r^q = (1,1, c_q, -s_q),
\label{eq:fermionrq}
\eeq
where the first two components are trivial because the first two generations do not mix with messengers and the last two are the $q_3'$ projections of the gauge eigenstates $q_3$ and $Q$, respectively, and similarly for $u$.  In Eq.~\eqref{eq:finolpup}, $i,j$ simply take values 2-3, and thus here we only need the first three components of $r^{q,u}$, but we present the complete vector which will be used below. $\widetilde{F}_{ij}$ is the coupling of the $i^{th}$ quark flavor to the $j^{th}$ flavino mass eigenstate, for example
\beq
\widetilde{F}^{q}_{ij} = F^q_{ik} \Gamma^{kj}_{\Phit}, \, k =1,2 ~,
\label{eq:qino}
\eeq
and we use a similar definition for $\tilde{\bar{f}}_{k}$ in the down sector.  The factor $H^{QU}_{ij}$ is the coupling of the $i^{th}$ $\qt$ and $j^{th}$ $\ut$ mass eigenstates to the light Higgs, which arises from summing over all six triple scalar couplings involving $Q,U$ smessengers and Higgses:
\be
H^{QU}_{ij} &=& -\mu_H \left(\lamu \cos \beta \, \Gamma^{4i}_{\qt} \Gamma^{4j}_{\ut} - \lamub^* \sin \beta \,\Gamma^{5i}_{\qt} \Gamma^{5j}_{\ut} \right) + \mu_Q \left(\lamu \sin \beta \, \Gamma^{5i}_{\qt} \Gamma^{4j}_{\ut} -\lamub^* \cos \beta \, \Gamma^{4i}_{\qt} \Gamma^{5j}_{\ut} \right) \nonumber \\
&& + \, \mu_U \left(\lamu \sin \beta \,  \Gamma^{4i}_{\qt} \Gamma^{5j}_{\ut} - \lamub^* \cos \beta \, \Gamma^{5i}_{\qt} \Gamma^{4j}_{\ut} \right).
\label{eq:sQUhiggs}
\ee

Because the fermion diagonalization is more involved, the down sector actually has two types of flavino diagrams. The first is the analogue of the up sector diagram:
\beq
y^d_{ij} = \lf S_{in} r^d_n r^q_j \, m_{\phit_k} \widetilde{F}^{d}_{nk} \widetilde{F}^{q}_{jk}  \Gamma^{{5l}^*}_{\qt} \Gamma^{{7m}^*}_{\dt} H^{QD}_{lm} \mathcal{F}(m_{\phit_k}, \,m_{\qt_l},\, m_{\dt_m}),
\label{eq:finolpdoa}
\eeq
where $l=1-5$ and $m =1-7$, 
\be
H^{QD}_{ij} &=& \mu_H \left(\lamd \sin \beta \, \Gamma^{5i}_{\qt} \Gamma^{5j}_{\dt} + \lamdb^* \cos \beta \,\Gamma^{5i}_{\qt} \Gamma^{7j}_{\dt} \right) - \mu_Q \left(\lamd \cos \beta \, \Gamma^{5i}_{\qt} \Gamma^{7j}_{\dt} +\lamdb^* \sin \beta \, \Gamma^{5i}_{\qt} \Gamma^{7j}_{\dt} \right) \nonumber \\
&& - \, \mu_D \left(\lamd \cos \beta \,  \Gamma^{5i}_{\qt} \Gamma^{7j}_{\dt} + \lamdb^* \sin \beta \, \Gamma^{5i}_{\qt} \Gamma^{7j}_{\dt} \right),
\label{eq:sQDhiggs}
\ee
\beq
r^d = (1,1, \Gamma^{11}_d, \Gamma^{21}_d, \Gamma^{31}_d).
\label{eq:fermionrd}
\eeq
The matrix $S_{ij}$ sums the relevant vertex over all possible gauge eigenstate fermion fields $j$ overlapping with mass eigenstate field $i
$ 
\be
S_{ij} = 
\begin{cases}
\delta_{ij}, & i \neq 3 \\
0, & i = 3, \, j < 3 \\ 
1, & i = 3, \, j \geq 3
\end{cases}
\label{eq:vertsum}
\ee
One could extend the definitions of $r^d$ and $S_{ij}$ to include the higher generation mass eigenstates ($d_4',D'$), but we do not consider the effects of mixing from these states, as we found them to be 1\% effects. Furthermore, the renormalization of the heavy fermion masses themselves is beyond our scope.  In Eq.~\eqref{eq:finolpdoa}, we take $n =1-4$. Since we also have the coupling $\bar{d} D \bar{\chi}$ in the down sector, there is also an additional contribution for $i = 3$, since we can take the $d_3'$ component of $D$:
\beq
y^d_{3j} = \lf  r^d_5 r^q_j  \, m_{\phit_k} \widetilde{F}^{q}_{jk} \tilde{\bar{f}}_{k} \, \Gamma^{{5l}^*}_{\qt} \Gamma^{{6m}^*}_{\dt} H^{QD}_{lm} \mathcal{F}(m_{\phit_k}, \,m_{\qt_l},\, m_{\dt_m}).
\label{eq:finolpdob}
\eeq

Next we consider the contributions from flavon loops; an example appears on the right-hand-side of Fig.~\ref{fig:charm}. There are two classes of flavon diagrams: one involves a mass insertion on each messenger line and the other does not. We begin with the mass-insertion type, and again start with the simpler up sector:
\be
y^u_{ij} &=& \frac{\lamu M_{Q'} M_{U'}}{32 \pi^2} c_u  c_q \, r^u_i r^q_j  \nonumber \\
 && \times  \left(\widehat{F}^{(u)Re}_{ik} + i \widehat{F}^{(u)Im}_{ik} \right) \left(\widehat{F}^{(q)Re}_{jk} + i \widehat{F}^{(q)Im}_{jk} \right)\mathcal{F}(m_{\phi_k}, \,M_{Q'},\, M_{U'}),
\label{eq:fonDlpup}
\ee 
where $M_{Q'}$, $M_{U'}$ are the messenger mass eigenvalues, i.e. $M_{Q'} = \sqrt{\mu_Q^2 + |F^q_{3i}\, \phi_i|^2}$ and similarly for $U'$, and $\widehat{F}^{Re}_{ij}$, $\widehat{F}^{Im}_{ij}$ is the coupling of the $i^{th}$ quark flavor to the $j^{th}$ flavon mass eigenstate, obtained by summing over the real and imaginary parts of the appropriate gauge eigenstate flavon doublet, respectively; for example
\begin{eqnarray}
\widehat{F}^{(q)Re}_{ij} &=& F^q_{ik} \Gamma^{kj}_{\Phi}, \, k = 1,2 \nonumber \\ 
\widehat{F}^{(q)Im}_{ij} &=& F^q_{ik} \Gamma^{kj}_{\Phi}, \, k = 11,12
\label{eq:qon}
\end{eqnarray}
and similarly for $u,\,d$, and $\hat{\bar f}_j$ in the down sector. The down sector again has two types of diagrams that correspond to this, with the first exactly analogous:
\be
y^d_{ij} &=& -\frac{\lamd M_{Q'} M^l_{D'}}{32 \pi^2} S_{im} r^d_m r^q_j  \, \Gamma^{3, l+1}_d \Gamma^{2, l}_{\bar{d}} c_q  \nonumber \\
 && \times \left(\widehat{F}^{(d)Re}_{mk} + i \widehat{F}^{(d)Im}_{mk} \right) \left(\widehat{F}^{(q)Re}_{jk} + i \widehat{F}^{(q)Im}_{jk} \right)  \mathcal{F}(m_{\phi_k}, \,M_{Q'},\, M^l_{D'}),
\label{eq:fonDlpdoa}
\ee
where the sum over $l = 1,2$ accounts for both heavy $d$-type fermions, and $m =1-4$. The second type again involves the additional $\bar{f}$ coupling:
\be
y^d_{3i} &=& -\frac{\lamd M_{Q'} M^k_{D'}}{32 \pi^2} r^d_5 r^q_i \, \Gamma^{3, k+1}_d \Gamma^{1, k}_{\bar{d}} c_q  \nonumber \\
 && \times \left(\widehat{F}^{(q)Re}_{ij} + i \widehat{F}^{(q)Im}_{ij} \right) \left(\hat{\bar f}^{Re}_{j} + i \hat{\bar f}^{Im}_{j} \right) \mathcal{F}(m_{\phi_j}, \,M_{Q'},\, M^k_{D'}),
\label{eq:fonDlpdob}
\ee

The second class of flavon diagrams does not have mass insertions, so these involve the $\lamub$ and $\lamdb$ couplings. Although the loop integral is formally divergent, for entries other than $y_{33}^{u,d}$ we need only consider the resulting finite log terms that are a function of the flavon masses. This is because the $F$-coupling piece vanishes when summed over all flavon mass eigenstates for $(i,j) \neq (3,3)$, a consequence of the fact that the first two generations do not couple to the Higgs at tree level. For the up sector
\beq
y^u_{ij} = -\frac{\lamub^*}{16 \pi^2} r^u_i r^q_j \cos \beta \left(\widehat{F}^{(u)Re}_{ik} + i \widehat{F}^{(u)Im}_{ik} \right) \left(\widehat{F}^{(q)Re}_{jk} + i \widehat{F}^{(q)Im}_{jk} \right)  \mathcal{G}(M_{Q'},\, M_{U'}, \,m_{\phi_k}),
\label{eq:fonlpup}
\eeq
\beq
\mathcal{G}(\mu, \,M,\, m) = \frac{m^4 \left(M^2-\mu ^2\right) \log \left(\frac{m^2}{\mu ^2}\right)-M^4 \left(m^2-\mu ^2\right) \log \left(\frac{M^2}{\mu ^2}\right)}{2\left(m^2-\mu ^2\right) \left(m^2-M^2\right) \left(M^2-\mu^2 \right)} .
\label{eq:fonfct}
\eeq
The analogous down sector formula is 
\beq
y^d_{ij} = -\frac{\lamdb^*}{16 \pi^2} S_{im} r^d_m r^q_j \sin \beta \, \Gamma^{2, l}_{\bar d} \Gamma^{{2, l}^*}_{\bar{d}} \left(\widehat{F}^{(d)Re}_{mk} + i \widehat{F}^{(d)Im}_{mk} \right) \left(\widehat{F}^{(q)Re}_{jk} + i \widehat{F}^{(q)Im}_{jk} \right)  \mathcal{G}(M_{Q'},\, M^l_{D'}, \,m_{\phi_k})
\label{eq:fonlpdoa}
\eeq
where $m =1-4$, and the diagram generated from the $\bar f$ coupling is
\beq
y^d_{3i} = -\frac{\lamdb^*}{16 \pi^2}  r^d_5 r^q_i \sin \beta \, \Gamma^{1, k}_{\bar d} \Gamma^{{2, k}^*}_{\bar{d}} \left(\widehat{F}^{(q)Re}_{ij} + i \widehat{F}^{(q)Im}_{ij} \right) \left(\hat{f}^{Re}_{j} + i \hat{f}^{Im}_{j} \right) \mathcal{G}(M_{Q'},\, M^k_{D'}, \,m_{\phi_j})
\label{eq:fonlpdob}
\eeq
\\
\indent \textbf{b. Gluino} \\
\\
The gluino loop, appearing in Fig.~\ref{fig:up}, is the dominant contribution to the entries in the first row and column of the Yukawa matrices, and controls the size of the 1$^{\rm st}$ generation mass. For the up sector,
\beq
y^u_{ij} = -\frac{\alpha_3}{2 \pi} S_{im} r^u_m S_{jn} r^q_n \Gamma^{mk}_{\ut} \Gamma^{nl}_{\qt} H^{QU}_{kl}  \mathcal{F}(m_{\tilde{g}},\, m_{\ut_k}, \,m_{\qt_l}),
\label{eq:ginolpup}
\eeq
where $m, n = 1-4$. For the down sector,
\beq
y^d_{ij} = -\frac{\alpha_3}{2 \pi} S_{im} r^d_m S_{jn} r^q_n \Gamma^{mk}_{\dt} \Gamma^{nl}_{\qt} H^{QD}_{kl}  \mathcal{F}(m_{\tilde{g}},\, m_{\dt_k}, \,m_{\qt_l}),
\label{eq:ginolpdo}
\eeq
where $m = 1-5$ and $n = 1-4$. There are also analogous diagrams with a bino. 

Approximating the threshold correction from messengers to the gaugino masses as a mass insertion is not strictly accurate at a PeV because other fields in this diagram are at that scale.  A more precise way to calculate would be be to blow up the mass insertion and include the two-loop effects of the messengers. This can be improved further still by resumming the effects of the thresholds that generate the gluino mass, and then including the full momentum dependence in the gluino propagator. We find that these effects only modify the contribution to the Yukawa coupling by $\mathcal{O}(1\%)$, so for computing our benchmark we use the simpler analytic formulas above. 
\\
\\
\indent \textbf{c. Gaugino-Higgsino} \\
\\
These diagrams, shown in Fig.~\ref{fig:GauginoHiggsino}, have already been discussed in Sec.~\ref{subsubsec:diag}. Here, we just present the complete formulas:
\beqa
y^u_{i3} &=&  \frac{\alpha_Y}{\pi} Q^Y_u Q^Y_{H_u} \, \lamu s_q \, S_{ik} \, r^u_k \, \Gamma^{4j}_{\ut} \, \Gamma^{{kj}^*}_{\ut}\, \mathcal{G}(\mu_H,\, m_{\tilde{B}}, \, m_{\ut_j}), \, k = 1 \textendash 4 \nonumber \\
y^d_{i3} &=&  -\frac{\alpha_Y}{\pi} Q^Y_d Q^Y_{H_d} \, \lamd s_q \, S_{ik} \, r^d_k \, \Gamma^{5j}_{\dt} \, \Gamma^{{kj}^*}_{\dt}\,  \mathcal{G}(\mu_H,\, m_{\tilde{B}}, \, m_{\dt_j}), \, k = 1\textendash5 \nonumber \\
y^u_{3i} &=& \lamu s_u \, S_{ik} \, r^q_k  \, \Gamma^{4j}_{\qt}\, \Gamma^{{kj}^*}_{\qt} \left[\frac{\alpha_Y}{\pi} Q^Y_q Q^Y_{H_u} \, \mathcal{G}(\mu_H,\, m_{\tilde{B}}, \, m_{\qt_j}) - \frac{\alpha_2}{\pi} C_F \, \mathcal{G}(\mu_H,\, m_{\tilde{W}}, \, m_{\qt_j})\right], \nonumber \\
y^d_{3i} &=&  \lamd \Gamma^{21}_{d} \, S_{ik} \, r^q_k \, \Gamma^{4j}_{\qt} \, \Gamma^{{kj}^*}_{\qt} \left[-\frac{\alpha_Y}{\pi} Q^Y_q Q^Y_{H_d} \, \mathcal{G}(\mu_H,\, m_{\tilde{B}}, \, m_{\qt_j}) \right. \nonumber \\
&& \left. + \frac{\alpha_2}{\pi} C_F \, \mathcal{G}(\mu_H,\, m_{\tilde{W}}, \, m_{\qt_j})\right],
\label{ghlpd3i}
\eeqa
where $k = 1\textendash4$ in the last two equations above, $Q^Y$ is the field hypercharge, and $C_F$ is the quadratic Casimir for the group, which is equal to $3/4$ and $4/3$ for $SU(2)_L$ and $SU(3)_C$, respectively. 

\subsection{Wavefunction Renormalization}
\label{app:wfrnlist}
\noindent Our procedure is outlined in Sec.~\ref{subsubsec:wfrn}, where the generic diagram appears and the contribution from the flavon sector is discussed. There are also loops involving gluino/squarks, Higgsino/squarks, and Higgs/quarks. We include all of these in the renormalization of the $q$ and $u$ fields, but retain only the gluino contribution for $d$, since the other loops are $y_b$ suppressed, or involve kinetic mixing, which we found to be an $\mo(1\%)$ effect on our SM model prediction, and thus neglect. 
In addition, since only the 3$^{\rm rd}$ generation couples to the Higgs at tree level, the diagrams involving the Higgs multiplet only contribute to the ``33'' entries. Unlike the contributions to $y_{33}^{u,d}$, we do include wavefunction renormalization of the 3$^{\rm rd}$ generation, as the K{\"a}ll{\'e}n-Lehmann representation theorem along with the positivity of quantum mechanics determines that all such contributions will increase the 3$^{\rm rd}$ generation masses \cite{Weinberg:1995mt}.  Taken together, we find shifts upwards of 10\%, and thus we cannot neglect them. We use the notation introduced in the previous section and take a renormalization scale $Q = 1000$ TeV,  the common scale at which the heavy states are integrated out. \\
\\
 \indent \textbf{a. Gluino}

\beq
\Sigma^q_{ij} = \frac{\alpha_3}{2\pi} C_F S_{il} r^{q^*}_l S_{jm} r^{q}_m \Gamma^{{lk}}_{\qt} \Gamma^{{mk}^*}_{\qt} \mathcal{W}(m_{\tilde{g}}, \, m_{\qt_k}, \, Q), \, l,m = 1\textendash4,
\label{eq:gwf}
\eeq

\beq
\mathcal{W}(M, \, m, \, Q) = \frac{M^4 \log \left(\frac{M^2}{Q^2} \right)}{2(M^2 - m^2)^2} + \frac{m^2 \,(m^2 - 2 M^2) \log \left(\frac{m^2}{Q^2} \right)}{2(M^2 - m^2)^2} + \frac{m^2-3M^2}{4(M^2 - m^2)} .
\label{eq:wfrnloop}
\eeq
Analogous formulas hold for $u$ and $d$; in the case of $d$, we take $l,m = 1-5$. There is also a wino contribution to $q$.\footnote{We neglect the contribution from the $U(1)_F$ gauge supermultiplet, which only changes the ``33'' entry from the overlap of the 3$^{\rm rd}$ generation massless eigenstate with the messenger gauge eigenstate.  This is expected to be a $\mo(1\%)$-level effect.} \\
\\
\indent \textbf{b. Flavino}

\beq
\Sigma^q_{ij} = \lf r^{q^*}_i r^q_j \widetilde{F}^{q^*}_{ik} \widetilde{F}^{q}_{jk} \Gamma^{{5l}^*}_{\qt} \Gamma^{5l}_{\qt}  \mathcal{W}(m_{\phit_k}, \,m_{\qt_l},\, Q),
\label{eq:finowf}
\eeq
and similarly for $u$. \\

\indent \textbf{c. Flavon}

\beq
\Sigma^q_{ij} = \frac{1}{32 \pi^2} r^{q^*}_i r^q_j \left(\widehat{F}^{{(q)Re}}_{ik} + i \widehat{F}^{{(q)Im}}_{ik} \right)^* \left(\widehat{F}^{(q)Re}_{jk} + i \widehat{F}^{(q)Im}_{jk} \right)  \mathcal{W}(M_{Q'},\, m_{\phi_k}, \, Q),
\label{eq:fonwf}
\eeq
and analogously for $u$. \\

\indent \textbf{d. Higgsino}

\beq
\Sigma^q_{33} = \frac{|s_q|^2}{16 \pi^2} \left(|\lamu|^2 \Gamma^{{4i}^*}_{\ut} \Gamma^{4i}_{\ut} \mathcal{W}(\mu_H,\, m_{\ut_i}, \, Q) + |\lamd|^2 \Gamma^{{5i}^*}_{\dt} \Gamma^{5i}_{\dt} \mathcal{W}(\mu_H,\, m_{\dt_i}, \, Q)\right),
\label{eq:hinowf}
\eeq
where the first term arises from putting $\tilde{H}_u$ in the loop and the second has $\tilde{H}_d$. The contribution to $\Sigma^u_{33}$ consists of only the $\tilde{H}_u$ piece and is obtained by taking $\ut \to \qt$. \\

\indent \textbf{e. Higgs} \\
\\
Since our renormalization scale is far above the scale of electroweak symmetry breaking, $SU(2)$ is approximately unbroken. Therefore, the tuning in the Higgs sector produces one light doublet that includes the SM Higgs, and one heavy doublet with degenerate scalars of mass $m_A$. In other words, the two Higgs doublets are in the extreme decoupling limit of the MSSM. 
The light doublet is given by $H_1 = -\cos \beta (i \sigma_2) H_d^* + \sin \beta H_u$, and the heavy doublet by $H_2 = \sin \beta (i \sigma_2) H_d^* + \cos \beta H_u$. The effects here describe loops with at least one heavy field (light Higgs or SM fermion masses are approximated to be 0 in the calculation), with purely light field effects taken into account in the RG evolution of the Yukawas from the high scale down to the weak scale. 

The effects of a heavy Higgs and fermion loop are given by
\be
\Sigma^q_{33}  &=& \frac{| s_q|^2}{16 \pi^2} \left[ |\lamu|^2 \cos^2 \beta \left(|s_u|^2 \, \mathcal{W}(0,\, m_A, \, Q) + c^2_u \mathcal{W}(M_{U'},\, m_A, \, Q) \right) \right. \nonumber \\
&& \left. {} + |\lamd|^2 \sin^2 \beta \left(\Gamma^{31}_{d} \Gamma^{{31}^*}_{d}  \mathcal{W}(0,\, m_A, \, Q) + \Gamma^{3i}_{d} \Gamma^{{3i}^*}_{d} \mathcal{W}(M^i_{D'},\, m_A, \, Q) \right) \right] \nonumber \\
\Sigma^{u}_{33}  &=& \frac{|\lamu|^2}{16 \pi^2}  |s_u|^2 \cos^2 \beta \left[|s_q|^2 \, \mathcal{W}(0,\, m_A, \, Q) + c^2_q \, \mathcal{W}(M_{Q'},\, m_A, \, Q)\right] ,
\label{eq:uhhwf}
\ee
while the effects of a light Higgs and a heavy messenger are given by
\be
\Sigma^q_{33} &=& \frac{|s_q|^2}{16 \pi^2} \left[ |\lamu|^2 \sin^2 \beta \, c^2_u \mathcal{W}(M_{U'},\, 0, \, Q) + |\lamd|^2 \cos^2 \beta \, \Gamma^{3i}_{d} \Gamma^{{3i}^*}_{d} \mathcal{W}(M^i_{D'},\, 0, \, Q) \right] \nonumber \\
\Sigma^{u}_{33}  &=& \frac{|\lamu|^2}{16 \pi^2} |s_u|^2 \, c^2_q  \, \sin^2 \beta  \, \mathcal{W}(M_{Q'},\, 0, \, Q).
\label{eq:uhlwf}
\ee

%%%
\section{Messenger Threshold Corrections to Gaugino Masses}
\label{app:mess}
%%%

Here we generalize the discussion of Sec.~\ref{sec:gaugino} to fully account for mixing. We organize the bookkeeping by introducing tensors $V^M$ and $W^{\bar M}$, which characterize the possible vertices. These are relatively simple for $Q$ and $U$, since $\bar Q$ and $\bar U$ are already mass eigenstates and there is only one nonzero eigenvalue:
\be
V^Q &=& (\Gamma^{{3 i}^*}_{\qt} s_q^*, \Gamma^{{4 i}^*}_{\qt} c_q) \nonumber \\
W^{\bar Q} &=& \Gamma^{5 i}_{\qt},
\label{eq:VWQ}
\ee
and analogously for $U$. Here $V$ is a matrix, with the row denoting whether $q_3$ or $Q$ is at the vertex, and we need to project out the component corresponding to the heavy fermionic mass eigenstate. Although we label them by $D$, the tensors for the down type messengers actually combine what were previously separate contributions from the $d_4$, $\bar d$ and $D$, $\bar D$ pairs:
\be
V^D  &=& \left(\left(\Gamma^{{3 i}^*}_{\dt} \Gamma^{12}_d, \Gamma^{{3 i}^*}_{\dt} \Gamma^{13}_d \right), \left(\Gamma^{{4 i}^*}_{\dt} \Gamma^{22}_d, \Gamma^{{4 i}^*}_{\dt} \Gamma^{23}_d\right), \left(\Gamma^{{6 i}^*}_{\dt} \Gamma^{32}_d, \Gamma^{{6 i}^*}_{\dt} \Gamma^{33}_d \right)\right) \nonumber \\
W^{\bar D} &=&  \left(\left(\Gamma^{5 i}_{\dt} \Gamma^{11}_{\bar{d}}, \Gamma^{5 i}_{\dt} \Gamma^{12}_{\bar{d}}\right), \left(\Gamma^{7 i}_{\dt} \Gamma^{21}_{\bar{d}}, \Gamma^{7 i}_{\dt} \Gamma^{22}_{\bar{d}}\right)\right),
\label{eq:VWD}
\ee
where in $V^D_{ijk}$ and $W^{\bar D}_{ijk}$, $i$ specifies which field is at the vertex, $j$ labels the fermionic eigenstate, and $k$ labels the scalar eigenstate. 

The threshold corrections to the gaugino masses can thus be expressed as 
\beq
\Delta m^Q_{\tilde{i}} = \sum_{j = 1}^{2} \frac{\alpha_i}{\pi} C^Q_i V^Q_{jk} W^{\bar Q}_k M_{Q'} \mathcal{J}(M_{Q'}, \, m_{\qt_k}), \qquad \mathcal{J}(M, \, m) = \frac{m^2}{M^2 - m^2} \log \frac{m^2}{M^2},
\label{eq:Qthresh}
\eeq 
with $M_{Q'}$ the physical mass of the heavy messenger. There is an analogous expression for $U$, while
\beq
\Delta m^D_{\tilde{i}} = \sum_{i = 1}^{3} \sum_{l = 1}^{2} \frac{\alpha_i}{\pi} C^D_i V^{D}_{ijk} W^{\bar D}_{ljk} M^j_{D'} \mathcal{J}(M^j_{D'}, \, m_{\dt_k}).
\label{eq:Dthresh}
\eeq 
The Dynkin indices weighted by degrees of freedom $C^M_i$ are given by
\be
C^Q &=& (1/6, 3/2, 1) \nonumber \\
C^U &=& (4/3, 0, 1/2) \nonumber \\
C^D &=& (1/3, 0 ,1/2)
\label{eq:Dynquarks}
\ee
for $\left( U(1)_Y, SU(2)_L, SU(3)_C \right)$. Our extra matter was introduced in complete representations of $SU(5)$, so we still need to account for $E$, $l_4$, and $L$. Since our flavor model does not discuss the leptonic sector in any detail, here we simply assume that the parameters for $E$ are the same as those for $U$, and identify $l_4$ with $d_4$, as well as $L$ with $D$. Although they live in the same GUT multiplets, the $C$ factors are of course different:
\be
C^E &=& (1, 0, 0) \nonumber \\
C^L &=& (1/2, 1/2, 0). 
\label{eq:Dynleps}
\ee

\section{Formulas for Select Flavor Observables}
\label{app:flavobs}

\indent \textbf{1. $B_s$ mixing: Wilson Coefficient of $\mathcal{O}_1$} \\
\\
In addition to the usual box diagram MSSM contributions to meson mixing involving squarks and gluinos, our model gives a contribution coming from analogous box diagrams with flavons/messengers and flavinos/smessengers. Since the 1$^{\rm st}$ generation fermions do not couple to the flavon sector, this additional contribution only exists for mesons that do not contain 1$^{\rm st}$ generation quarks. The most precisely measured of these is $B_s$. We therefore calculate the box diagram for $B_s$ mixing and extract the coefficient of the effective operator $\mathcal{O}_1$, given in Eq.~\eqref{eq:o1}, which we then use to calculate the contribution to the mass splitting and $CP$-violating phase in the $B_s$ system. \\
\\
\indent \textbf{a. Flavino}  \\
\beq
C_1 = \frac{c^2_q}{128 \pi^2} \widetilde{F}^{q^*}_{2i} \widetilde{F}^{q}_{3i} \widetilde{F}^{q^*}_{2j} \widetilde{F}^{q}_{3j}  \Gamma^{5k}_{\qt} \Gamma^{{5k}^*}_{\qt} \Gamma^{5l}_{\qt} \Gamma^{{5l}^*}_{\qt}\mathcal{B}_1(m_{\phit_i}, \, m_{\phit_j}, \, m_{\qt_k}, \, m_{\qt_l}),
\label{eq:coneflav}
\eeq
\be
\mathcal{B}_1(M_1, \,M_2,\, m_1, \, m_2) &=& \frac{1}{(m_1^2-m_2^2)(m_1^2-M_1^2)(m_1^2-M_2^2)(m_2^2-M_1^2)(m_2^2-M_2^2)(M_1^2-M_2^2)} \nonumber \\
&& \times \left(m_1^4 m_2^4 (M_1^2 - M_2^2) \log \frac{m_1^2}{m_2^2} + m_1^4 M_1^4 (M_2^2 - m_2^2) \log \frac{m_1^2}{M_1^2} \right. \nonumber \\ 
&& \left. {} + m_1^4 M_2^4 (m_2^2-M_1^2) \log \frac{m_1^2}{M_2^2} + m_2^4 M_1^4 (m_1^2 - M_2^2) \log \frac{m_2^2}{M_1^2}  \right. \nonumber \\ 
&& \left. {} + m_2^4 M_2^4 (M_1^2-m_1^2) \log \frac{m_2^2}{M_2^2} + M_1^4 M_2^4 (m_1^2 - m_2^2) \log \frac{M_1^2}{M_2^2}\right) .
\ee
\indent \textbf{b. Flavon} \\
\\
\be
C_1 &=& \frac{c^2_q}{512 \pi^2} \left(\widehat{F}^{(q)Re}_{2i} + i \widehat{F}^{(q)Im}_{2i} \right)^* \left(\widehat{F}^{(q)Re}_{3i} + i \widehat{F}^{(q)Im}_{3i} \right) \nonumber \\
&& \times \left(\widehat{F}^{(q)Re}_{2j} + i \widehat{F}^{(q)Im}_{2j} \right)^* \left(\widehat{F}^{(q)Re}_{3j} + i \widehat{F}^{(q)Im}_{3j} \right) \mathcal{B}_2(m_{\phi_i}, \, m_{\phi_j}, \, M_{Q'}),
\label{eq:coneino}
\ee

\beq
\mathcal{B}_2(m_1, \, m_2, \, M) =  \frac{m_1^4 \log \frac{m_1^2}{M^2}}{(M^2-m_1^2)^2 (m_1^2-m_2^2)} - \frac{m_2^4 \log \frac{m_2^2}{M^2}}{(M^2-m_2^2)^2 (m_1^2-m_2^2)} + \frac{M^2}{(M^2-m_1^2)(M^2-m_2^2)} ,
\eeq
where the total Wilson coefficient for $\mathcal{O}_1$ is the sum of Eqs.~\eqref{eq:coneflav} and~\eqref{eq:coneino}. \\
\\
\indent \textbf{2. CEDM} \\
\\
\indent \textbf{a. Flavino}  \\
\\
Our model gives additional contributions to chromo-EDMs for 2$^{\rm nd}$ and 3$^{\rm rd}$ generation quarks. The diagrams involve flavinos and flavons ({\it cf.}~Fig.~\ref{fig:edm}) and are constructed from the ones shown in Fig.~\ref{fig:charm} by attaching a gluon line to a member of the messenger multiplet. The flavino loops are typically the dominant contribution, and they give:
\beq
\tilde{d}_{in} = \frac{v \sin \beta \, m_{\phit_j}}{16 \pi^2} r^q_i r^u_n Im \{ \widetilde{F}^{q}_{ij} \widetilde{F}^{u}_{nj}  \Gamma^{{5 k}^*}_{\qt} \Gamma^{{5 l}^*}_{\ut} H^{QU}_{kl} \} \left( \mathcal{D} (m_{\phit_j}, \, m_{\qt_k}, \, m_{\ut_l}) +  \mathcal{D} (m_{\phit_j}, \, m_{\ut_l}, \, m_{\qt_k}) \right),
\label{eq:cedm}
\eeq
where $v = 174$ GeV is the Higgs vev; the first term corresponds to gluon emission from the $\qt$ and the second from $\ut$, and
\beq
\mathcal{D} (M, \, \mu, \, m) = \frac{1}{2 M^4} \int_{0}^{1} dw \,w \int_{0}^{1-w} dx \int_{0}^{1-x-w} dy \left[ w + (x+y) \frac{\mu^2}{M^2} + (1-w-x-y) \frac{m^2}{M^2} \right]^{-2} .
\eeq
\\
\indent \textbf{b. Flavon}  \\
\\
We also get loops with flavons and smessengers. Just as for the loops generating Yukawa couplings, there are contributions with and without mass insertions on the messenger lines.  For the dipole calculation though, the latter are manifestly finite.
Those without mass insertions are typically larger, giving
\\
\be
\tilde{d}_{ij} &=& -\frac{v \cos \beta}{32 \pi^2} r^q_i r^u_j Im \left\{\lamub^* \left(\widehat{F}^{(q)Re}_{ik} + i \widehat{F}^{(q)Im}_{ik} \right) \left(\widehat{F}^{(u)Re}_{jk} + i \widehat{F}^{(u)Im}_{jk} \right)  \right\} \nonumber \\
&&\times \big(\mathcal{D}_{\phi}^{\rm no-MI}(M_{Q'},\, M_{U'}, \,m_{\phi_k}) + \mathcal{D}_{\phi}^{\rm no-MI}(M_{U'},\, M_{Q'}, \,m_{\phi_k}) \big),
\ee 
where 
\be
\mathcal{D}_{\phi}^{\rm no-MI} (M, \, \mu, \, m) &=& \frac{1}{2 \, m^2} \int_{0}^{1} dw \int_{0}^{1-w} dy \int_{0}^{1-y-w} dz \nonumber \\
&&\times \left( \frac{3(y+z)-1}{[(1-w-y-z) + (w+z) M^2/m^2 + y \, \mu^2/m^2]} \right. \nonumber \\ 
&& \left. - \frac{w M^2}{m^2 \, [(1-w-y-z) + (w+z) M^2/m^2 + y \, \mu^2/m^2]^2} \right) .
\ee
Lastly, we come to the flavon loop with mass insertions:
\be
\tilde{d}_{ij} &=& \frac{v \sin \beta \, M_{Q'} M_{U'}}{32 \pi^2} c_u  c_q \, r^q_i r^u_j Im \left\{\lamu \left(\widehat{F}^{(q)Re}_{ik} + i \widehat{F}^{(q)Im}_{ik} \right) \left(\widehat{F}^{(u)Re}_{jk} + i \widehat{F}^{(u)Im}_{jk} \right) \right\} \nonumber \\
&& \times \big(\mathcal{D}_{\phi}^{\rm MI}(M_{Q'},\, M_{U'}, \,m_{\phi_k}) + \mathcal{D}_{\phi}^{\rm MI}(M_{U'},\, M_{Q'}, \,m_{\phi_k}) \big),
\ee
with 
\be
\mathcal{D}_{\phi}^{\rm MI} (M, \, \mu, \, m) &=& \frac{1}{2 \, m^4}\int_{0}^{1} dw \int_{0}^{1-w} dy \int_{0}^{1-y-w} dz \nonumber \\
&&\times  \frac{(w+y+z)}{[(1-w-y-z) + (w+z) M^2/m^2 + y \, \mu^2/m^2]^2} .
\ee

%%%%%%%%%%%%%%
% References
%%%%%%%%%%%%%%

\bibliography{lit}

\end{document}